\documentclass[11pt]{article}

\usepackage{arxiv}

\usepackage{graphicx}
\usepackage{amsmath,amssymb,amsthm}
\usepackage[hidelinks]{hyperref}
\usepackage{xparse}
\usepackage{tikz}
\usetikzlibrary{arrows}
\usetikzlibrary{matrix,backgrounds}
\pgfdeclarelayer{myback}
\pgfsetlayers{myback,background,main}
\tikzset{mycolor/.style = {line width=1bp,color=#1}}
\tikzset{myfillcolor/.style = {draw,fill=#1,#1,rounded corners}}

\NewDocumentCommand{\fhighlight}{O{blue!40} m m}{%
\draw[myfillcolor=#1] (#2.north west)rectangle (#3.south east);
}
\NewDocumentCommand{\fhighlightL}{O{blue!40} m m}{%
\draw[myfillcolor=#1] (#2.south west)rectangle (#3.south east);
}

\newcommand{\nums}[1]{\sisetup{round-mode=places,round-precision=5,group-separator={}}\num{#1}}

\usepackage{siunitx}
\usepackage{mathtools}
\usepackage{subcaption}
\usepackage{enumitem}
\usepackage{xfrac}
\usepackage{booktabs}

\usepackage{xcolor}
\definecolor{azure}{rgb}{0.0, 0.5, 1.0}
\definecolor{asparagus}{rgb}{0.53, 0.66, 0.42}
\definecolor{cadetgrey}{rgb}{0.57, 0.64, 0.69}
\definecolor{awesome}{rgb}{1.0, 0.13, 0.32}

\def\intdomd{\Omega_\delta}
\def\uelast{\underline{u}}
\def\uperid{u}

\newcommand\edit[1]{{\color{black}{#1}}}

\graphicspath{{./Figures/}}

\newcommand{\orcid}[1]{\href{https://orcid.org/#1}{\includegraphics[height=10pt]{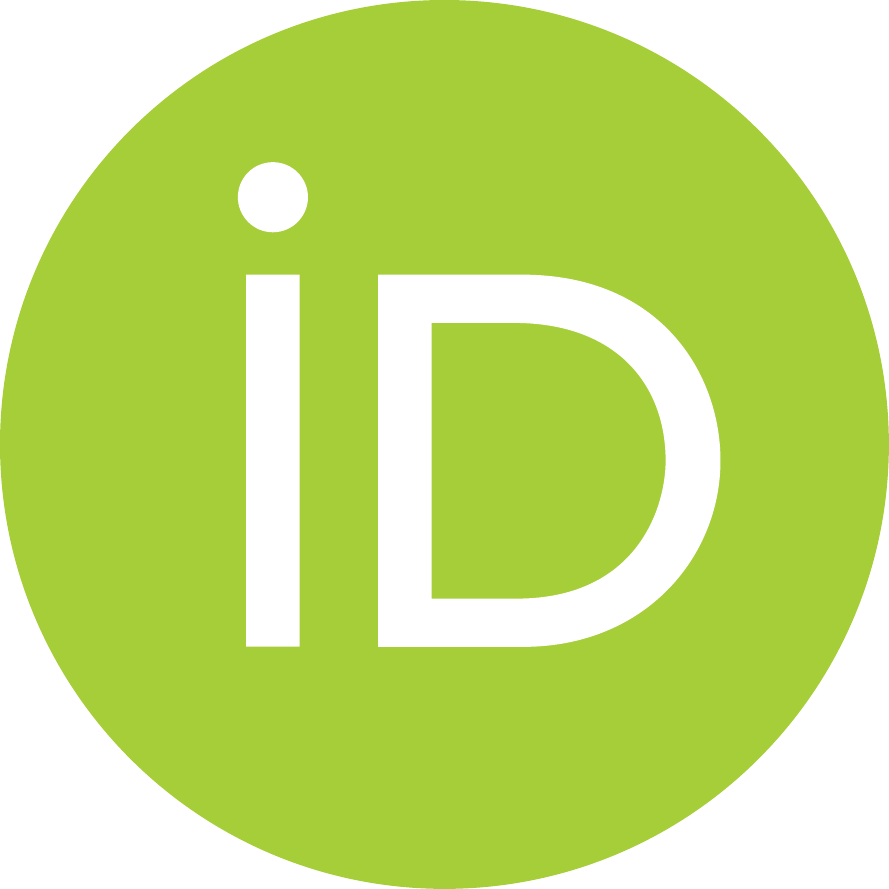}}}

\begin{document}





\title{Coupling approaches for classical linear elasticity and \edit{bond-based} peridynamic models}

\author{Patrick Diehl\orcid{0000-0003-3922-8419} \\ LSU Center for Computation and Technology, Louisiana State University, \\
Digital Media Center, 340 E.\ Parker Blvd, Baton Rouge, LA 70803, USA \\ \textit{pdiehl@cct.lsu.edu}  \AND Serge Prudhomme \\ Department of Mathematics and Industrial Engineering, Polytechnique Montr\'eal, \\
C.P. 6079, succ.\ Centre-ville, Montr\'eal, Qu\'ebec H3C~3A7, Canada \\
\textit{serge.prudhomme@polymtl.ca}}



\maketitle   

\begin{abstract}
Local-nonlocal coupling approaches provide a means to combine the computational efficiency of local models and the accuracy of nonlocal models. This paper studies the continuous and discrete formulations of three \edit{existing} approaches for the coupling of classical linear elasticity and bond-based peridynamic models, namely 1) a method that enforces matching displacements in an overlap region, 2) a variant that enforces a constraint on the stresses instead, and 3) a method that considers a variable horizon in the vicinity of the interfaces. The performance of the three coupling approaches is compared on a series of one-dimensional numerical examples that involve cubic and quartic manufactured solutions. Accuracy of the proposed methods is measured in terms of the difference between the solution to the coupling approach and the solution to the classical linear elasticity model, which can be viewed as a modeling error. \edit{The objective of the paper is to assess the quality and performance of the discrete formulation for this class of force-based coupling methods.}
\end{abstract}




\newpage

\section{Introduction}
\label{Sect:introduction}

There has been, in the past few years, a great interest for the development of local-nonlocal coupling methods in order to take advantage, on the one hand, of the computational efficiency of local models and, on the other hand, of the accuracy of nonlocal models. We actually refer to the recent survey~\cite{d2019review} for an overview and classification of generic local-nonlocal coupling methods. It is clear that the coupling of local and nonlocal models potentially have the following two benefits, namely the reduction of the computational cost in nonlocal modeling and a means to apply boundary conditions for nonlocal models. Indeed, nonlocal models such as peridynamics, molecular dynamics, or smoothed particle hydrodynamics, can be very computationally intensive, see e.g.~\cite{diehl2021comparative}. However, they are often needed only in small regions, as it is the case when simulating cracks in computational mechanics, suggesting that one can consider a less expensive local model in the remainder of the domain. Moreover, a recent review~\cite{Diehl2019} on the validation of peridynamics against actual experiments has revealed that one major challenge for using peridynamics lies in the treatment of the nonlocal boundary conditions~\cite{du2016nonlocal, madenci2018state, madenci2018weak, gu2018revisit, Prudhomme-Diehl-2020,you2020asymptotically,d2021prescription,d2020physically}. One can therefore imagine using local models all along the boundaries of the domain, for which boundary conditions can be unambiguously prescribed, and retaining the nonlocal models only in the interior of the domain by employing a coupling approach.

\edit{The paper focuses on approaches that couple the bond-based peridynamic model with the classical linear elasticity continuum model using a force-based coupling formulation. We refer again the reader to to~\cite{d2019review} for the description of other types of coupling approaches such as optimization-based or energy-based methods.} We adopt here, in order to establish the coupling methods, a deductive approach based on the coupling formulation of classical linear elasticity models in one dimension. This allows us to determine the minimal requirements when identifying the necessary constraints for matching the models. We thus \edit{identify the following} three coupling methods, namely a method that matches the displacement in an overlap region of size given by the horizon of the peridynamic model and that we refer to as MDCM, a variant that matches stresses instead and that we call MSCM, and a method that introduces a variable horizon and that we refer to as VHCM. The first coupling method, i.e.\ MDCM, is in fact similar to the majority of the existing coupling approaches available in the literature~\cite{doi:10.1080/15376494.2019.1602237, kilic2010coupling, madenci2018coupling, sun2019superposition, bie2018coupling, liu2012coupling, FANG201989, GALVANETTO201641, Zaccariotto-CMAME-2018, zaccariotto2017enhanced}. MSCM matches the stress field produced by the two models over a region of size defined by the horizon, as before, and features some similarities with the approach suggested in~\cite{silling2020Couplingstresses}, \edit{apart from the discretization of the stress at the coupling interface}. By contrast, VHCM, unlike MDCM and MSCM, avoids the introduction of an overlapping domain by scaling the horizon to zero when approaching the coupling interface, in a manner similar to~\cite{silling2015variable, NIKPAYAM2019308}. In all three cases, we formulate  first the coupling methods at the continuous level and then propose discrete formulations based on the finite differences method, for the classical linear elasticity model, and the collocation method~\cite{parks2008implementing}, for the peridynamic model. The advantages in doing so are twofold: first, to allow one to distinguish the discretization error from the modeling error and second, to be able to choose compatible numerical methods when discretizing the two models and the matching constraints. We note that coupling methods proposed in the literature are often introduced at the discrete level using the finite element method for the local model and a collocation approach for the nonlocal model. 

We assess and compare the performance of the three coupling approaches on several numerical experiments dealing with the simulation of a one-dimensional bar. We show that the three methods produce the same error-free solution, as expected, on problems involving manufactured cubic polynomial solutions. However, the numerical results provided by the three methods are different when the manufactured solution for the linear elasticity problem is given by a polynomial function of degree at least four. We then measure the accuracy of the proposed methods in terms of the modeling error, which is defined as the difference between the solution to the classical linear elasticity problem and the solution to each coupled problem. We then perform some $\delta$-convergence and $m$-convergence analyses in order to evaluate the potential of each coupling approach.

The paper is organized as follows. Section~\ref{Sect:modelproblem} introduces the model problem and some preliminaries. Section~\ref{Sect:CF-ElasticityModels} describes the continuous formulation of the coupling between classical linear elasticity models in one dimension. Section~\ref{Sect:CF-CouplingMethods} presents the three coupling approaches for the classical linear elasticity and peridynamic models at the continuous level. Section~\ref{Sect:discretization} provides the discrete formulations and detailed algorithms of the coupling methods based on the finite differences and collocation methods. Section~\ref{Sect:numericalexamples} describes the numerical examples including $\delta$-convergence and $m$-convergence studies. Finally, Section~\ref{Sect:conclusions} concludes the paper.

\section{Model problem and preliminaries}
\label{Sect:modelproblem}

The model problem will consist in studying the static equilibrium of a bar held fixed at one end and subjected to a longitudinal traction at the other end. We shall suppose here that the deformations in the bar are infinitesimally small and can be adequately described by the theory of linear elasticity. Moreover, we assume for simplicity that the bar has a unit cross-sectional area $A=1$. We present below the local model based on classical elasticity theory and the non-local model counterpart derived from the linearized bond-based peridynamic theory~\cite{Silling-JMPS-2000}.

\subsection{Classical linear elasticity model}
\label{Sect:CLEM}

Let \(\Omega = (0,\ell) \subset \mathbb R\) and \(\overline{\Omega}\) be the closure of \(\Omega\), i.e.\ \(\Omega=[0,\ell]\). The continuum local problem consists in finding the displacement \(\uelast\in\overline{\Omega}\) such that:
\begin{align}
\label{eq:1dlinearelasticity}
- E \uelast''(x) = f_b(x), &\quad \forall x \in \Omega, \\
\label{eq:Dirichlet}
\uelast(x) = 0, &\quad \text{at}\ x=0,\\
\label{eq:Neumann}
E\uelast'(x) = g, &\quad \text{at}\ x=\ell,
\end{align}
where \(E\) is the constant modulus of elasticity of the bar, \(f_b=f_b(x)\) is a scalar function describing the external body force density (per unit length), and \(g \in \mathbb R\) is the traction force applied at end point \(x=\ell\). 
We will work with the  mixed boundary value problem for most of the theoretical development to be as general as possible, but will also consider in the numerical examples some problems with homogeneous Dirichlet conditions at both ends, replacing the Neumann boundary condition~\eqref{eq:Neumann} by the Dirichlet boundary condition \(u(\ell)=0\). 
We also suppose that \(f_b\) is chosen sufficiently smooth so that regularity in the solution is not an issue when comparing the solutions from the two models. 
Note that we use the notation \(\uelast\) for the solution to the classical elasticity problem in order to emphasize that it may be different from the peridynamic solution \(\uperid\) introduced below.

\subsection{Peridynamic model}

\begin{figure}[tbp]
\centering
\small
\begin{tikzpicture}
\draw (0,0) -- (10.0,0);
\draw (0,-0.1) -- (0,0.1);
\draw (1,-0.1) -- (1,0.1);
\draw (2,-0.1) -- (2,0.1);
\draw (3.5,-0.1) -- (3.5,0.1);
\draw (4.5,-0.1) -- (4.5,0.1);
\draw (5.5,-0.1) -- (5.5,0.1);
\draw (8.0,-0.1) -- (8.0,0.1);
\draw (9.0,-0.1) -- (9.0,0.1);
\draw (10.0,-0.1) -- (10.0,0.1);
\node[above] at (0.0,-0.55) {$0$};
\node[above] at (1.0,-0.59) {$a-\delta$};
\node[above] at (2.0,-0.55) {$a$};
\node[above] at (3.5,-0.59) {$x-\delta$};
\node[above] at (4.5,-0.55) {$x$};
\node[above] at (5.5,-0.59) {$x+\delta$};
\node[above] at (8.0,-0.55) {$b$};
\node[above] at (9.0,-0.59) {$b+\delta$};
\node[above] at (10.0,-0.55) {$\ell$};
\draw[arrows=<->, >=stealth]  (0.0,1.05) -- (10.0,1.05);
\draw  (0.0,0.25) -- (0.0,1.15);
\draw  (10.0,0.25) -- (10.0,1.15);
\node[above] at (5.00,1.05) {$\Omega$};
\draw[arrows=<->, >=stealth]  (2.0,0.45) -- (8.0,0.45);
\draw  (2.0,0.25) -- (2.0,0.55);
\draw  (8.0,0.25) -- (8.0,0.55);
\node[above] at (5.00,0.45) {$\intdomd$};
\draw[arrows=<->, >=stealth]  (3.5,-0.75) -- (5.5,-0.75);
\draw  (3.5,-0.65) -- (3.5,-0.85);
\draw  (5.5,-0.65) -- (5.5,-0.85);
\node[below] at (4.5,-0.8) {$H_\delta(x)$};
\end{tikzpicture}
\caption{Definition of the computational domains for the peridynamic model.}
\label{Fig:peridynamicsdomains}
\end{figure}
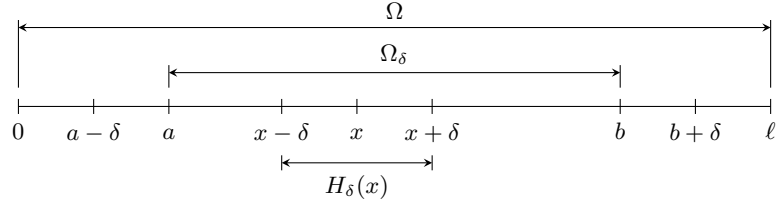

Our objective will be to replace the local model by the peridynamic model in a subregion of $\Omega$.
Let $\delta > 0$ denote the so-called horizon of the peridynamic model and let $H_\delta(x) = (x-\delta, x+\delta)$ be the subdomain of the neighboring particles within the horizon. Because of the nonlocal nature of the model, we can only consider subregions $\intdomd = (a,b)$ such that $a > \delta$ and $b < \ell-\delta$. In that case, we observe that for any given point $x$ in the interval $\intdomd$, we have that $H_\delta(x) \subset \Omega$, as show in Figure~\ref{Fig:peridynamicsdomains}. The general formulation, in one or higher dimension, of the linearized microelastic model~\cite{Silling-JMPS-2000} is given by:
\begin{equation}
\label{eq:linearizedperidynamics}
- \int_{H_\delta(x)} \kappa \frac{\xi \otimes \xi}{\| \xi \|^3} (\uperid(y) - \uperid(x)) dy = f_b(x), 
\end{equation}
where $\kappa$ is the parameter that characterizes the stiffness of the ``bonds'' between point $x$ and the neighboring points $y \in H_\delta(x)$, $\xi$ is the vector between two material points in the reference configuration, i.e.\ $\xi = y-x$, $\| \xi \|$ is the Euclidean norm of vector $\xi$, and $\uperid(x)$ is the displacement of $x$ in the deformed configuration. In the case of the one-dimensional bar, the above integral at a point $x\in \intdomd$ can be rewritten as:
\begin{equation}
\label{eq:1dperidynamics}
- \int_{x-\delta}^{x+\delta} \kappa \frac{\uperid(y) - \uperid(x)}{|y-x|} dy = f_b(x).
\end{equation}

We now identify the material parameter $\kappa$ such that the solution of the linearized microelastic peridynamic model is compatible with that of the continuum local model. One may do so by matching the strain energy of the two models~\cite{Silling-JMPS-2000}. Alternatively, one can recover equation~\eqref{eq:1dlinearelasticity} from~\eqref{eq:1dperidynamics} by taking the limit $\delta \rightarrow 0$. Supposing that the displacement $\uperid(x)$ is sufficiently smooth, the Taylor expansion of $u(y)$ around $x$ yields, for $y \in \intdomd$,
\begin{equation}
\label{eq:Taylor}
\uperid(y) - \uperid(x) 
= \uperid'(x) (y-x) 
+ \frac{1}{2} \uperid''(x) (y-x)^2
+ \frac{1}{6} \uperid'''(x) (y-x)^3 
+ \frac{1}{24} \uperid''''(x) (y-x)^4
+ \ldots
\end{equation}
Substituting~\eqref{eq:Taylor} for $u(y) - u(x)$ in the integral in~\eqref{eq:1dperidynamics}, one gets:
\begin{equation}
\label{eq:forceperidynamics}
\int_{x-\delta}^{x+\delta} \kappa \frac{\uperid(y) - \uperid(x)}{|y-x|} dy = \frac{\kappa\delta^2}{2} \Big(\uperid''(x) + \frac{\delta^2}{24} u''''(x) + \ldots \Big),
\end{equation}
so that~\eqref{eq:1dperidynamics} becomes:
\begin{equation}
\label{eq:expansion}
- \frac{\kappa\delta^2}{2} \Big(\uperid''(x) +  \frac{\delta^2}{24} u''''(x) + \ldots \Big) = f_b(x), \quad \forall x \in \intdomd.
\end{equation}
By taking the limit when $\delta \rightarrow 0$, one then recovers the differential equation~\eqref{eq:1dlinearelasticity} pointwise whenever $\kappa$ is chosen as:
\begin{equation}
\label{eq:kappa}
\frac{\kappa\delta^2}{2} = E, \quad \text{that is}\
\kappa = \frac{2E}{\delta^2},
\end{equation}
in agreement with~\cite{Seleson-Du-Parks-CMAME-2016}. However, if $\delta$ is kept finite, the value $\kappa=2E/\delta^2$ will always induce a modeling error between the solutions of the peridynamic and linear elasticity models, unless all derivatives $u^{(k)}$ of order $k \geq 4$ of the displacement field $u$ vanish for all $x \in \intdomd$. In other words, this is a well known fact that the peridynamic and local models are fully compatible if $u(x)$ is a polynomial function of degree at most three. We shall say that the peridynamic model provides an approximation of the linear elasticity model with a degree of precision equal to three with respect to the parameter $\delta$. Our goal will be to build coupling methods whose degree of precision will also be three.

The challenging issue is that the degree of precision decreases as soon as the integral in~\eqref{eq:1dperidynamics} is evaluated over subdomains of $H_\delta(x)$, which is the case when $x$ approaches boundaries or interfaces. This is related to the so-called skin effect~\cite{bobaru2009-1D}. Indeed, truncation of the set $H_\delta(x)$ close to a boundary induces forces involving the first derivative of $\uperid$. For example, following~\cite{Silling-JMPS-2000,Ongaro-Seleson-CMAME-2021}, we can introduce the stress at a point $x$ in the domain as:
\begin{equation}
\label{eq:sigma1}
\sigma^\pm(u)(x) = \int_{x-\delta}^{x} \int_{x}^{z\pm\delta} \kappa \frac{u(y) - u(z)}{|y-z|} dydz,
\end{equation}
and show, using the Taylor expansion~\eqref{eq:Taylor} as before and substituting $E$ for $\kappa\delta^2/2$, that:
\[
\sigma^\pm(u)(x) = Eu'(x) + \frac{E\delta^2}{24} u'''(x) + \mathcal O(\delta^3).
\]
We thus observe that these integral quantities are second-order approximations of the stress $Eu'(x)$ with respect to $\delta$ with a degree of precision of two. In other words, $\sigma^\pm(u)$ provide the exact value of $Eu'$ at point $x$ whenever $u$ is quadratic in the neighborhood of $x$. In order to obtain approximations with a degree of precision of three, one needs to include the higher-order term such that: 
\begin{equation}
\label{eq:sigma2}
\sigma^\pm(u)(x)
= \frac{\delta}{2} 
\int_{x-\delta}^{x} \int_{x}^{z\pm\delta} \kappa \frac{u(y) - u(z)}{|y-z|} dydz
- \frac{\kappa \delta^4}{48} u'''(x).
\end{equation}
Now, if $u$ is cubic in the neighborhood of $x$, $\sigma^\pm(u)$ lead to the exact value of $Eu'$ at $x$. We will see how these results will become useful when building the coupling methods. We also note that one could consider alternative definitions of the stress at a point $x$ computed from the peridynamic model.

\section{Coupling of linear elasticity models}
\label{Sect:CF-ElasticityModels}

The purpose of this section is to review the general coupling formulation in the case of two classical linear elasticity models and to study the particular setting where the classical equation~\eqref{eq:1dlinearelasticity} is replaced by~\eqref{eq:expansion}. We will also introduce a modified formulation in which the pointwise interfaces are replaced by overlapping regions between the two models.

\subsection{General formulation}
\label{Sect:generalformulation}

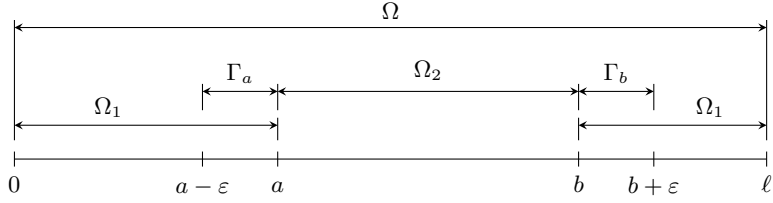
\begin{figure}[tbp]
\centering
\small
\begin{tikzpicture}
\draw (0,0) -- (10.0,0);
\draw (0,-0.1) -- (0,0.1);
\draw (2.5,-0.1) -- (2.5,0.1);
\draw (3.5,-0.1) -- (3.5,0.1);
\draw (7.5,-0.1) -- (7.5,0.1);
\draw (8.5,-0.1) -- (8.5,0.1);
\draw (10.0,-0.1) -- (10.0,0.1);
\node[above] at (0.0,-0.55) {$0$};
\node[above] at (2.5,-0.59) {$a-\varepsilon$};
\node[above] at (3.5,-0.55) {$a$};
\node[above] at (10.0,-0.55) {$\ell$};
\node[above] at (7.5,-0.55) {$b$};
\node[above] at (8.5,-0.59) {$b+\varepsilon$};
\draw[arrows=<->, >=stealth]  (0.0,1.75) -- (10.0,1.75);
\draw  (0.0,0.25) -- (0.0,1.85);
\draw  (10.0,0.25) -- (10.0,1.85);
\node[above] at (5.00,1.75) {$\Omega$};
\draw[arrows=<->, >=stealth]  (0.0,0.45) -- (3.5,0.45);
\draw  (3.5,0.25) -- (3.5,0.55);
\node[above] at (1.25,0.45) {$\Omega_1$};
\draw[arrows=<->, >=stealth]  (2.5,0.9) -- (3.5,0.9);
\draw  (2.5,0.65) -- (2.5,1.0);
\node[above] at (3.0,0.9) {$\Gamma_a$};
\draw[arrows=<->, >=stealth]  (3.5,0.9) -- (7.5,0.9);
\draw  (3.5,0.65) -- (3.5,1.0);
\node[above] at (5.5,0.94) {$\Omega_2$};
\draw  (7.5,0.65) -- (7.5,1.0);
\draw[arrows=<->, >=stealth]  (7.5,0.9) -- (8.5,0.9);
\draw  (8.5,0.65) -- (8.5,1.0);
\node[above] at (8.0,0.9) {$\Gamma_b$};
\draw[arrows=<->, >=stealth]  (7.5,0.45) -- (10.0,0.45);
\draw  (7.5,0.25) -- (7.5,0.55);
\node[above] at (9.25,0.45) {$\Omega_1$};
\end{tikzpicture}
\caption{Definition of the computational domains for the coupling of two linear elasticity models.}
\label{Fig:couplingelasticity}
\end{figure}

We suppose here that we are interested in coupling linear elasticity models. We thus partition domain $\Omega$ into the subdomains $\Omega_1=(0,a)\cup(b,\ell)$ and $\Omega_2 = (a,b)$, where we consider the first model in $\Omega_1$, with modulus of elasticity $E_1$, and the second model in $\Omega_2$, with $E_2$. The configuration is shown in Figure~\ref{Fig:couplingelasticity}. It is well-known that the coupled model consists in this case in solving for $\uelast_1$ in $\overline{\Omega}_1$ and $\uelast_2$ in $\overline{\Omega}_2$ such that:
\begin{equation}
\label{eq:coupling_elast_1}
\begin{aligned}
- E_1 \uelast_1''(x) &= f_b(x), \quad \forall x \in \Omega_1, \\
- E_2 \uelast_2''(x) &= f_b(x), \quad \forall x \in \Omega_2,
\end{aligned}
\end{equation}
with boundary conditions:
\begin{equation}
\begin{aligned}
\uelast_1(x) &= 0, \quad \text{at}\ x = 0, \\
E_1 \uelast_1'(x) &= g, \quad \text{at}\ x = \ell,
\end{aligned}
\end{equation}
and interface conditions:
\begin{equation}
\label{eq:coupling_elast_interface}
\begin{aligned}
\uelast_1(x) - \uelast_2(x) &= 0, \quad \text{at}\ x=a\ \text{and}\ x=b, \\
E_1 \uelast_1'(x) - E_2 \uelast_2'(x) &= 0, \quad \text{at}\ x=a\ \text{and}\ x=b,
\end{aligned}
\end{equation}
where the last two equations correspond to the continuity of the displacement and the stress at the pointwise interfaces.

\subsection{Examples}
\label{Sect:CF-examples}

\begin{figure}[tbp]
\centering
\includegraphics[height=0.35\textwidth]{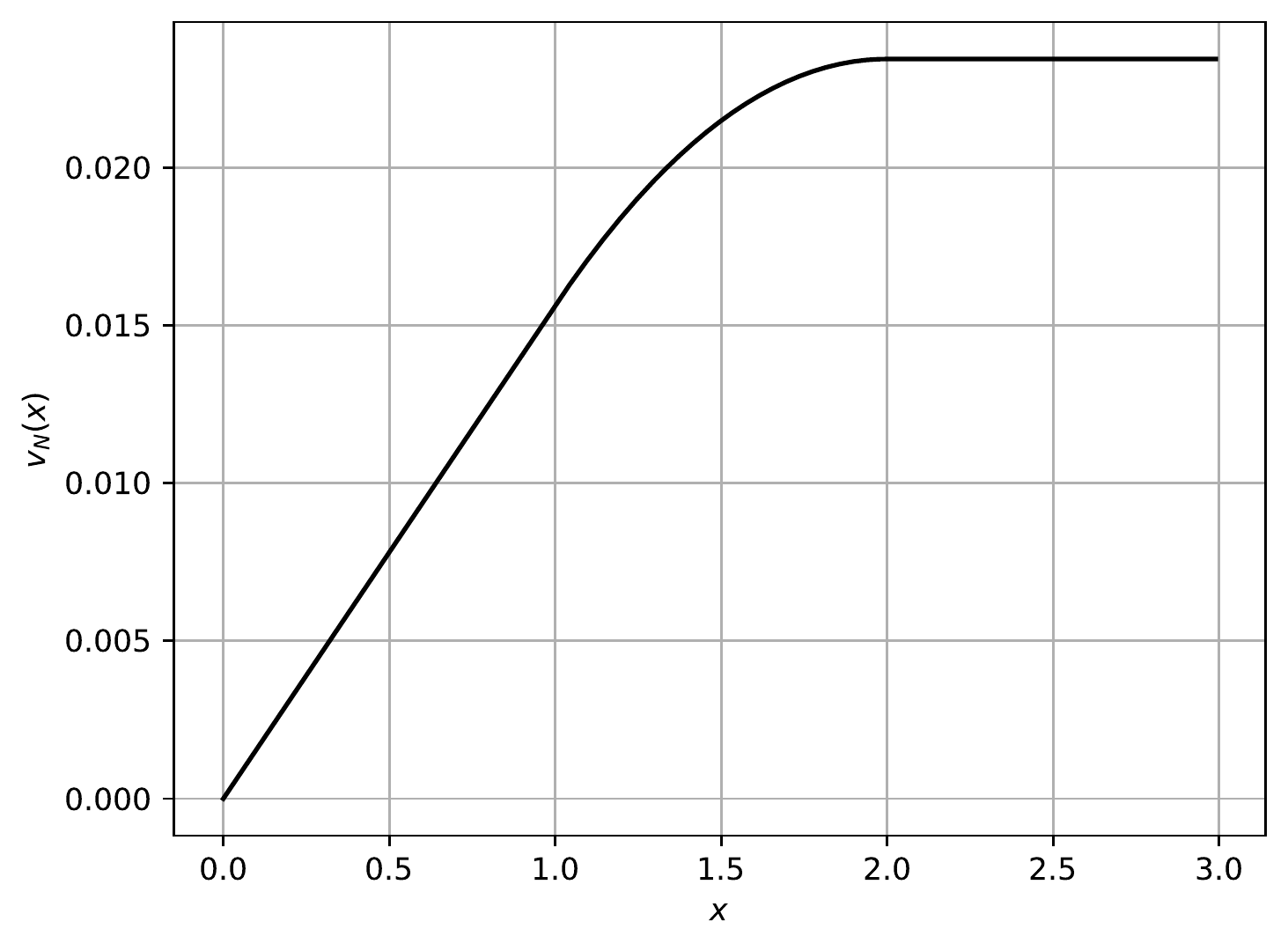}
\quad
\includegraphics[height=0.35\textwidth]{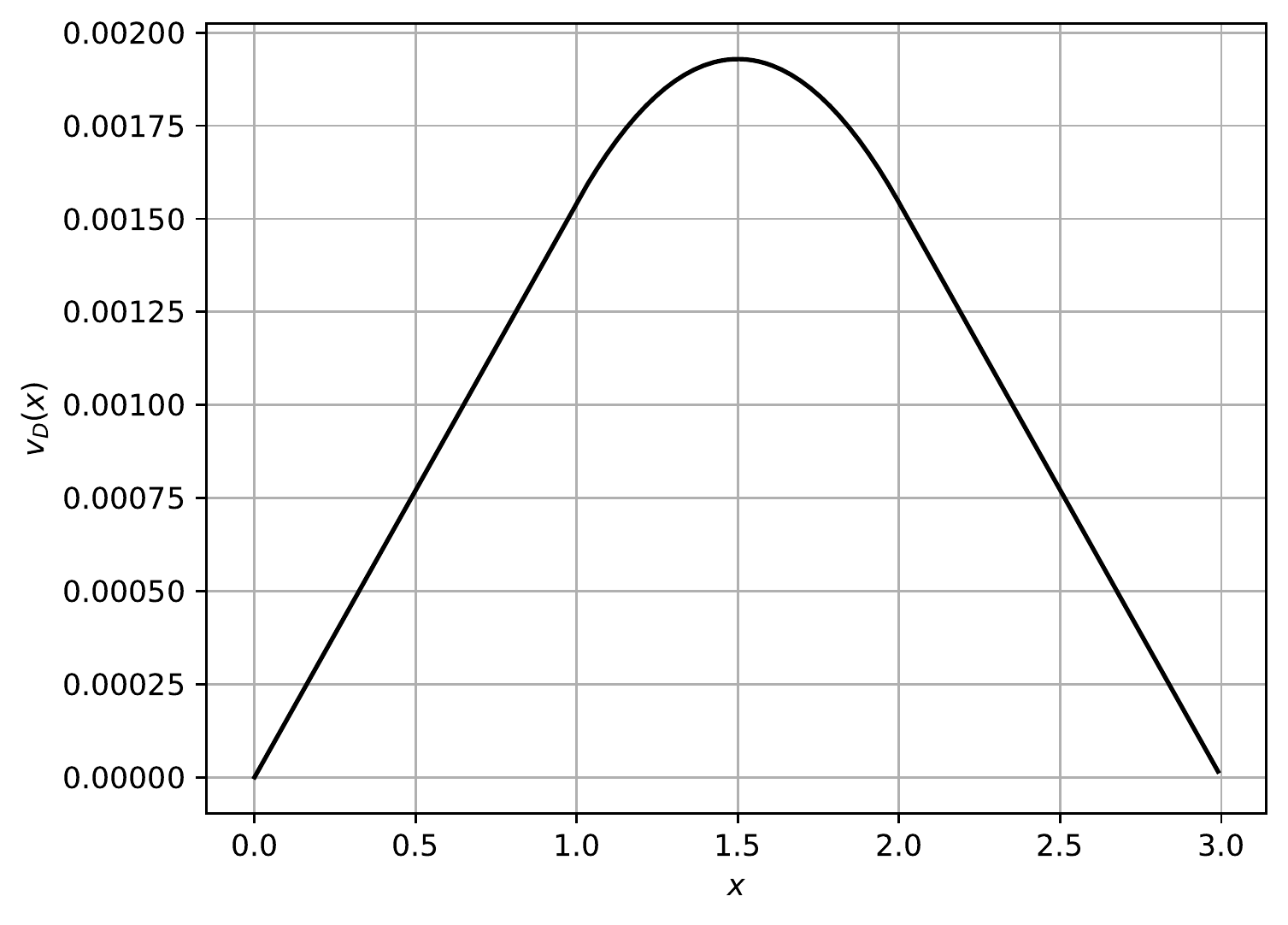}
\caption{Illustration of the solutions $V_N$~\eqref{eq:vNeumann} and $v_D$~\eqref{eq:vDirichlet} with $a=1$, $b=2$, $b=3$, and $\delta=1/8$. For $V_N$, shown on the left, $\uelast(x) = x^4$, implying $\lambda=24$, and for~$V_D$, $\uelast(x) = x^2(144-96x+16x^2)/81$ so that $\lambda=128/27$.}
\label{Fig:examplesofv}
\end{figure}

In this section, we analyze the solution to the above coupling problem when the model equation in $\Omega_2$ is actually replaced by the differential equation~\eqref{eq:expansion} arising from the peridynamic model. We consider the special case when the exact solution is a quartic polynomial with constant fourth derivative equal to $\lambda$. For the sake of simplicity, we shall slightly modify the notation as $\uelast_1(x)=\uelast(x)$, $E_1=E$, $\uelast_2(x)=\uperid(x)$, $E_2=E=\kappa \delta^2/2$, so that the problem reads:
\begin{equation}
\label{eq:CMexample}
\begin{aligned}
- E \uelast''(x) = f_b(x), &\quad \forall x \in \Omega_1, \\
- E \uperid''(x) - E\frac{\delta^2}{24} \lambda = f_b(x), &\quad \forall x \in \Omega_2, \\
\uelast(x) = 0, &\quad \text{at}\ x = 0, \\
E \uelast'(x) = g, &\quad \text{at}\ x = \ell, \\
\uelast(x) - \uperid(x) = 0, &\quad \text{at}\ x=a\ \text{and}\ x=b, \\
E\uelast'(x) - E\uperid'(x) = 0, &\quad \text{at}\ x=a\ \text{and}\ x=b.
\end{aligned}
\end{equation}
Since all equations are linear here, the above problem can be recast, using the superposition principle, as that of searching for the function $v=v(x)$, for all $x \in \overline{\Omega}$, such that:
\begin{equation}
\label{eq:function-v}
\begin{aligned}
- E v''(x) = 0, &\quad \forall x \in \Omega_1, \\
- E v''(x) = E\frac{\delta^2}{24} \lambda, &\quad \forall x \in \Omega_2, \\
v(x) = 0, &\quad \text{at}\ x = 0, \\
E v'(x) = 0, &\quad \text{at}\ x = \ell, \\
v(x^-) - v(x^+) = 0, &\quad \text{at}\ x=a\ \text{and}\ x=b, \\
Ev'(x^-) - Ev'(x^+) = 0, &\quad \text{at}\ x=a\ \text{and}\ x=b.
\end{aligned}
\end{equation}
where $x^- = \lim_{\epsilon \to 0} (x-\epsilon)$ and $x^+ = \lim_{\epsilon \to 0} (x+\epsilon)$, with $\epsilon >0$. The function $v$ corresponds to the difference between the solution pair $(\uelast,u)$ of the coupling problem~\eqref{eq:CMexample} and the solution $\uelast$ to the linear elasticity problem~\eqref{eq:1dlinearelasticity} over the whole domain $\Omega$. The general solution is then given by:
\begin{equation}
v(x) = \left\{
\begin{aligned}
&C_1x + C_2, &&\quad \forall x \in [0,a] \\
&- \frac{\lambda\delta^2}{48} x^2 + C_3x + C_4, &&\quad \forall x \in [a,b] \\
&C_5x + C_6, &&\quad \forall x \in [b,\ell]
\end{aligned}
\right.
\end{equation}
Upon applying the boundary and interface conditions, one obtains the solution:
\begin{equation}
\label{eq:vNeumann}
v_N(x) = \left\{
\begin{aligned}
&\frac{\lambda\delta^2}{24} (b-a) x, &&\quad \forall x \in [0,a] \\
&\frac{\lambda\delta^2}{48} \big(b^2 - a^2 - (b-x)^2 \big), &&\quad \forall x \in [a,b] \\
&\frac{\lambda\delta^2}{48} (b^2 - a^2), &&\quad \forall x \in [b,\ell]
\end{aligned}
\right.
\end{equation}
We also consider the problem with homogeneous Dirichlet conditions at both ends, replacing the Neumann condition $E\uelast'(\ell) = g$ by $\uelast(\ell) = 0$, all other equations in~\eqref{eq:CMexample} remaining the same. In that case, the solution $v(x)$ reads:
\begin{equation}
\label{eq:vDirichlet}
v_D(x) = \left\{
\begin{aligned}
&\frac{\lambda\delta^2}{48} \frac{(b-a)}{\ell} \big( 2\ell - (a+b) \big) x, &&\quad \forall x \in [0,a] \\
&\frac{\lambda\delta^2}{48} \Big( \big(b^2 - a^2 \big)\frac{\ell-x}{\ell} - (b-x)^2 \Big), &&\quad \forall x \in [a,b] \\
&\frac{\lambda\delta^2}{48} \big(b^2 - a^2 \big)\frac{\ell-x}{\ell}, &&\quad \forall x \in [b,\ell]
\end{aligned}
\right.
\end{equation}
We illustrate in Figure~\ref{Fig:examplesofv} the solutions~\eqref{eq:vNeumann} and~\eqref{eq:vDirichlet} for particular values of the data, i.e.\ $a=1$, $b=2$, $b=3$, and $\delta=1/8$. Moreover, for~\eqref{eq:vNeumann}, we suppose that the exact solution to the classical linear elasticity problem is given by $\uelast(x) = x^4$, so that $\lambda=24$, while for~\eqref{eq:vDirichlet}, we take $\uelast(x) = x^2(144-96x+16x^2)/81$, so that $\lambda=128/27$.
We observe in Figure~\ref{Fig:examplesofv} that the solution~\eqref{eq:vNeumann} in the interval $[b,\ell]=[2,3]$ remains constant, which corresponds to the maximum value of $v_N$ over the whole domain $[0,\ell]=[0,3]$:
\begin{equation}
\label{eq:vmaxNeumann}
v_{N,\max} = \frac{\lambda\delta^2}{48} (b^2 - a^2) = \frac{3}{2} \delta^2.
\end{equation}
As far as the solution $v_D$~\eqref{eq:vDirichlet} is concerned, it reaches its maximum value in the interval $[a,b]$, and more specifically at $x=b-(b^2-a^2)/(2\ell)$. Then
\begin{equation}
\label{eq:vmaxDirichlet}
v_{D,\max} = \frac{\lambda\delta^2}{48} (b^2 - a^2) \frac{4\ell^2-4b\ell+ b^2-a^2}{4\ell^2} = \frac{10}{81} \delta^2.
\end{equation}
We will use these quantities to assess the the efficiency of the coupling methods between classical linear elasticity and peridynamic models in the numerical examples of Section~\ref{Sect:numericalexamples}.

\subsection{Modified formulation}
\label{Sect:modifiedformulation}

In the particular case where $E_1 = E_2 = E$, the condition of stress continuity at the pointwise interfaces $x=a$ and $x=b$ can be replaced by a constraint involving the displacements on overlap regions between the two domains $\Omega_1$ and $\Omega_2$. In order to do so, we consider the overlap regions $\Gamma_a$ and $\Gamma_b$, both of size $\varepsilon$, as shown in Figure~\ref{Fig:couplingelasticity}. The above problem can then be recast as finding $\uelast_1$ in $\overline{\Omega}_1$ and $\uelast_2$ in $\overline{\Omega_2 \cup \Gamma_a \cup \Gamma_b}$ such that:
\begin{equation}
\label{eq:coupling_elast_2}
\begin{aligned}
- E \uelast_1''(x) = f_b(x), &\quad \forall x \in \Omega_1, \\
- E \uelast_2''(x) = f_b(x), &\quad \forall x \in \text{Int} \big( \overline{\Omega_2 \cup \Gamma_a \cup \Gamma_b} \big) = (a-\varepsilon,b+\varepsilon), \\
\uelast_1(x) = 0, &\quad \text{at}\ x = 0, \\
E \uelast_1'(x) = g, &\quad \text{at}\ x = \ell, \\
\uelast_1(x) - \uelast_2(x) = 0,  &\quad \text{at}\ x=a\ \text{and}\ x=b \\
\uelast_1(x-\varepsilon) - \uelast_2(x-\varepsilon) = 0,  &\quad \text{at}\ x=a\\
\uelast_1(x+\varepsilon) - \uelast_2(x+\varepsilon) = 0,  &\quad \text{at}\ x=b.
\end{aligned}
\end{equation}
The above problem, and in particular the last two equations, ensure that $\uelast_1(x) = \uelast_2(x)$, for all $x \in \Gamma_a \cup \Gamma_b$. This naturally implies that $\uelast_1'(a) = \uelast_2'(a)$ and $\uelast_1'(b) = \uelast_2'(b)$, which allows one to recover the interface conditions $E \uelast_1'(x) - E \uelast_2'(x) = 0$, at $x=a$ and $x=b$, as required by~\eqref{eq:coupling_elast_interface}. However, this coupling problem would clearly provide an incorrect solution if $E_1 \neq E_2$. This observation will become relevant when we attempt to couple classical linear elasticity and peridynamic models, since the latter is not exactly equivalent to the former in the case of finite horizon $\delta$. 

\section{Coupling of classical linear elasticity and peridynamic models}
\label{Sect:CF-CouplingMethods}

\begin{figure}[tbp]
\centering
\small
\begin{tikzpicture}
\draw (0,0) -- (10.0,0);
\draw (0,-0.1) -- (0,0.1);
\draw (2.5,-0.1) -- (2.5,0.1);
\draw (3.5,-0.1) -- (3.5,0.1);
\draw (7.5,-0.1) -- (7.5,0.1);
\draw (8.5,-0.1) -- (8.5,0.1);
\draw (10.0,-0.1) -- (10.0,0.1);
\node[above] at (0.0,-0.55) {$0$};
\node[above] at (2.5,-0.59) {$a-\delta$};
\node[above] at (3.5,-0.55) {$a$};
\node[above] at (10.0,-0.55) {$\ell$};
\node[above] at (7.5,-0.55) {$b$};
\node[above] at (8.5,-0.59) {$b+\delta$};
\draw[arrows=<->, >=stealth]  (0.0,1.75) -- (10.0,1.75);
\draw  (0.0,0.25) -- (0.0,1.85);
\draw  (10.0,0.25) -- (10.0,1.85);
\node[above] at (5.00,1.75) {$\Omega$};
\draw[arrows=<->, >=stealth]  (0.0,0.45) -- (3.5,0.45);
\draw  (3.5,0.25) -- (3.5,0.55);
\node[above] at (1.25,0.45) {$\Omega_e$};
\draw[arrows=<->, >=stealth]  (2.5,0.9) -- (3.5,0.9);
\draw  (2.5,0.65) -- (2.5,1.0);
\node[above] at (3.0,0.9) {$\Gamma_a$};
\draw[arrows=<->, >=stealth]  (3.5,0.9) -- (7.5,0.9);
\draw  (3.5,0.65) -- (3.5,1.0);
\node[above] at (5.5,0.94) {$\intdomd$};
\draw  (7.5,0.65) -- (7.5,1.0);
\draw[arrows=<->, >=stealth]  (7.5,0.9) -- (8.5,0.9);
\draw  (8.5,0.65) -- (8.5,1.0);
\node[above] at (8.0,0.9) {$\Gamma_b$};
\draw[arrows=<->, >=stealth]  (7.5,0.45) -- (10.0,0.45);
\draw  (7.5,0.25) -- (7.5,0.55);
\node[above] at (9.25,0.45) {$\Omega_e$};
\end{tikzpicture}
\caption{Definition of the computational domains for the coupling between a linear elasticity model and a peridynamic model.}
\label{Fig:couplingLinearElastPeryd}
\end{figure}
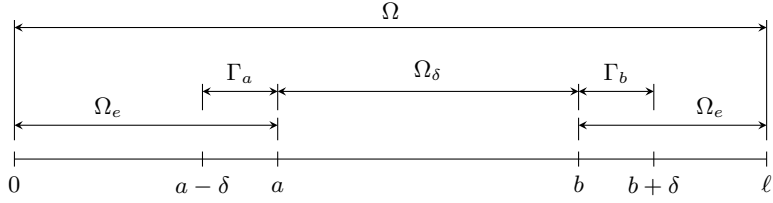

The main challenge when employing a peridynamic model of constant horizon~$\delta$ is that it is necessary to extend the domain~$\Omega_\delta$ by extra layers of size $\delta$ in order to be able to correctly estimate the integral term in the vicinity of the boundaries. One objective in coupling peridynamics with classical linear elasticity is therefore to recover,  from the linear elasticity solution, compatible displacements in those layers that can be used in the peridynamic model.
As before, we introduce two points $a$ and $b$ inside $\Omega=(0,\ell)$ such that $a < b$ and define the subdomains $\Omega_e=(0,a) \cup (b,\ell)$ and $\intdomd=(a,b)$, as shown in Figure~\ref{Fig:couplingLinearElastPeryd}. We also consider the so-called overlapping domains $\Gamma_a = (a-\delta,a)$ and $\Gamma_b=(b,b+\delta)$ of size $\delta$. Following the discussion of the previous section, we propose two approaches to construct compatible displacements from the linear elasticity solution over the overlapping regions $\Gamma_a$ and $\Gamma_b$. Alternatively, in order to avoid extending the peridynamic region~$\Omega_\delta$, one could consider a peridynamic model with variable horizon such that $\delta$ goes to zero when approaching the boundaries of the domain. This will be the basis of a third coupling approach presented below.

\subsection{Coupling method with matching displacements (MDCM)}

In the first approach, we choose to constrain the displacements from the two models to match in $\Gamma_a$ and $\Gamma_b$. The problem consists in finding $\uelast \in \overline{\Omega}_e$ and $\uperid \in \overline{\Omega_\delta \cup \Gamma_a \cup \Gamma_b}$ such that:
\begin{equation}
\label{eq:CM-displacement}
\begin{aligned}
- E \uelast''(x) = f_b(x), 
&\quad \forall x \in \Omega_e, \\
- \int_{x-\delta}^{x+\delta} \kappa \frac{\uperid(y) - \uperid(x)}{|y-x|} dy = f_b(x), 
&\quad \forall x \in \overline{\Omega}_\delta, \\
\uelast(x) = 0, 
&\quad \text{at}\ x = 0, \\
E \uelast'(x) = g, 
&\quad \text{at}\ x = \ell, \\
\uperid(x) - \uelast(x) = 0, 
&\quad \forall x \in \overline{\Gamma_a \cup \Gamma_b}.
\end{aligned}
\end{equation}
The above formulation has a degree of precision of three with respect to the parameter~$\delta$ and should provide the exact solution given by the classical linear elasticity model as long as the solution is a polynomial function of degree up to three. However, the two models are not compatible anymore when the solution is a polynomial of higher degree and one should expect errors due to the matching of the displacements in the overlapping regions. This approach is similar in essence, up to some slight variations, to the methods proposed in~\cite{DElia-Bochev-2021,Zaccariotto-CMAME-2018} and is inspired by~\eqref{eq:coupling_elast_2}.

\subsection{Coupling method with matching stresses (MSCM)}

In the second approach, we choose to constrain the stresses from the two models in the overlapping regions. We then search for $\uelast \in \overline{\Omega}_e$ and $\uperid \in \overline{\Omega_\delta \cup \Gamma_a \cup \Gamma_b}$ such that:
\begin{equation}
\label{eq:CM-stress}
\begin{aligned}
- E \uelast''(x) = f_b(x), 
&\quad \forall x \in \Omega_e, \\
- \int_{x-\delta}^{x+\delta} \kappa \frac{\uperid(y) - \uperid(x)}{|y-x|} dy = f_b(x), 
&\quad \forall x \in \Omega_\delta, \\
\uelast(x) = 0, 
&\quad \text{at}\ x = 0, \\
E \uelast'(x) = g, 
&\quad \text{at}\ x = \ell, \\
\uperid(x) - \uelast(x) = 0, 
&\quad \text{at}\ x = a\ \text{and}\ x=b, \\
\sigma^+(u)(x) - E\uelast'(x) = 0, 
&\quad \forall x \in \overline{\Gamma}_a, \\
\sigma^-(u)(x) - E\uelast'(x) = 0, 
&\quad \forall x \in \overline{\Gamma}_b,
\end{aligned}
\end{equation}
where \(\sigma^\pm(u)\) could be given by either~\eqref{eq:sigma1} or~\eqref{eq:sigma2}. However, in order to preserve the degree of precision of three for the coupling approach, one needs to approximate the stresses $\sigma^\pm(u)$ by~\eqref{eq:sigma2}. 
We remark that this approach is more akin to the general formulation~\eqref{eq:coupling_elast_1}-\eqref{eq:coupling_elast_interface} for coupling classical linear elasticity models, in the sense that the continuity of the stress at the interfaces should be better approximated than in the first coupling method~\eqref{eq:CM-displacement}. This will be indeed confirmed by the numerical examples. \edit{This approach is inspired by~\cite{silling2020Couplingstresses}.}

\subsection{Coupling method with variable horizon (VHCM)}

The objective here is to introduce a formulation of the peridynamic model in which the horizon $\delta$ is allowed to vary within the domain~$\intdomd$, in such a way that the horizon should tend to zero as one approaches the interfaces $x=a$ and $x=b$ while remaining constant inside $\intdomd$ sufficiently far from the interfaces. In that way, the peridynamic model would naturally converge to the classical linear elasticity model when approaching the coupling region and it would be unnecessary to consider overlapping domains. 

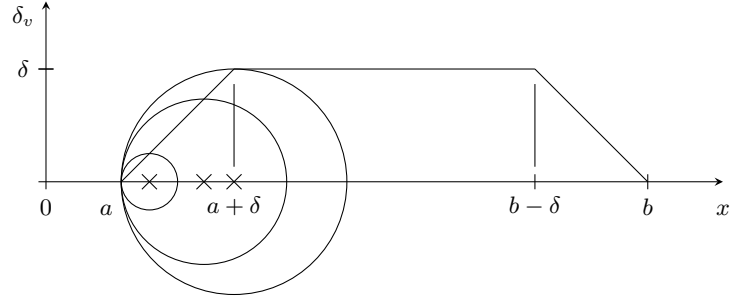
\begin{figure}
\centering
\small
\begin{tikzpicture}
\draw[arrows=->, >=stealth] (-0.1,0) -- (9.0,0);
\draw[arrows=->, >=stealth] (0,-0.1) -- (0,2.4);
\draw (1.0,0.0) -- (2.5,1.5);
\draw (2.5,1.5) -- (6.5,1.5);
\draw (6.5,1.5) -- (8.0,0.0);
\draw (2.5,0.2) -- (2.5,1.3);
\draw (6.5,0.2) -- (6.5,1.3);
\draw (6.5,-0.1) -- (6.5,0.1);
\draw (8.0,-0.1) -- (8.0,0.1);
\draw (-0.1,1.5) -- (0.1,1.5);
\draw (-0.1,1.5) -- (0.1,1.5);
\node at (-0.3,2.2) {$\delta_v$};
\node at (-0.3,1.5) {$\delta$};
\node[above] at (0.0,-0.55) {$0$};
\node[above] at (0.8,-0.55) {$a$};
\node[above] at (2.5,-0.56) {$a+\delta$};
\node[above] at (6.5,-0.56) {$b-\delta$};
\node[above] at (8.0,-0.55) {$b$};
\node[above] at (9.0,-0.55) {$x$};
\draw (2.5,0) circle (1.5);
\draw (2.4,-0.1) -- (2.6,0.1);
\draw (2.4,0.1) -- (2.6,-0.1);
\draw (2.1,0) circle (1.1);
\draw (2.0,-0.1) -- (2.2,0.1);
\draw (2.0,0.1) -- (2.2,-0.1);
\draw (1.375,0) circle (0.375);
\draw (1.275,-0.1) -- (1.475,0.1);
\draw (1.275,0.1) -- (1.475,-0.1);
\end{tikzpicture}
\caption{Example of a variable horizon function $\delta_v(x)$. The circles centered at points $x\in (a,a+\delta)$ are representations of the associated domains $H_\delta(x)$ in terms of $\delta_v(x)$.}
\label{Fig:variablehorizon}
\end{figure}

Following~\cite{Prudhomme-Diehl-2020}, given $\delta \in \mathbb R^+$, the variable horizon $\delta_v(x)$ should thus satisfy the following requirements:
\begin{equation}
0 \leq \delta_v(x) \leq \delta, \qquad
\delta_v(x) \leq x-a, \qquad
\delta_v(x) \leq b-x, \qquad \forall x \in \intdomd = (a,b).
\end{equation} 
The choice of the function $\delta_v(x)$ is obviously not unique. However, the simplest continuous function that fulfill these requirements is the piecewise linear function shown in Figure~\ref{Fig:variablehorizon}:
\begin{equation}
\label{eq:deltafn}
\delta_v(x) = \left\{ 
\begin{array}{ll} 
x-a, & \quad a < x \leq a+\delta, \\ 
\delta, & \quad a+\delta < x \leq b - \delta, \\ 
b - x, & \quad b-\delta < x < b. 
\end{array}
\right.
\end{equation}
Alternative functions, in particular smoother functions, could also be considered as long as they satisfy the above requirements. In view of~\eqref{eq:kappa}, the horizon being a function of $x$ implies that the material parameter $\kappa$ should also depend on $x$, i.e.\ $\kappa=\bar\kappa(x)$, if the non-local model is to be compatible with the classical linear elasticity model, so that~\eqref{eq:1dperidynamics} should now read
\begin{equation}
\label{eq:1dperidynamics-VHM}
- \int_{x-\delta_v(x)}^{x+\delta_v(x)} \bar\kappa(x) \frac{\uperid(y) - \uperid(x)}{|y-x|} dy
= f_b(x), \quad \forall x \in \intdomd.
\end{equation}
Compatibility of the two models everywhere with a convergence of order $\mathcal O(\delta^2)$ implies that the quantity $\bar\kappa(x) \delta_v^2(x)$ should remain constant for all $x \in \intdomd$, i.e.\ with $\kappa$ satisfying~\eqref{eq:kappa}, one has
\begin{equation}
\label{eq:productkappadelta}
\bar\kappa(x) \delta_v^2(x) = \kappa \delta^2, \quad \forall x \in \intdomd.
\end{equation}

The coupling method with variable horizon is based on the general formulation~\eqref{eq:coupling_elast_1}-\eqref{eq:coupling_elast_interface} and consists in finding $\uelast \in \overline{\Omega}_e$ and $\uperid \in \overline{\Omega}_\delta$ such that:
\begin{equation}
\label{eq:CM-variablehorizon}
\begin{aligned}
- E \uelast''(x) = f_b(x), 
&\quad \forall x \in \Omega_e, \\
- \int_{x-\delta_v(x)}^{x+\delta_v(x)} \bar\kappa(x) \frac{\uperid(y) - \uperid(x)}{|y-x|} dy = f_b(x), 
&\quad \forall x \in \intdomd, \\
\uelast(x) = 0, 
&\quad \text{at}\ x = 0, \\
E \uelast'(x) = g, 
&\quad \text{at}\ x = \ell, \\
\uperid(x) - \uelast(x) = 0, 
&\quad \text{at}\ x = a\ \text{and}\ x=b, \\
\sigma^+(u)(x) - E\uelast'(x) = 0, 
&\quad \text{at}\ x=a, \\
\sigma^-(u)(x) - E\uelast'(x) = 0, 
&\quad \text{at}\ x=b.
\end{aligned}
\end{equation}
The main advantage of this approach in regards to the other two coupling methods is that it does not involve any overlapping domains, which should simplify its implementation. Moreover, its degree of precision should be three as long as the stresses $\sigma^\pm(u)$ are approximated by~\eqref{eq:sigma2}. \edit{This approach is inspired by~\cite{silling2015variable, NIKPAYAM2019308}.}

\section{Discretization}
\label{Sect:discretization}

We have introduced in the previous section the continuous formulation of three coupling methods based on the coupling of elasticity models as described in Section~\ref{Sect:CF-ElasticityModels}. We are now in a position to provide the discrete formulations of those coupling problems. For the sake of simplicity, we choose here to discretize the peridynamic model using a collocation approach. It seems therefore natural to approximate the classical linear elasticity model by a finite difference scheme, rather than by the finite element method as it would require to derive the weak formulation of the strong problem and to make specific adjustments when coupling the peridynamic and classical linear elasticity models. The goal will be to ensure that the discrete methods all preserve the degree of precision of three with respect to both the horizon and the  discretization parameter.

\subsection{Discretization of the computational domain}

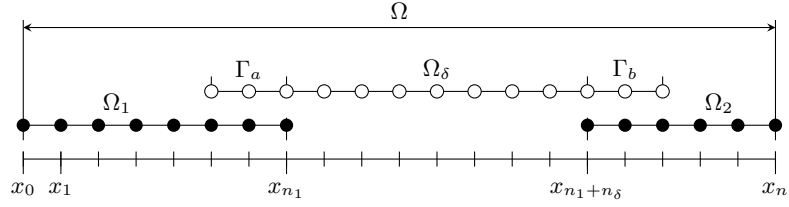
\begin{figure}[tbp]
\centering
\small
\begin{tikzpicture}
\draw (0,0) -- (10.0,0);
\foreach \i in {0,...,20}
{\draw (0.5*\i,-0.1) -- (0.5*\i,0.1);}
\foreach \x in {0.0,0.5,3.5,7.5,10.0}
{\draw (\x,-0.2) -- (\x,0.1);}
\node[below] at (0.0,-0.2)  {$x_0$};
\node[below] at (0.5,-0.2)  {$x_1$};
\node[below] at (3.5,-0.2)  {$x_{n_1}$};
\node[below] at (7.5,-0.2)  {$x_{n_1+n_\delta}$};
\node[below] at (10.0,-0.2) {$x_n$};
\draw[arrows=<->, >=stealth] (0.0,1.75) -- (10.0,1.75);
\draw (0.0,0.25) -- (0.0,1.85);
\draw (10.0,0.25) -- (10.0,1.85);
\node[above] at (5.00,1.75) {$\Omega$};
\draw (0.0,0.45) -- (3.5,0.45);
\node[above] at (1.25,0.5) {$\Omega_1$};
\draw (2.5,0.9) -- (3.5,0.9);
\node[above] at (3.0,0.95) {$\Gamma_a$};
\draw (3.5,0.9) -- (7.5,0.9);
\node[above] at (5.5,0.95) {$\intdomd$};
\draw (7.5,0.9) -- (8.5,0.9);
\node[above] at (8.0,0.95) {$\Gamma_b$};
\draw (7.5,0.45) -- (10.0,0.45);
\node[above] at (9.25,0.5) {$\Omega_2$};
\foreach \x in {3.5,7.5}
{\draw (\x,0.25) -- (\x,0.55);}
\foreach \x in {2.5,3.5,7.5,8.5}
{\draw (\x,0.9) -- (\x,1.1);}
\foreach \i in {0,...,7}
{\node[circle,color=black,fill=black,inner sep=0pt,minimum size=5pt,label=below:{}] at (0.5*\i,0.45) {};}
\foreach \i in {5,...,17}
{\node[circle,draw=black,fill=white,inner sep=0pt,minimum size=5pt,label=below:{}] at (0.5*\i,0.90) {};}
\foreach \i in {15,...,20}
{\node[circle,color=black,fill=black,inner sep=0pt,minimum size=5pt,label=below:{}] at (0.5*\i,0.45) {};}
\end{tikzpicture}
\caption{Definition of the grid points and degrees of freedom (represented by \(\bullet\) for the degrees of freedom associated with the classical linear elasticity model and by \(\circ\) for the degrees of freedom associated with the peridynamic model) for the coupling methods with overlaps.}
\label{Fig:discretization}
\end{figure}

For the sake of simplicity in the notation, we first decompose $\Omega_e$ into the two subdomains $\Omega_1=(0,a)$ and $\Omega_2=(b,\ell)$, as shown in Figure~\eqref{Fig:discretization}.
For a given $\delta$, we introduce a uniform grid spacing $h$ chosen such that, as customarily done in the literature~\cite{Silling-Askari-CS-2005,Parks-Lehoucq-CPC-2008}, $\delta$ be a multiple of $h$, i.e.\ $\delta/h = m$, with $m$ a positive integer. Moreover, we choose the grid size $h$ to be the same in each of the subregions of the computational domain. In other words, the numbers of intervals \(n_\delta\), \(n_1\), and \(n_2\) in \(\Omega_\delta\), \(\Omega_1\), and \(\Omega_2\), respectively, are taken such that
\begin{equation}
h = \frac{b-a}{n_\delta} = \frac{a}{n_1} = \frac{\ell-b}{n_2}.
\end{equation}
There are therefore a total of \(n+1\) grid points, with \(n=n_1+n_\delta+n_2\), uniformly distributed as:
\begin{equation}
x_k = k h, \quad k = 0,1,\ldots, n.
\end{equation}

\subsection{Algorithm for the coupling method with matching displacements}

We need to consider a different numbering of the degrees of freedom associated with the discrete displacement fields as they need to be duplicated over the overlapping domains \(\overline{\Gamma}_a\) and \(\overline{\Gamma}_b\) for the first two coupling methods. For convenience, we denote by \(N_1=n_1+1\), \(N_\delta= n_\delta+1+2m\), and \(N_2=n_2+1\), the numbers of degrees of freedom in $\overline{\Omega}_1$, $\overline{\intdomd\cup \Gamma_a \cup \Gamma_b}$, and $\overline{\Omega}_2$, respectively, so that the total number of degrees of freedom is given by \(N=N_1+N_\delta+N_2\). We then adopt the following correspondence between the numbering of the degrees of freedom and the numbering of the grid points, see Figure~\ref{Fig:discretization}:
\begin{equation}
\label{eq:dofnumbering}
\begin{aligned}
u_i &= \underline{u}(x_k), \quad \forall i=1,\ldots,N_1,  && k=i-1, \\
u_i &= u(x_k), \quad \forall i=N_1+1,\ldots,N_1+N_\delta, && k=i-2-m,\\
u_i &= \underline{u}(x_k), \quad \forall i=N_1+N_\delta+1,\ldots,N, && k=i-3-2m.
\end{aligned}
\end{equation}
The second derivative in~\eqref{eq:1dlinearelasticity} is approximated by the second-order central difference stencil.
Approximation of the integral in~\eqref{eq:1dperidynamics} is obtained by classical quadrature formula using the grid points $x_k$. Note that other more advanced quadrature rules, especially developed for non-local models, could alternatively be used, see e.g.~\cite{Seleson-CMAME-2014,Trask-You-CMAME-2019}. We use here the second-order trapezoidal integration rule, shown below in the particular case that $m=2$. Moreover, we approximate the first derivative in the Neumann boundary condition by the one-sided third-order finite difference formula, which ensures an approximation with a degree of precision of three with respect to $h$, and hence to $\delta$.

\begin{figure}[tb]
\centering
\scriptsize
\begin{tikzpicture}
  \matrix (m)[
    matrix of math nodes, 
    nodes in empty cells,
    nodes={text width={width(999)}, 
    align=right},
    right delimiter=\rbrack,left delimiter=\lbrack
  ] {
1 \\
-1 & 2 & -1 & & \\
& & \ddots & & & & \\
& & -1 & 2 & -1 \\
& & 1 & & & -1 & & & \\
& & & 1 & & & -1 & & & \\
& & & & 1 & & & -1 & & \\
& & & & & -1 & -4 & 10 & -4 & -1 \\
& & & & & & & & \ddots & \\
& & & & & & & -1 & -4 & 10 & -4 & -1 \\
& & & & & & & & & 1 & & & -1 & & \\
& & & & & & & & & & 1 & & & -1 \\
& & & & & & & & & & & 1 & & & -1 \\
& & & & & & & & & & & & -1 & 2 & -1 \\
& & & & & & & & & & & & & & \ddots \\
& & & & & & & & & & & & & & -1 & 2 & -1 \\
& & & & & & & & & & & & & -2 & 9 & -18 & 11 \\
} ;
\begin{pgfonlayer}{myback}
\fhighlight[azure!20]{m-2-1}{m-4-5}
\fhighlightL[azure!20]{m-13-13}{m-16-17}
\fhighlightL[awesome!20]{m-7-6}{m-10-12}
\fhighlightL[cadetgrey!30]{m-4-3}{m-7-8}
\fhighlightL[cadetgrey!30]{m-10-10}{m-13-15}
\fhighlight[asparagus!30]{m-1-1}{m-1-1}
\fhighlightL[asparagus!30]{m-16-14}{m-17-17}
\end{pgfonlayer}
\end{tikzpicture}
\caption{Sketch of the assembled stiffness matrix for the coupling method with matching stresses (MDCM). The first and the last rows, shown in {green}, correspond to the Dirichlet boundary condition at $x_0$ and the Neumann boundary condition at $x_n$, respectively. The two blocks in {blue} correspond to $\Omega_1$ and $\Omega_2$ while the block in {red} corresponds to $\Omega_\delta$. The two blocks in {grey} correspond to $\Gamma_a$ and $\Gamma_b$. Note that, for the sake of simplicity, the zero entries are not shown and the parts of the coefficients involving $E$ or $\kappa\delta^2/2$ and the denominators in $h$ or $h^2$ are omitted.}
\label{fig:MDCMmatrix}
\end{figure}
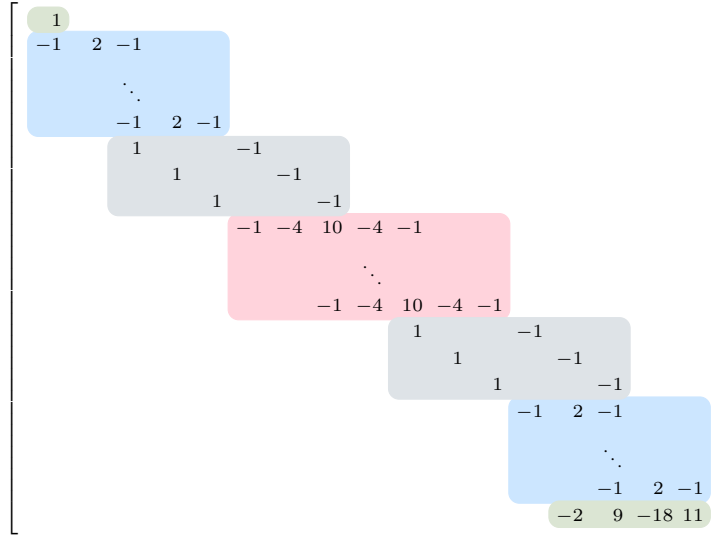

The discretization of Problem~\eqref{eq:CM-displacement} leads to the following system of equations:
\begin{enumerate}[itemsep=0pt,topsep=4pt,parsep=0pt,leftmargin=28pt]
\item 
Dirichlet boundary condition at \(x=0\):
\begin{equation}
u_1 = 0.
\end{equation}
\item 
In \(\Omega_1\): 
\(\forall i=2,\ldots,N_1-1\), and \(k=i-1\):
\begin{equation}
- E \frac{u_{i-1}-2u_i+u_{i+1}}{h^2} = f_b(x_k).
\end{equation}
\item 
In \(\overline{\Gamma}_a\): 
\(\forall i=N_1,\ldots,N_1+m\):
\begin{equation}
u_{i-m} - u_{i+1} = 0.
\end{equation}
\item 
In \(\overline{\Omega}_\delta\): 
\(\forall i=N_1+1+m,\ldots,N_1+N_\delta-m\), and \(k=i-2-m\):
\begin{equation}
- \frac{\kappa\delta^2}{2} \frac{u_{i-2}+4u_{i-1}-10u_i+4u_{i+1}+u_{i+2}}{8h^2} = f_b(x_k).
\end{equation}
\item 
In \(\overline{\Gamma}_b\): 
\(\forall i=N_1+N_\delta+1-m,\ldots,N_1+N_\delta+1\):
\begin{equation}
u_{i-1} - u_{i+m} = 0.
\end{equation}
\item 
In \(\Omega_2\): 
\(\forall i=N_1+N_\delta+2,\ldots,N-1\), and \(k=i-3-2m\):
\begin{equation}
- E \frac{u_{i-1}-2u_i+u_{i+1}}{h^2} = f_b(x_k).
\end{equation}
\item 
Neumann boundary condition at \(x=\ell\):
\begin{equation}
E \frac{-2u_{N-3}+9u_{N-2}-18u_{N-1}+11u_N}{6h} = g.
\end{equation}
\end{enumerate}
We show the structure of the resulting stiffness matrix in Figure~\ref{fig:MDCMmatrix}. It is clear the matrix could be simplified by eliminating the duplicated nodes in the interfaces. In doing so, we would remove the two blocks in grey in Figure~\ref{fig:MDCMmatrix}. However, we prefer to keep the matrix as is for a better comparison of its structure with those of the other two methods. 

\subsection{Algorithm for the coupling method with matching stresses}

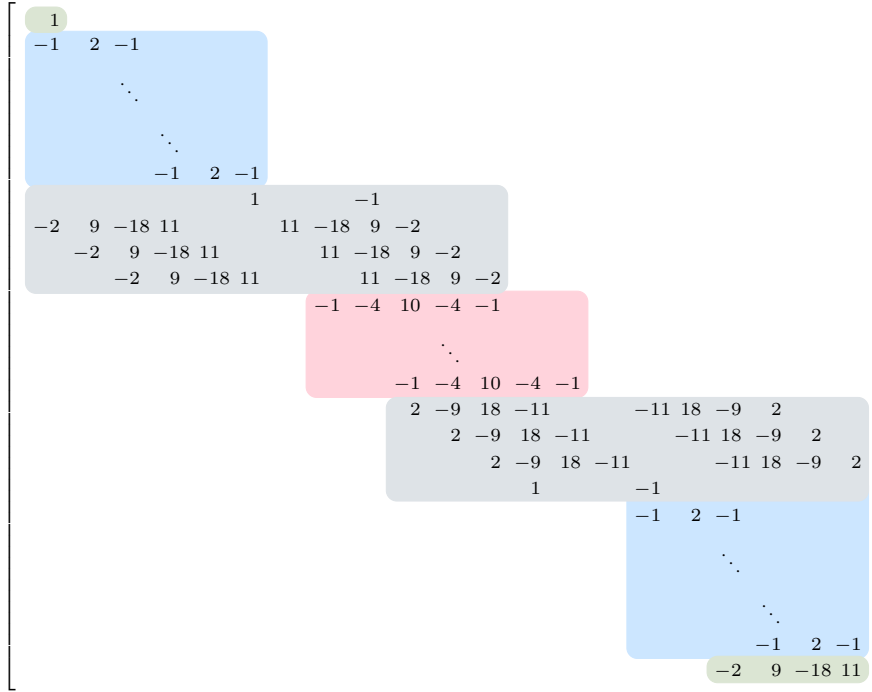
\begin{figure}[tb]
\centering
\scriptsize
\begin{tikzpicture}
  \matrix (m)[
    matrix of math nodes, 
    row sep=0pt, 
    nodes in empty cells,
    nodes={text width={width(999)}, 
    align=right},
    right delimiter=\rbrack,left delimiter=\lbrack
  ] {
1 \\
-1 & 2 & -1 & & \\
& & \ddots & & & & \\
& & & \ddots & & & \\
& & & -1 & 2 & -1 \\
& & & & & 1 & & & -1 \\
-2 & 9 & -18 & 11 & & & 11 & -18 & 9 & -2 \\
& -2 & 9 & -18 & 11 & & &  11 & -18 & 9 & -2 \\
& & -2 & 9 & -18 & 11 & & &  11 & -18 & 9 & -2 \\
& & & & & & & -1 & -4 & 10 & -4 & -1 \\
& & & & & & & & & & \ddots & \\
& & & & & & & & & -1 & -4 & 10 & -4 & -1 \\
& & & & & & & & & 2 & -9 & 18 & -11 & & & -11 & 18 & -9 & 2 & & \\
& & & & & & & & & & 2 & -9 & 18 & -11 & & & -11 & 18 & -9 & 2 & \\
& & & & & & & & & & & 2 & -9 & 18 & -11 & & & -11 & 18 & -9 & 2 \\
& & & & & & & & & & & & 1 & &  & -1 & & & & & \\
& & & & & & & & & & & & & & & -1 & 2 & -1 & & \\
& & & & & & & & & & & & & & & & & \ddots & & \\
& & & & & & & & & & & & & & & & & & \ddots & \\
& & & & & & & & & & & & & & & & & & -1 & 2 & -1 \\
& & & & & & & & & & & & & & & & & -2 & 9 & -18 & 11 \\
} ;
\begin{pgfonlayer}{myback}
\fhighlight[azure!20]{m-2-1}{m-5-6}
\fhighlight[azure!20]{m-16-16}{m-20-21}
\fhighlightL[awesome!20]{m-9-8}{m-12-14}
\fhighlightL[cadetgrey!30]{m-5-1}{m-9-12}
\fhighlightL[cadetgrey!30]{m-12-10}{m-16-21}
\fhighlight[asparagus!30]{m-1-1}{m-1-1}
\fhighlightL[asparagus!30]{m-20-18}{m-21-21}
\end{pgfonlayer}
\end{tikzpicture}
\caption{Sketch of the assembled stiffness matrix for the coupling method with matching stresses (MSCM). The first and the last rows, shown in {green}, correspond to the Dirichlet boundary condition at $x_0$ and the Neumann boundary condition at $x_n$, respectively. The two blocks in {blue} correspond to $\Omega_1$ and $\Omega_2$ while the block in {red} corresponds to $\Omega_\delta$. The two blocks in {grey} correspond to $\Gamma_a$ and $\Gamma_b$. Note that, for the sake of simplicity, the zero entries are not shown and the parts of the coefficients involving $E$ or $\kappa\delta^2/2$ and the denominators in $h$ or $h^2$ are omitted.}
\label{fig:MSCMmatrix}
\end{figure}

We consider here the same numbering of the degrees of freedom, see~\eqref{eq:dofnumbering}, as in the previous case. We note that the peridynamic equation is not solved, contrary to the previous algorithm, at $x=a$ nor at $x=b$, as we apply at those points the constraint on the stresses. The first derivative in these constraints are also approximated by one-sided third-order finite differences stencils as in the Neumann boundary condition. The discretization of Problem~\eqref{eq:CM-stress} then leads to the system of equations:
\begin{enumerate}[itemsep=0pt,topsep=4pt,parsep=0pt,leftmargin=28pt]
\item 
Dirichlet boundary condition at \(x=0\):
\begin{equation}
u_1 = 0.
\end{equation}
\item 
In \(\Omega_1\): 
\(\forall i=2,\ldots,N_1-1\), and \(k=i-1\):
\begin{equation}
- E \frac{u_{i-1}-2u_i+u_{i+1}}{h^2} = f_b(x_k).
\end{equation}
\item 
Constraint on displacement at \(x=a\):
\begin{equation}
u_{N_1} - u_{N_1+1+m} = 0.
\end{equation}
\item 
Constraint on stress in \(\overline{\Gamma}_a\):
\(\forall i=N_1+1,\ldots,N_1+1+m\), and \(k=i-2-m\):
\begin{equation}
\label{eq:constraintG1}
\sigma_{h}^+(x_{k}) - E \frac{-2u_{i-4-m}+9u_{i-3-m}-18u_{i-2-m}+11u_{i-1-m}}{6h} = 0.
\end{equation}
\item 
In \(\Omega_\delta\): 
\(\forall i=N_1+2+m,\ldots,N_1+N_\delta-1-m\), and \(k=i-2-m\):
\begin{equation}
- \frac{\kappa\delta^2}{2} \frac{u_{i-2}+4u_{i-1}-10u_i+4u_{i+1}+u_{i+2}}{8h^2} = f_b(x_k).
\end{equation}
\item 
Constraint on stress in \(\overline{\Gamma}_b\): 
\(\forall i=N_1+N_\delta-m,\ldots,N_1+N_\delta\), and \(k=i-2-m\):
\begin{equation}
\label{eq:constraintG2}
\sigma_{h}^-(x_{k}) - E \frac{-11u_{i+1+m}+18u_{i+2+m}-9u_{i+3+m}+2u_{i+4+m}}{6h} = 0.
\end{equation}
\item 
Constraint on displacement at \(x=b\):
\begin{equation}
u_{N_1+N_\delta-m} - u_{N_1+N_\delta+1} = 0.
\end{equation}
\item In \(\Omega_2\): 
\(\forall i=N_1+N_\delta+2,\ldots,N-1\), and \(k=i-3-2m\):
\begin{equation}
- E \frac{u_{i-1}-2u_i+u_{i+1}}{h^2} = f_b(x_k).
\end{equation}
\item 
Neumann boundary condition at \(x=\ell\):
\begin{equation}
E \frac{-2u_{N-3}+9u_{N-2}-18u_{N-1}+11u_N}{6h} = g.
\end{equation}
\end{enumerate}
Finally, the stresses \(\sigma_{h}^\pm(x_{k})\), introduced in~\eqref{eq:constraintG1} and~\eqref{eq:constraintG2}, are constructed as approximations given by~\eqref{eq:sigma2} with a degree of precision of three with respect to $h$ and $\delta$. Assuming that the solution $u(x)$ is sufficiently regular in the overlapping regions $\Gamma_a$ and $\Gamma_b$, the stresses can be approximated using one sided third-order finite differences stencils of the first derivative. It follows that, \(\forall i=N_1+1,\ldots,N_1+1+m\), and \(k=i-2-m\):
\[
\sigma_{h}^+(x_{k})
= \frac{\kappa\delta^2}{2} \frac{-11u_{i}+18u_{i+1}-9u_{i+2}+2u_{i+3}}{6h}.
\]
In the same way, we get, \(\forall i=N_1+N_\delta-m,\ldots,N_1+N_\delta\), and \(k=i-2-m\):
\[
\sigma_{h}^-(x_{k})
= \frac{\kappa\delta^2}{2} \frac{-2u_{i-3}+9u_{i-2}-18u_{i-1}+11u_i}{6h}.
\]
The structure of the resulting stiffness matrix is shown in Figure~\ref{fig:MSCMmatrix}.

\subsection{Algorithm for the coupling method with variable horizon}

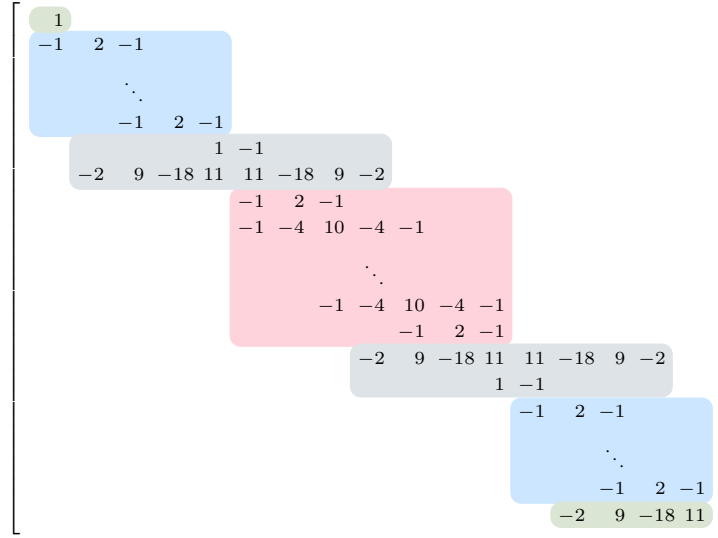
\begin{figure}[tb]
\centering
\scriptsize
\begin{tikzpicture}
  \matrix (m)[
    matrix of math nodes,
    nodes in empty cells,
    nodes={text width={width(999)}, 
    align=right},
    right delimiter=\rbrack,left delimiter=\lbrack
  ] {
1 \\
-1 & 2 & -1 & & \\
& & \ddots \\
& & -1 & 2 & -1 \\
& & & & 1 & -1 \\
& -2 & 9 & -18 & 11 & 11 & -18 & 9 & -2 \\
& & & & & -1 & 2 & -1 & \\
& & & & & -1 & -4 & 10 & -4 & -1 \\
& & & & & & & & \ddots & \\
& & & & & & & -1 & -4 & 10 & -4 & -1 \\
& & & & & & & & & -1 & 2 & -1 \\
& & & & & & & & -2 & 9 & -18 & 11 & 11 & -18 & 9 & -2 \\
& & & & & & & & & & & 1 & -1 & & &  \\
& & & & & & & & & & & & -1 & 2 & -1 & & \\
& & & & & & & & & & & & & & \ddots & & \\
& & & & & & & & & & & & & & -1 & 2 & -1 \\
& & & & & & & & & & & & & -2 & 9 & -18 & 11 \\
} ;
\begin{pgfonlayer}{myback}
\fhighlight[azure!20]{m-2-1}{m-4-5}
\fhighlight[azure!20]{m-14-13}{m-16-17}
\fhighlight[awesome!20]{m-7-6}{m-11-12}
\fhighlightL[cadetgrey!30]{m-4-2}{m-6-9}
\fhighlightL[cadetgrey!30]{m-11-9}{m-13-16}
\fhighlight[asparagus!30]{m-1-1}{m-1-1}
\fhighlight[asparagus!30]{m-17-14}{m-17-17}
\end{pgfonlayer}
\end{tikzpicture}
\caption{Sketch of the assembled stiffness matrix for the coupling method with variable horizon (VHCM). The first and the last rows, shown in {green}, correspond to the Dirichlet boundary condition at $x_0$ and the Neumann boundary condition at $x_n$, respectively. The two blocks in {blue} correspond to $\Omega_1$ and $\Omega_2$ while the block in {red} corresponds to $\Omega_\delta$. The two blocks in {grey} correspond to $\Gamma_a$ and $\Gamma_b$. Note that, for the sake of simplicity, the zero entries are not shown and the parts of the coefficients involving $E$ or $\kappa\delta^2/2$ and the denominators in $h$ or $h^2$ are omitted.}
\label{fig:VHCMmatrix}
\end{figure}

In this approach, we need to modify the numbering of the degrees of freedom as these should be duplicated only at the interface points $x=a$ and $x=b$. In fact, the numbering can easily be obtained from~\eqref{eq:dofnumbering} by setting $m=0$ in those equations. We then have \(N_1=n_1+1\), \(N_\delta= n_\delta+1\), and \(N_2=n_2+1\) so that the total number of degrees of freedom is given by \(N=N_1+N_\delta+N_2\), and 
\begin{equation}
\label{eq:dofnumbering-nooverlap}
\begin{aligned}
u_i &= \underline{u}(x_k), \quad \forall i=1,\ldots,N_1,  && k=i-1, \\
u_i &= u(x_k), \quad \forall i=N_1+1,\ldots,N_1+N_\delta, && k=i-2,\\
u_i &= \underline{u}(x_k), \quad \forall i=N_1+N_\delta+1,\ldots,N, && k=i-3.
\end{aligned}
\end{equation}
The discretization of Problem~\eqref{eq:CM-variablehorizon} then leads to the system of equations:
\begin{enumerate}[itemsep=0pt,topsep=4pt,parsep=0pt,leftmargin=28pt]
\item 
Dirichlet boundary condition at \(x=0\):
\begin{equation}
u_1 = 0.
\end{equation}
\item 
In \(\Omega_1\): 
\(\forall i=2,\ldots,N_1-1\), and \(k=i-1\):
\begin{equation}
- E \frac{u_{i-1}-2u_i+u_{i+1}}{h^2} = f_b(x_k).
\end{equation}
\item 
Constraint on displacement at \(x=a\):
\begin{equation}
u_{N_1} - u_{N_1+1} = 0.
\end{equation}
\item 
Constraint on stress at \(x=a\):
\begin{equation}
\sigma_{h}^+(x_{n_1}) - 
E \frac{-2u_{N_1-3}+9u_{N_1-2}-18u_{N_1-1}+11u_{N_1}}{6h} = 0.
\end{equation}
\item 
In \(\Omega_\delta\): 
\(i=N_1+2\) and \(k=i-2\) (i.e.\ $m=1$):
\begin{equation}
- \frac{\kappa\delta^2}{2} \frac{u_{i-1}-2u_i+u_{i+1}}{h^2} = f_b(x_k).
\end{equation}
\item 
In \(\Omega_\delta\): 
\(\forall i=N_1+3,\ldots,N_1+N_\delta-2\), and \(k=i-2\) (i.e.\ $m=2$):
\begin{equation}
- \frac{\kappa\delta^2}{2} \frac{u_{i-2}+4u_{i-1}-10u_i+4u_{i+1}+u_{i+2}}{8h^2} = f_b(x_k).
\end{equation}
\item 
In \(\Omega_\delta\): 
\(i=N_1+N_\delta-1\) and \(k=i-2\) (i.e.\ $m=1$):
\begin{equation}
- \frac{\kappa\delta^2}{2} \frac{u_{i-1}-2u_i+u_{i+1}}{h^2} = f_b(x_k).
\end{equation}
\item 
Constraint on stress at \(x=b\): 
\begin{equation}
\sigma_{h}^-(x_{n_1+n_\delta}) - 
E \frac{-11u_{N_1+N_\delta+1}+18u_{N_1+N_\delta+2}-9u_{N_1+N_\delta+3}+2u_{N_1+N_\delta+4}}{6h} = 0.
\end{equation}
\item 
Constraint on displacement at \(x=b\):
\begin{equation}
u_{N_1+N_\delta} - u_{N_1+N_\delta+1} = 0.
\end{equation}
\item In \(\Omega_2\): 
\(\forall i=N_1+N_\delta+2,\ldots,N-1\), and \(k=i-3\):
\begin{equation}
- E \frac{u_{i-1}-2u_i+u_{i+1}}{h^2} = f_b(x_k).
\end{equation}
\item Neumann boundary condition at \(x=\ell\):
\begin{equation}
E \frac{-2u_{N-3}+9u_{N-2}-18u_{N-1}+11u_N}{6h} = g.
\end{equation}
\end{enumerate}
Moreover, the approximate stresses $\sigma_{h}^+(x_{n_1})$ and $\sigma_{h}^-(x_{n_1+n_\delta})$ will be given by:
\begin{align}
& \sigma_{h}^+(x_{n_1}) 
= E 
\frac{-11u_{N_1+1}+18u_{N_1+2}-9u_{N_1+3}+2u_{N_1+4}}{6h},\\
& \sigma_{h}^-(x_{n_1+n_\delta}) 
= E
\frac{-2u_{N_1+N_\delta-3}+9u_{N_1+N_\delta-2}-18u_{N_1+N_\delta-1}+11u_{N_1+N_\delta}}{6h}.
\end{align}
Finally, the stiffness matrix corresponding to the coupling method with variable horizon is shown in Figure~\ref{fig:VHCMmatrix}.

\section{Numerical examples}
\label{Sect:numericalexamples}

The objective of this section is to present several numerical examples in order to compare the solutions of the three coupling approaches, namely the coupling method with matching displacements (MDCM), the coupling method with matching stresses (MSCM), and the coupling method with variable horizon (VHCM). 
For simplicity, but without loss of generality, we shall consider in all experiments a one-dimensional bar with material property $E=1$, so that $\kappa=2/\delta^2$.
Some examples will involve a bar with prescribed mixed boundary conditions, i.e.\ with a homogeneous Dirichlet boundary at $x=0$ and a Neumann boundary condition at $x=\ell$ while other problems will consider homogeneous Dirichlet boundary conditions at both extremities. 
Since the three coupling methods have all a degree of precision of three, we shall first consider a series of examples with manufactured solutions involving polynomials of degree up to three to confirm that the methods all lead to the exact solution in that case. We shall subsequently study problems with manufactured solutions made of polynomial functions of degree four.

\subsection{Problems with cubic solutions}

\begin{figure}[tb!]
\centering
\subfloat[]{\includegraphics[height=0.38\textwidth]{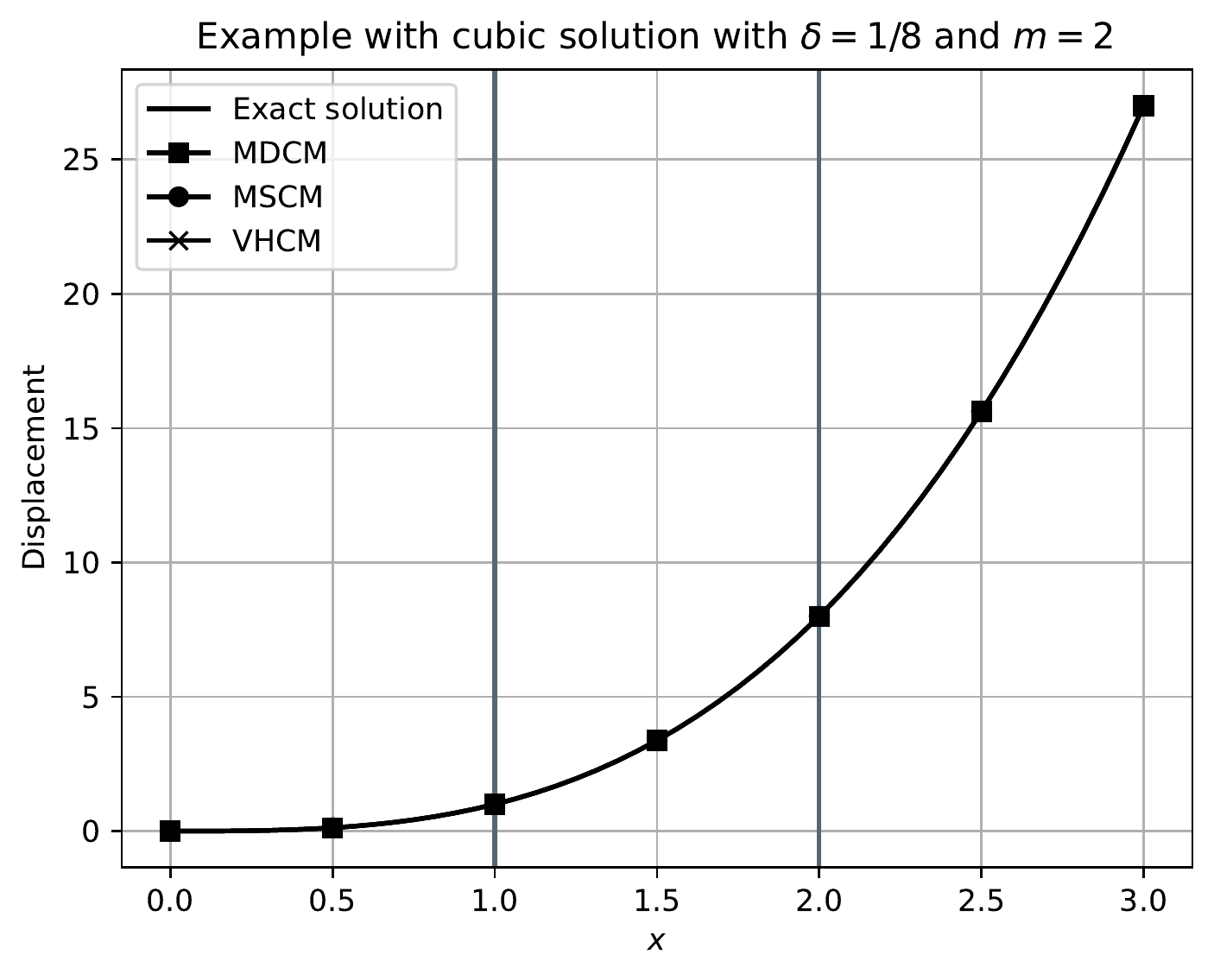}}
\hfill
\subfloat[]{\includegraphics[height=0.38\textwidth]{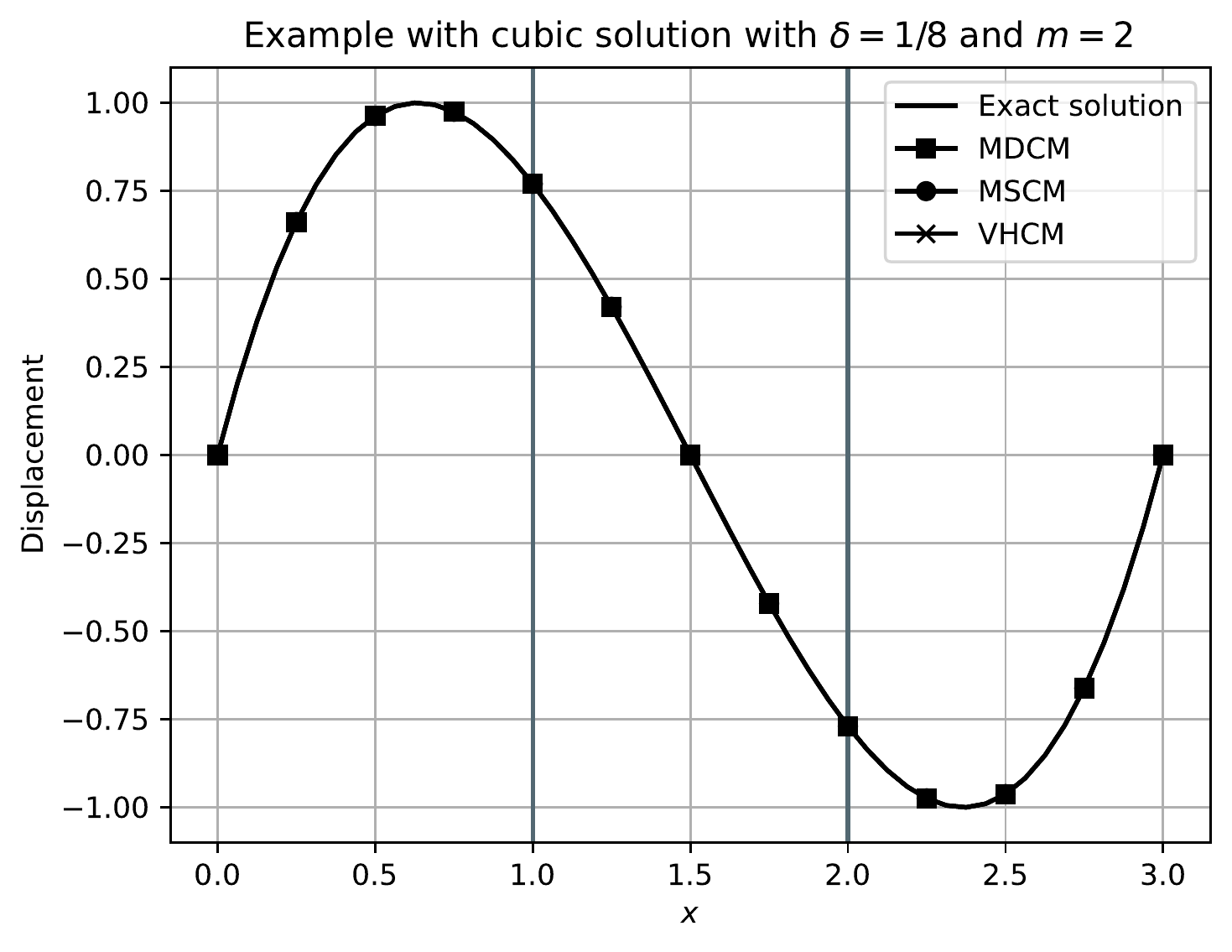}}
\caption{Examples with cubic manufactured solutions~\eqref{eq:cubicMixedBC} and~\eqref{eq:cubicDirichletBC} for the problems (a) with mixed boundary conditions and (b) with Dirichlet boundary conditions at both extremities, respectively. The numerical solutions are shown here with horizon $\delta = 1/8$ and grid size $h=\delta/m=1/16$.}
\label{Fig:cubicsolutions}
\end{figure}

We consider the configuration of Figure~\ref{Fig:couplingLinearElastPeryd} with $\ell=3$ and interface locations $a=1$ and $b=2$. In order to have nodes at $x=1$ and at $x=2$, the domain is partitioned into sub-intervals of size $h=1/n$, with $n$ given, so that the grid consists of $3n$ elements. The horizon is taken as a multiple of $h$, i.e.\ $\delta=mh$. The first example illustrates the linear elasticity problem with mixed boundary conditions for which the manufactured solution is chosen as
\begin{equation}
\label{eq:cubicMixedBC}
\uelast(x) = x^3,
\end{equation}
so that the data is given by
\[
f_b(x)= -\uelast''(x) = -6 x, \qquad g=\uelast'(\ell) = 3\ell^2=27.
\]
In the second example, we consider homogeneous Dirichlet boundary conditions at both extremities. The manufactured solution is constructed in this case as 
\begin{equation}
\label{eq:cubicDirichletBC}
\uelast(x) 
= \frac{2}{3\sqrt{3}}\ x(3-2x)(3-x), 
\end{equation}
so that the displacement is zero at $x=0$ and $x=\ell$ and the corresponding loading term is given by
\[
f_b(x) = - \uelast''(x) = - \frac{2}{\sqrt{3}} (-6 + 4x).
\]
The solutions obtained with the three coupling methods are shown in Figure~\ref{Fig:cubicsolutions} for the two cases. Unsurprisingly, the graphs confirm that each approach is able to reproduce the exact solution without error. The numerical solutions are shown here only in the case where $\delta=1/8$ and $m=2$ but they of course coincide with the exact solution with any other consistent values of these parameters.

\subsection{Problems with quartic solutions}

We consider in this section several problems whose manufactured solutions are given by polynomials of degree four. We know that the coupling methods will necessarily provide approximations with non-vanishing errors when compared to the solution to the classical elasticity problem. One difficulty in this case is to determine a measure to assess and compare the performance of the three coupling methods. In order to so, we propose to evaluate, for a given discretization $h$ of the domain, the difference $\Delta(x)$ between the discrete solution $u_{h,\text{\itshape CM}}$ of each coupling method and the finite difference solution $\uelast_{h,\text{\itshape FDM}}$ of the classical linear elasticity problem described in Section~\ref{Sect:CLEM}, that is
\begin{equation}
\label{eq:error}
    \Delta(x) = u_{h,\text{\itshape CM}}(x) - \uelast_{h,\text{\itshape FDM}}(x), 
    \quad \forall x \in \overline{\Omega}.
\end{equation}
We have seen in Section~\ref{Sect:CF-examples} that $\Delta(x)$ corresponds to an approximation of the function $v(x)$ (denoted $=v_N(x)$ in the case of mixed boundary conditions and $v_D(x)$ in the case of Dirichlet boundary conditions) satisfying Problem~\eqref{eq:function-v} that admits a maximal value $v_{\max}$ in $\overline{\Omega}$. We thus introduce in each case the error $\Delta_\text{max}$ as
\begin{equation}
\label{eq:errormax}
    \Delta_\text{max} = \max_{x \in \overline{\Omega}} \Delta(x),
\end{equation}
and compute the relative error 
\begin{equation}
\label{eq:relativeerror}
    \mathcal E_r = \bigg| \frac{\Delta_\text{max} - v_{\max}}{v_{\max}} \bigg|
\end{equation}
in order to assess the quality of the \edit{discretized} coupling methods. We note that the value of $v_{\max}$ depends on the exact solution $\uelast$ of the linear elasticity problem, on the type of boundary conditions, as well as on the profile of the horizon $\delta$ over the domain in terms of the interface locations $a$ and $b$.

In the examples below, we will treat both the $\delta$-convergence~(see e.g.~\cite{seleson2016convergence,bobaru2009-1D}), for which $m$ is fixed while $\delta$ goes to zero, and the $m$-convergence, for which $\delta$ is fixed while $m$ increases, that is $h$ goes to zero.

\subsubsection{Case with mixed boundary conditions}

\begin{figure}[tb!]
    \centering
    \includegraphics[width=0.485\textwidth]{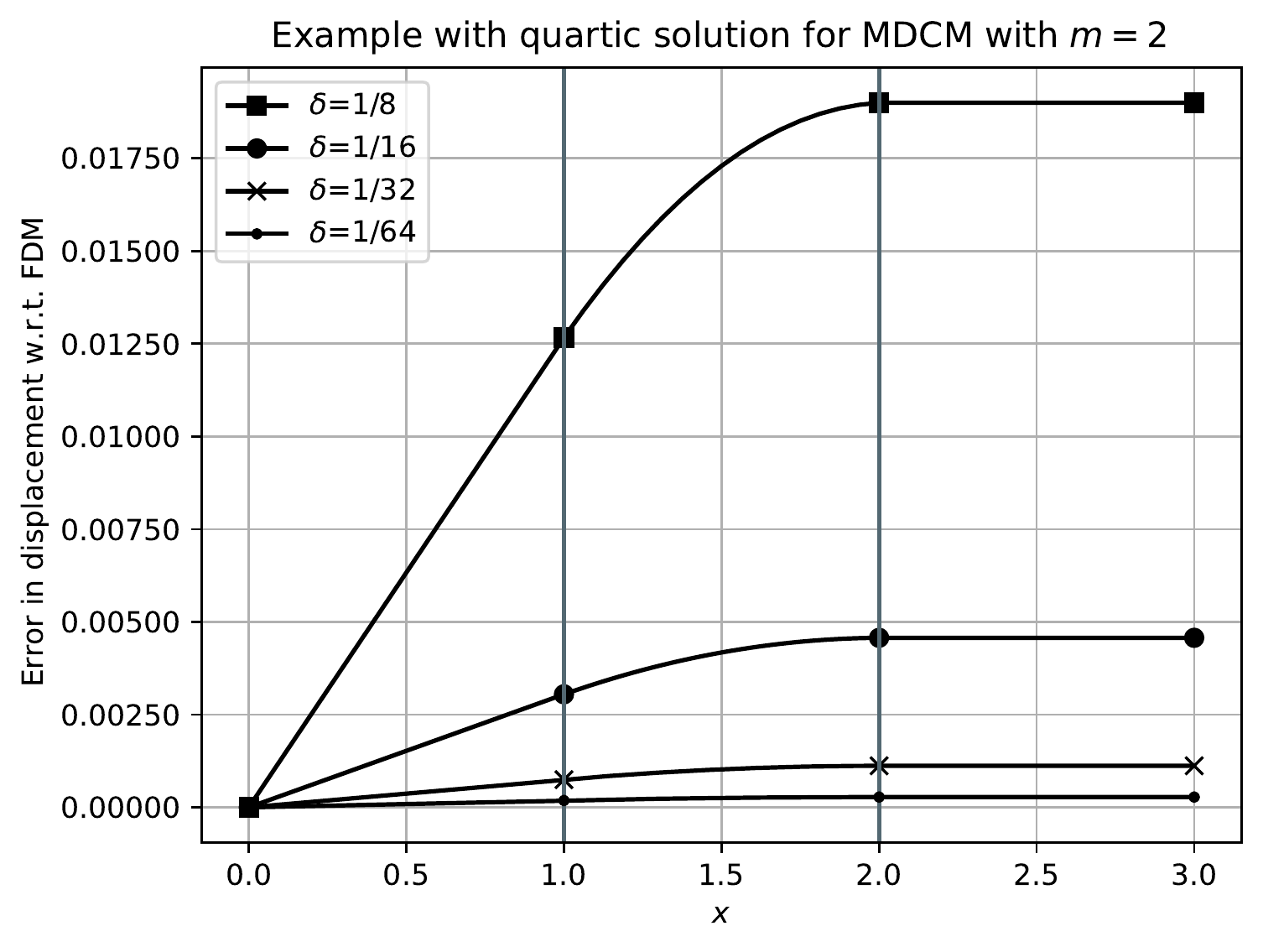}
    \hfill
    \includegraphics[width=0.485\textwidth]{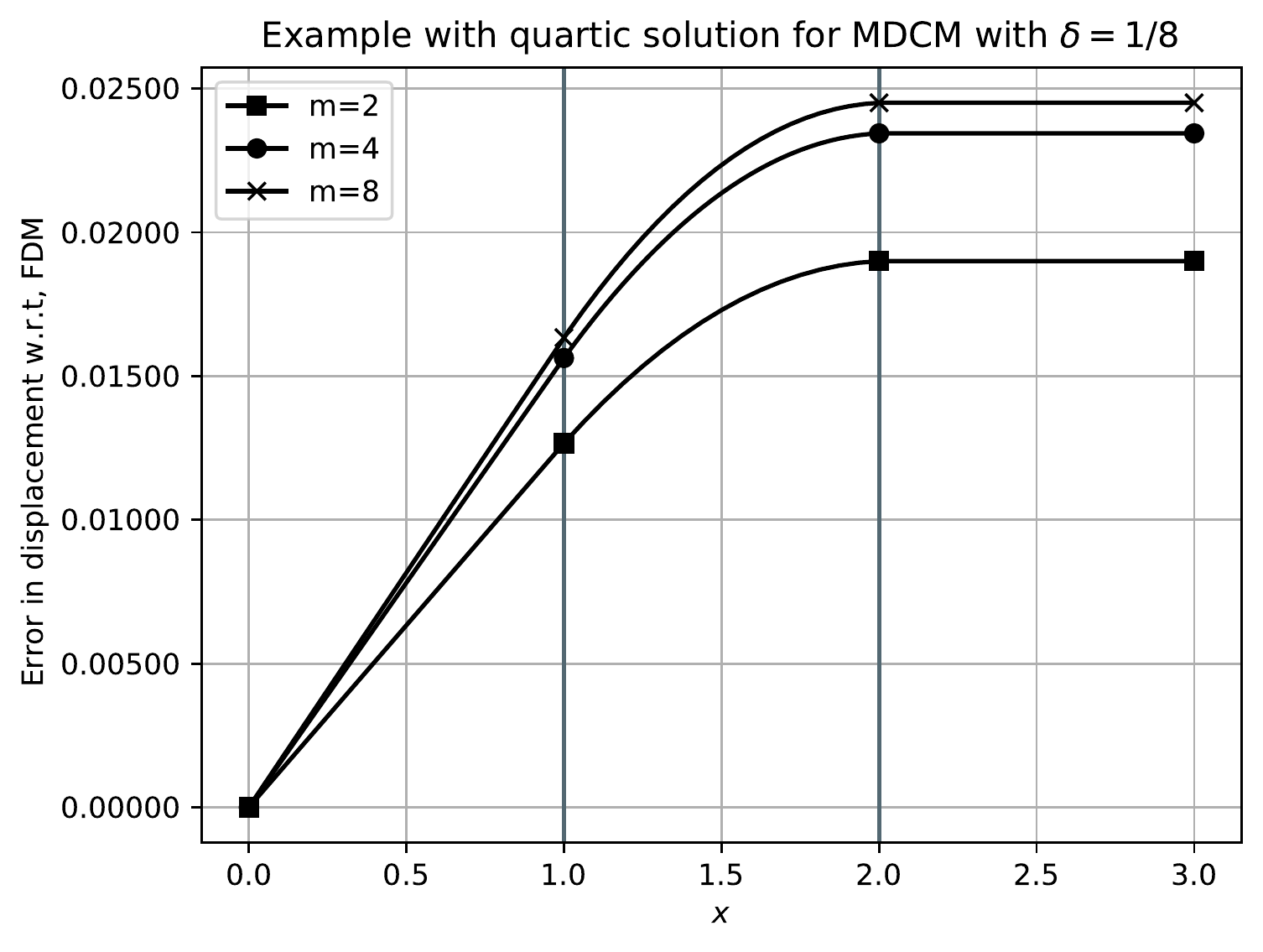}
    \\
    \includegraphics[width=0.485\textwidth]{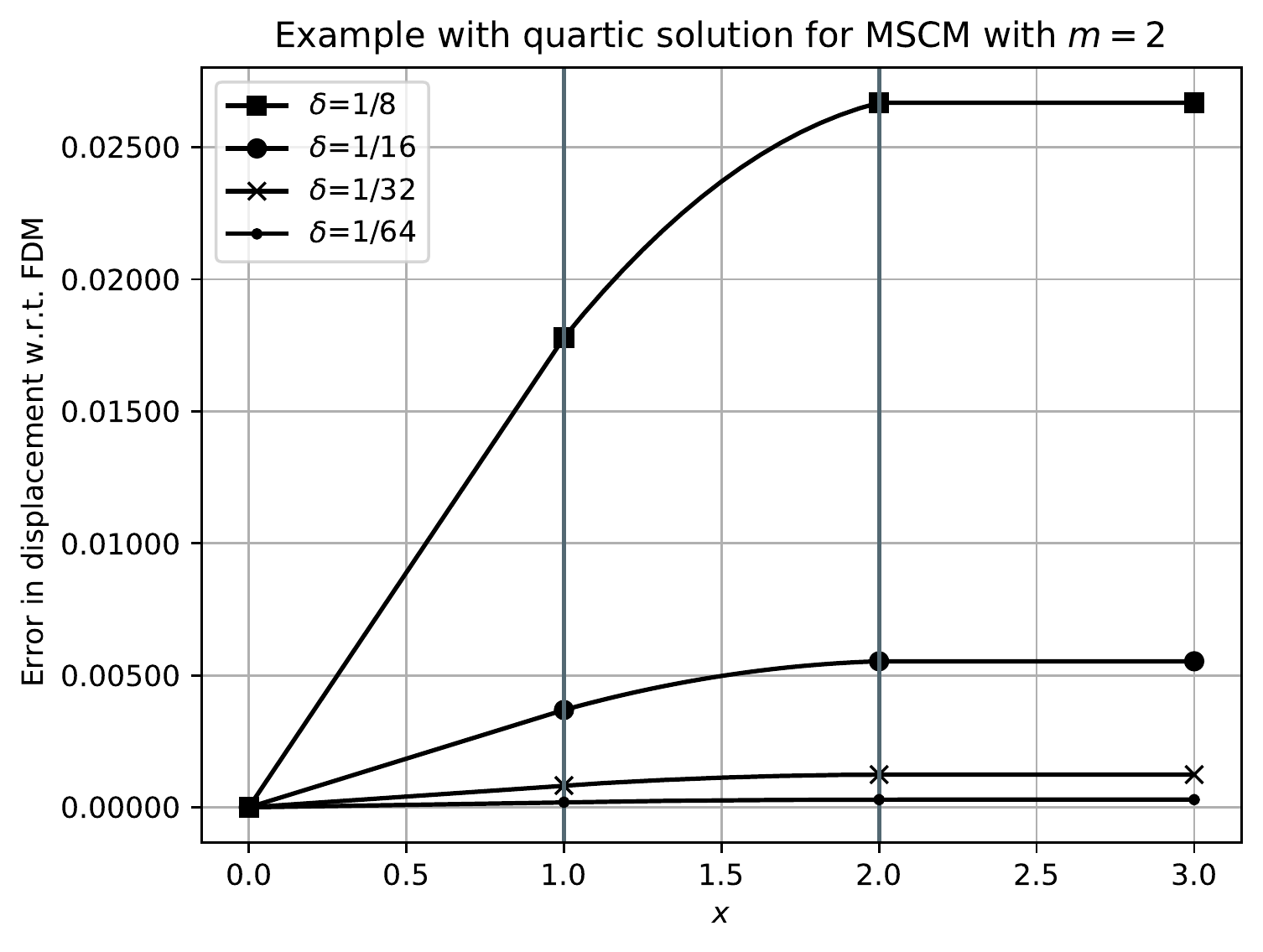}
    \hfill
    \includegraphics[width=0.485\textwidth]{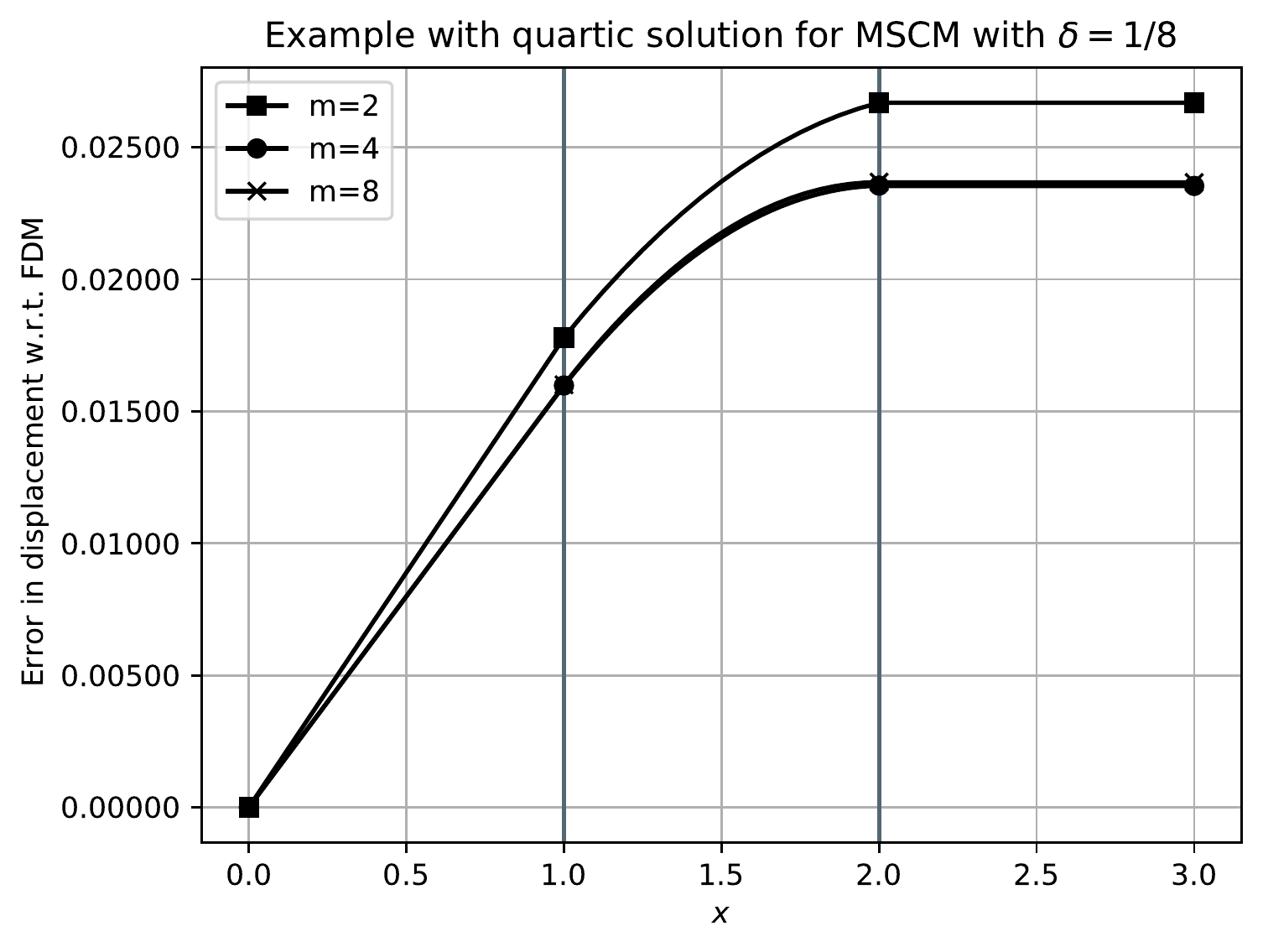}    
    \\
    \includegraphics[width=0.485\textwidth]{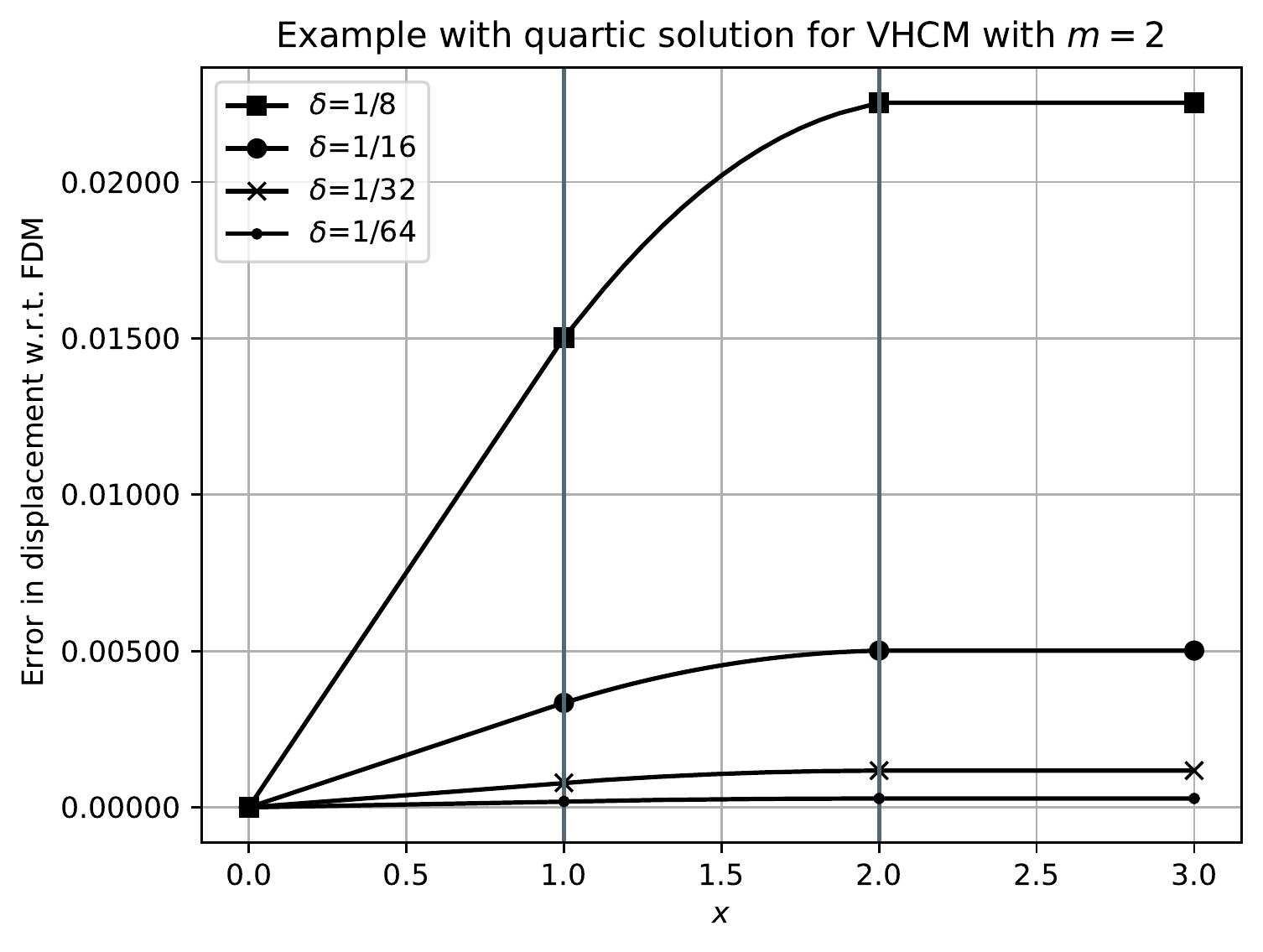}
    \hfill
    \includegraphics[width=0.485\textwidth]{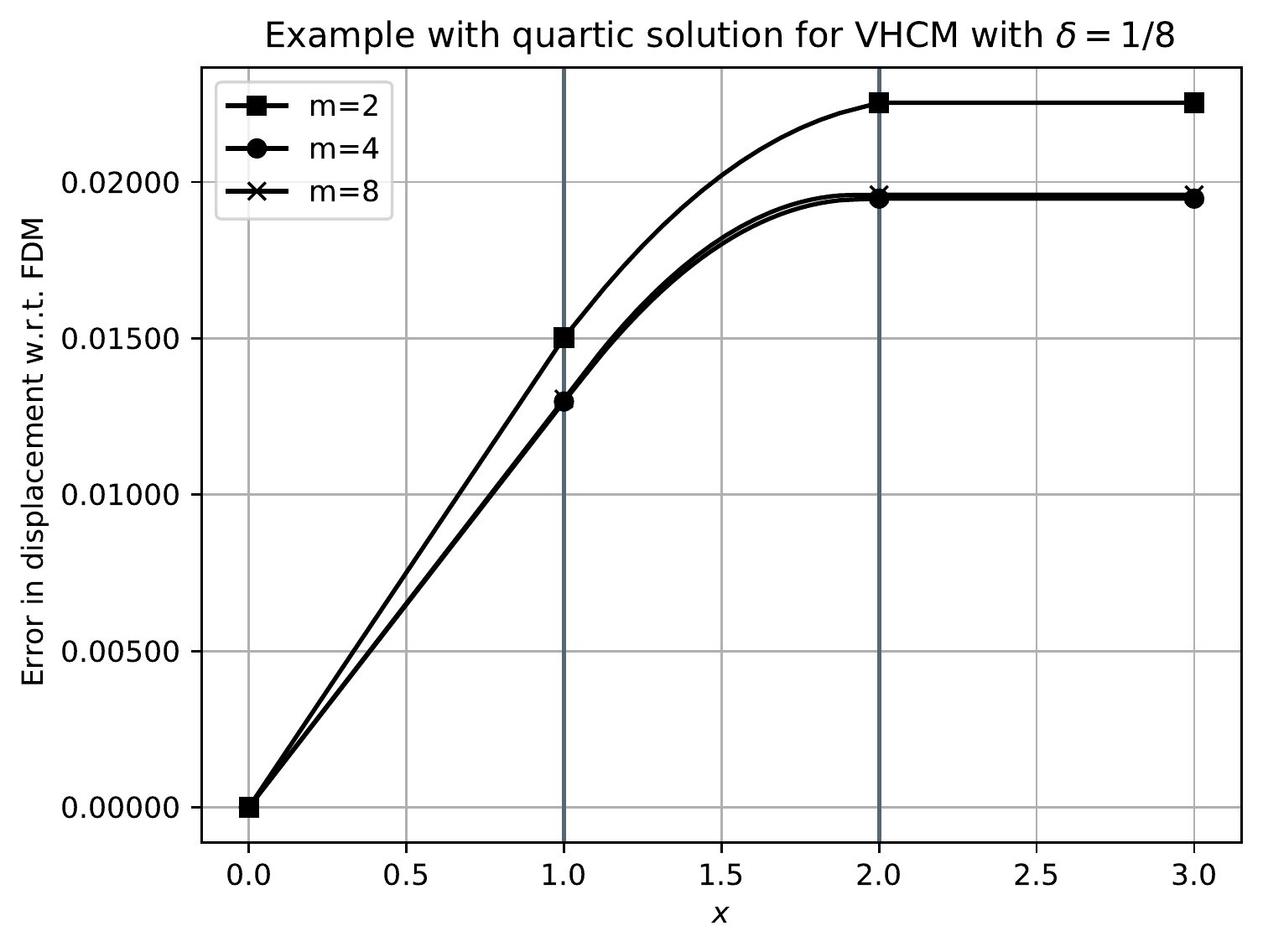}
    \caption{Error $\Delta(x)$ obtained by the three coupling methods (top row: MDCM, middle row: MSCM, and bottow row: VHCM) in the case of the quartic solution for the problem with mixed boundary conditions and interface locations $a=1$ and $b=2$. Left column: $\delta$-convergence with $m=2$. Right column: $m$-convergence with $\delta=1/8$.}
    \label{fig:SNBC-quartic-error}
\end{figure}

As before, we define the computational domain with $\ell=3$ and interface locations $a=1$ and $b=2$. We start with the simplest quartic manufactured solution, i.e.\ the monomial
\begin{equation}
    \uelast(x) = x^4.
\end{equation}
In the case of mixed boundary conditions, the data of the problem with $E=1$ are then given by 
\begin{equation}
    f_b(x) = - \uelast''(x) = - 12x^2, \qquad g = \uelast'(\ell) = 4 \ell^3.
\end{equation}
For MDCM and MSCM, the maximal value of $v_N(x)$ is attained for any point in the interval $[2,3]$ and is given by~\eqref{eq:vmaxNeumann}. For VHCM, the value needs to be corrected to take into account the fact that the horizon $\delta$ varies in the proximity of the interfaces. Omitting the lengthy calculations, one gets:
\begin{equation}
    v_{N,\max} 
    = \frac{\lambda\delta^2}{48} (b^2 - a^2) - \frac{\lambda}{36} (a+b) \delta^3
    = \frac{3}{2} \delta^2 - 2 \delta^3,
\end{equation}
where it is reminded that $\lambda = \uelast''''=24$ and $\delta$ corresponds to the constant provided in~\eqref{eq:deltafn}. Values of $v_{N,\max}$ for various horizons and for the three coupling methods are reported in Table~\ref{tab:SNBC-quartic-error}. We also compile in this table the numerical values of $\Delta_\text{max}$ and $\mathcal E_r$ obtained by the three methods and several values of~$\delta$ and~$m$. Plots of the function $\Delta(x)$ are shown in Figure~\ref{fig:SNBC-quartic-error} for a corresponding subset of~$\delta$ and~$m$.

We observe from these results that the error $\Delta_\text{max}$ exhibits quadratic convergence with respect to the parameter $\delta$ for the three coupling methods, as expected, meaning that their solutions all converge to the classical linear elasticity solution as $\delta$ tends to zero. \edit{First, we see for $m=8$ that the errors $\Delta_\text{max}$ are all very close to the respective value of $v_{N,\text{max}}$ for the three approaches, even more so when $\delta$ is small. However, we would like to point out that the behavior of MDCM clearly differs from that of MSCM and VHCM in terms of $m$-convergence, and in particular for $\delta=1/8$, see for instance the right column of Figure~\ref{fig:SNBC-quartic-error}. First, for $m=2$ and each value of~$\delta$, we observe that the relative error $\mathcal E_r$ associated with MDCM is consistently larger than that obtained by the other two methods. We also see that $\mathcal E_r$ is larger for MDCM than for MSCM for $m=4$ when $\delta=1/16$, $1/32$, and $1/64$. In the case $m=4$ and $\delta=1/8$, the relative error for MDCM is coincidentally much smaller than that for MSCM, but increases again a lot when $m$ is increased to $8$ due to the fact that $\Delta_\text{max}$ now overshoots $v_{N,\text{max}}$. In fact, we regularly observe larger variations in $\Delta_\text{max}$ for MDMC when the value of $m$ is increased. On the other hand, the behaviors of the MSCM and VHCM solutions are similar. For these reasons, the latter two methods seem to exhibit a more stable behavior for the coupling of peridynamic and classical linear elasticity models.}



\begin{table}[tb!]
    \sisetup{round-mode=places,round-precision=7,group-separator={}}
    \centering
    \begin{tabular}{cc|ccc|ccc}
       & &  \multicolumn{3}{c|}{Error $\Delta_\text{max}$} & \multicolumn{3}{c}{Relative error $\mathcal E_r$} \\
       \midrule
       $\delta$ & $m$ & MDCM & MSCM & VHCM & MDCM & MSCM & VHCM  \\ 
       \midrule
       $\sfrac{1}{8}$ & 2  & \num{0.0189985} & \num{0.0266890} & \num{0.0225350} &  \nums{0.18939563036353016} & \nums{0.13872936965799454} & \nums{0.1537912607587714} \\
       & 4  & \num{0.0234427} & \num{0.0252280} & \num{0.0194695} &  \nums{0.00022319627836016784} & \nums{0.07639507127320637} & \nums{0.0031632654630811886} \\
       & 8  & \num{0.0245057} & \num{0.0250035} & \num{0.0195917} &  \nums{0.045576525052941484} & \nums{0.06681675947735737} & \nums{0.003092732426011935} \\
       \cmidrule{2-5}
       \multicolumn{2}{r|}{$v_{N,\max}$} & \num{0.0234375} & \num{0.0234375} & \num{0.0195312} \\
       \midrule
       $\sfrac{1}{16}$ & 2  & \num{0.0045721} & \num{0.0055334} & \num{0.0050141} &  \nums{0.21969781457543527} & \nums{0.05563531454269347} & \nums{0.06645851712197658} \\
       & 4  & \num{0.0056769} & \num{0.0059001} & \num{0.0051803} &  \nums{0.031138402603877086} & \nums{0.006947535851698679} & \nums{0.03552875692680986} \\
       & 8  & \num{0.0059471} & \num{0.0060094} & \num{0.0053329} &  \nums{0.014975760039912226} & \nums{0.025595878730503802} & \nums{0.007116942243142562} \\
       \cmidrule{2-5}
       \multicolumn{2}{r|}{$v_{N,\max}$} & \num{0.0058594} & \num{0.0058594} & \num{0.0053711} \\
       \midrule
       $\sfrac{1}{32}$ & 2  & \num{0.0011208} & \num{0.0012410} & \num{0.0011761} &  \nums{0.23484890571368547} & \nums{0.15281765506369993} & \nums{0.16221928860708748} \\
       & 4  & \num{0.0013963} & \num{0.0014242} & \num{0.0013342} &  \nums{0.04681920610892121} & \nums{0.027776237957975052} & \nums{0.04960071388632059} \\
       & 8  & \num{0.0014644} & \num{0.0014721} & \num{0.0013876} &  \nums{0.0003245869350697224} & \nums{0.0049854588869493455} & \nums{0.011555923639958643} \\
       \cmidrule{2-5}
       \multicolumn{2}{r|}{$v_{N,\max}$} & \num{0.0014648} & \num{0.0014648} & \num{0.0014038} \\
       \midrule
       $\sfrac{1}{64}$ & 2  & \num{0.0002774} & \num{0.0002925} & \num{0.0002843} &  \nums{0.24242442335040928} & \nums{0.20140880160033703} & \nums{0.20704345492289422} \\
       & 4  & \num{0.0003462} & \num{0.0003497} & \num{0.0003384} &  \nums{0.05465993508308505} & \nums{0.04513808493114387} & \nums{0.05618760598070444} \\
       & 8  & \num{0.0003633} & \num{0.0003643} & \num{0.0002925} &  \nums{0.00797462997919259} & \nums{0.005320158602747445} & \nums{0.013634200316873636} \\
       \cmidrule{2-5}
       \multicolumn{2}{r|}{$v_{N,\max}$} & \num{0.0003662} & \num{0.0003662} & \num{0.0003586} \\
       \midrule
    \end{tabular}
    \caption{Error $\Delta_\text{max}$ and relative error $\mathcal E_r$ with respect to parameters $\delta$ and $m$ obtained by the three coupling methods in the case of the quartic solution for the problem with mixed boundary conditions and interface locations $a=1$ and $b=2$.}
    \label{tab:SNBC-quartic-error}
\end{table}

\subsubsection{Case with Dirichlet boundary conditions at both extremities}

\begin{figure}[tb!]
    \centering
    \includegraphics[width=0.485\textwidth]{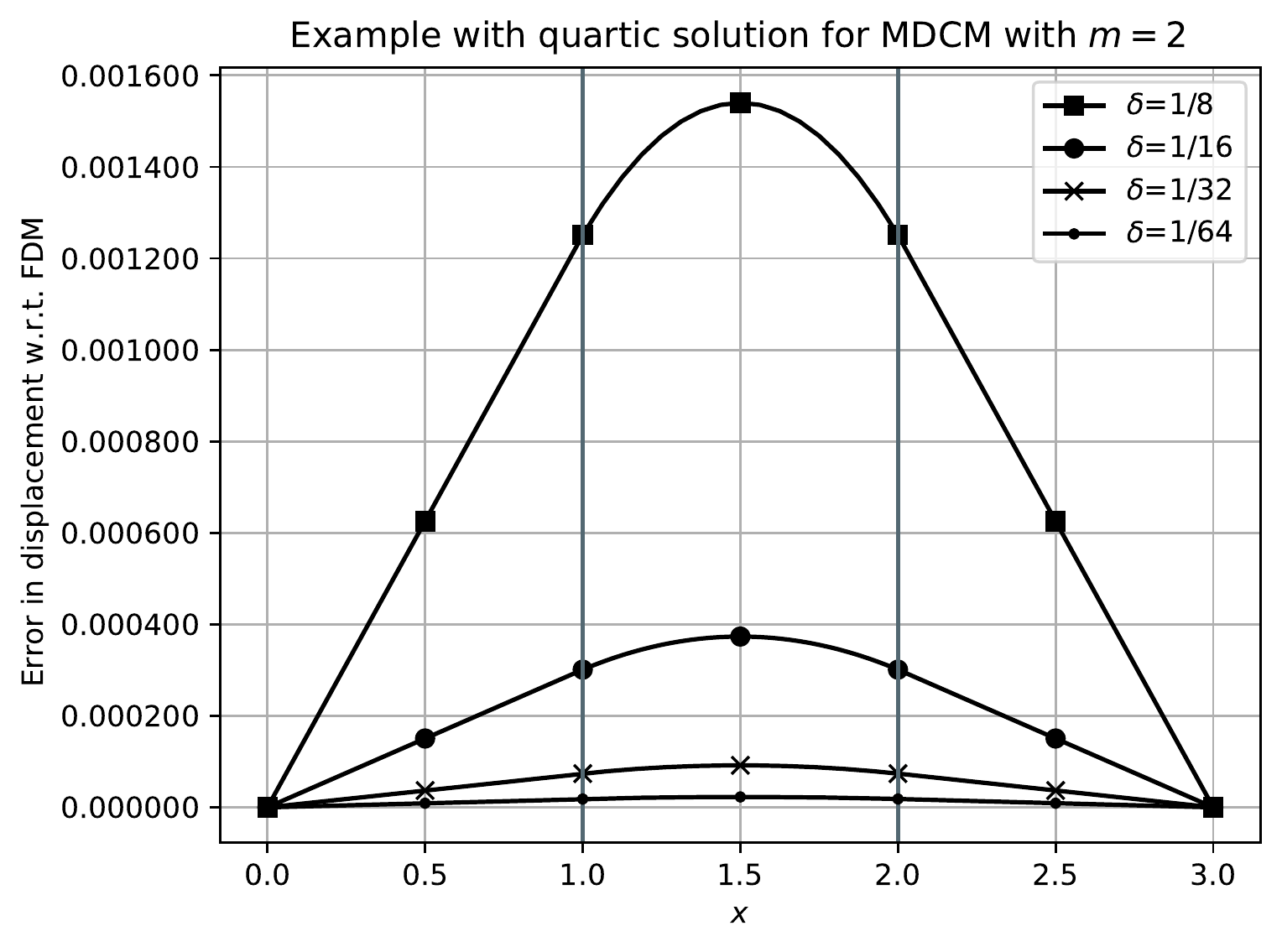}
    \hfill
    \includegraphics[width=0.485\textwidth]{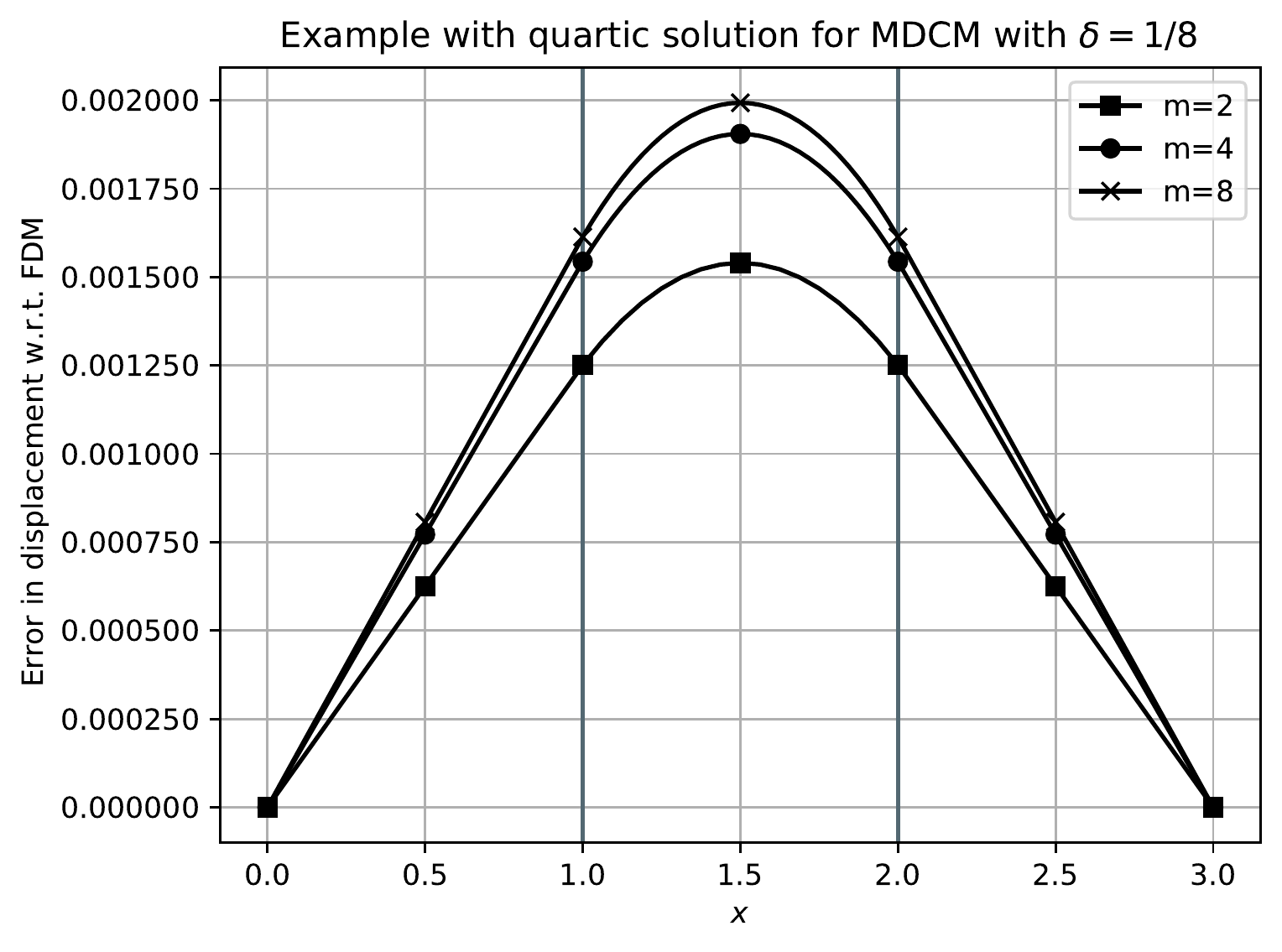} 
    \\
    \includegraphics[width=0.485\textwidth]{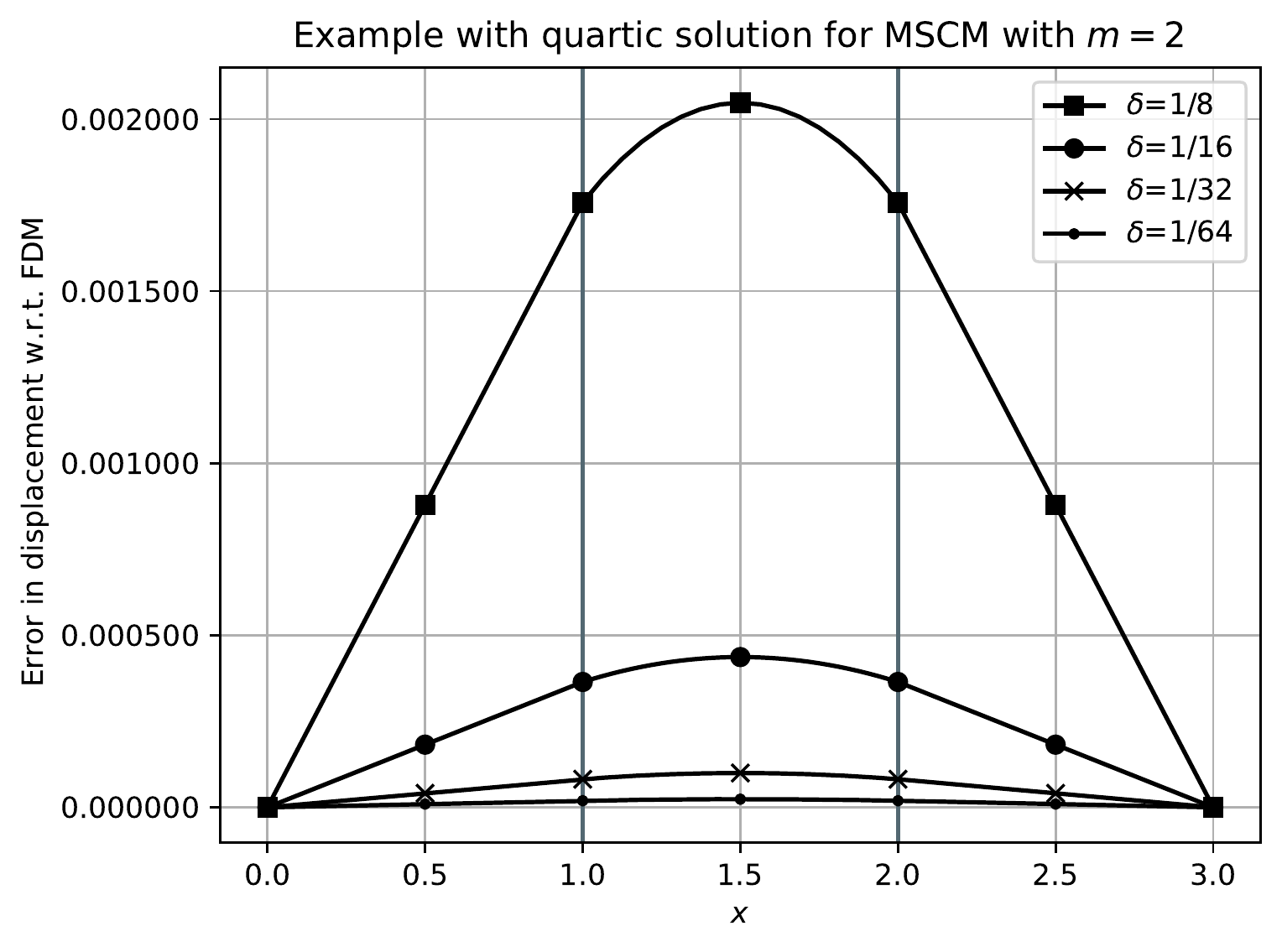}
    \hfill
    \includegraphics[width=0.485\textwidth]{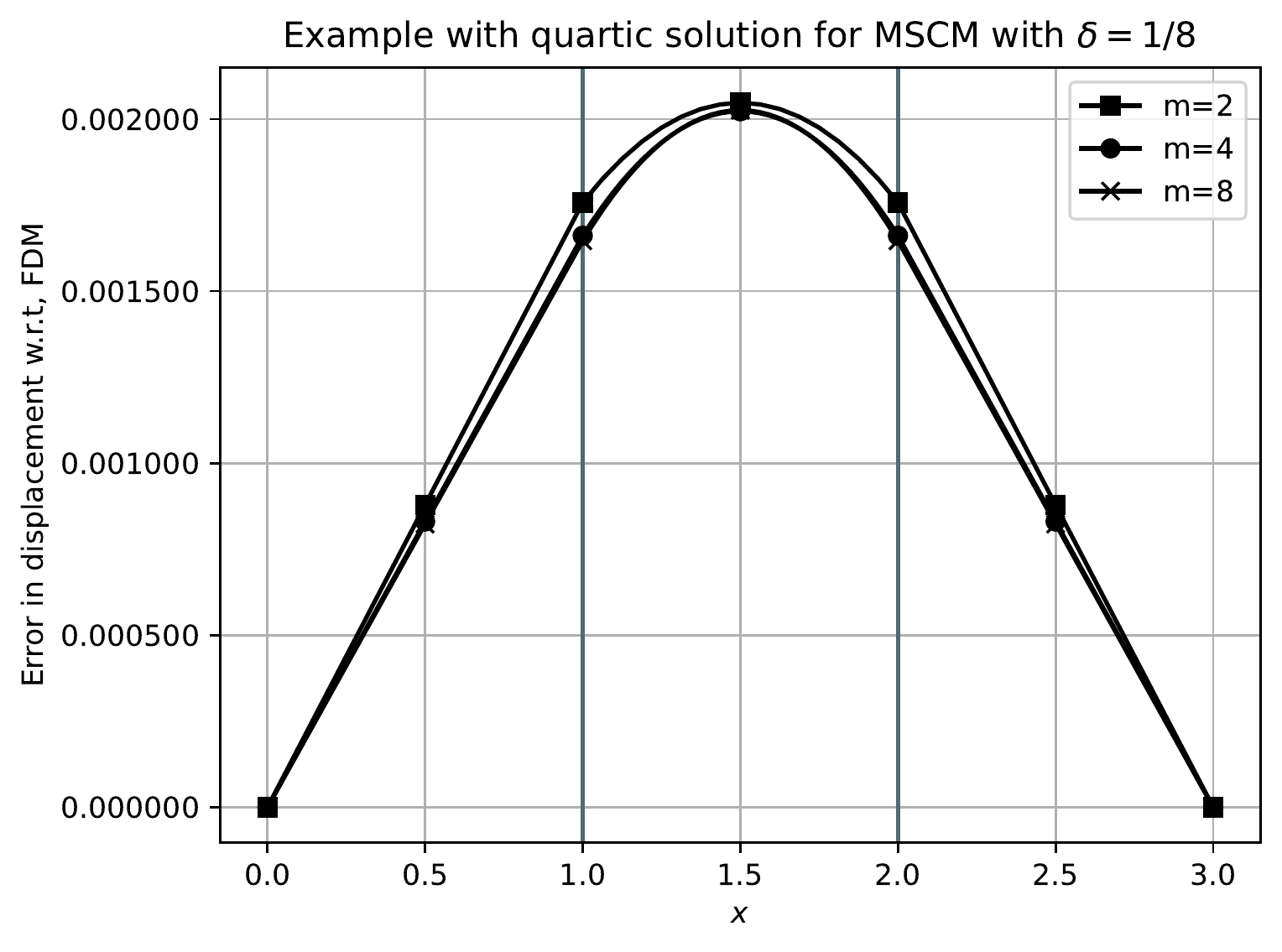}
    \\
    \includegraphics[width=0.485\textwidth]{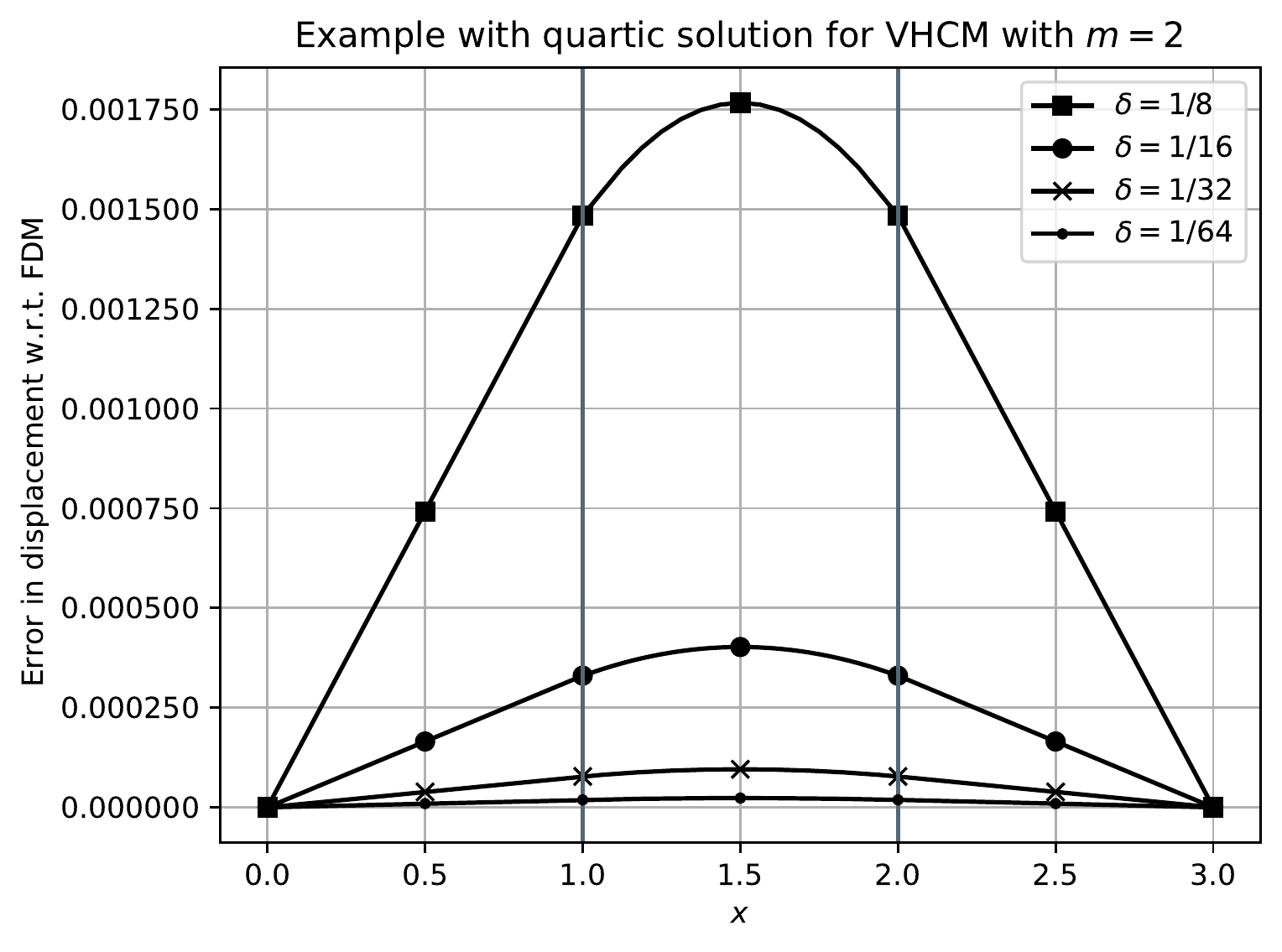}
    \hfill
    \includegraphics[width=0.485\textwidth]{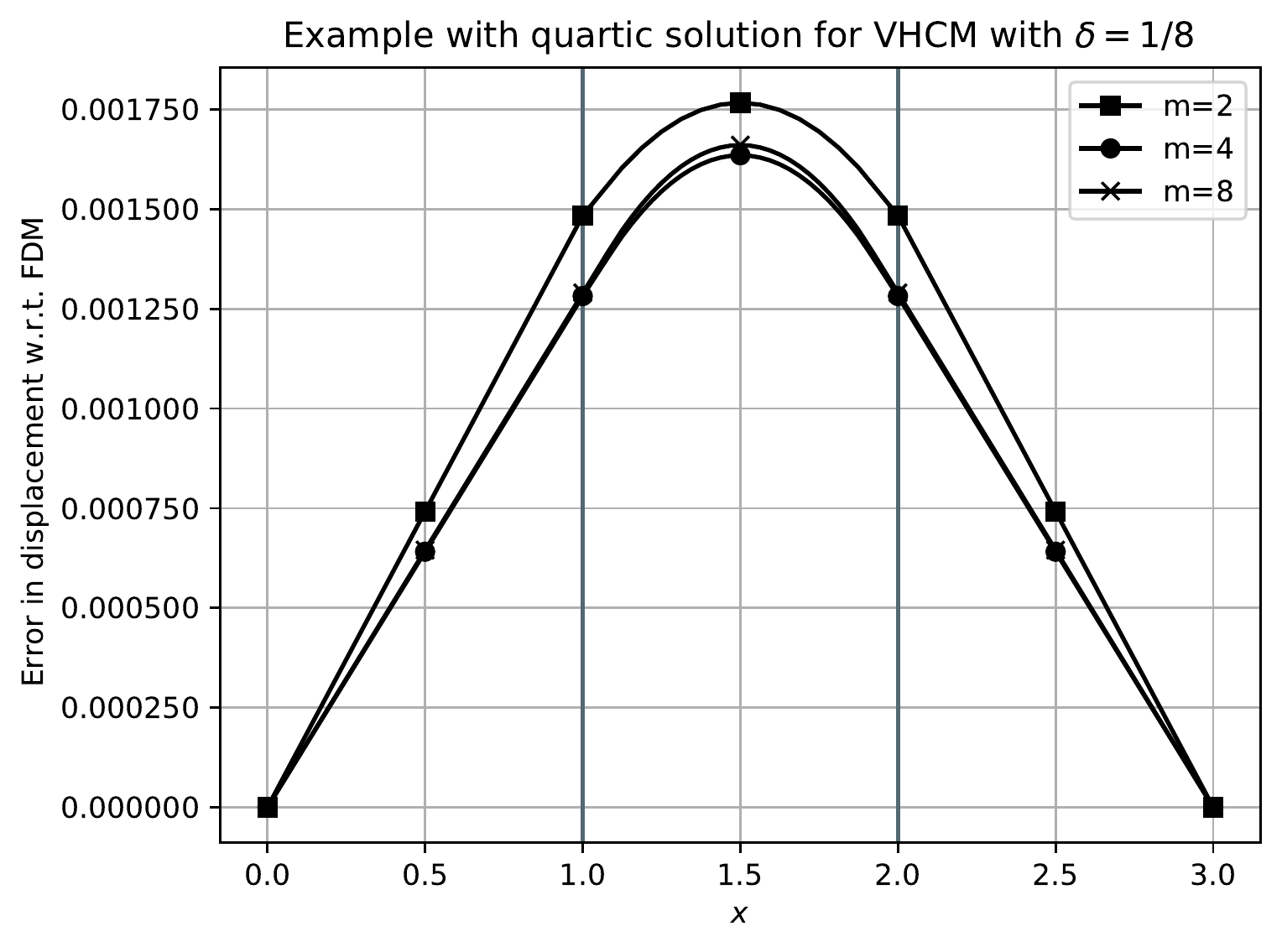}    
    \caption{Error $\Delta(x)$ obtained by the three coupling methods (top row: MDCM, middle row: MSCM, and bottow row: VHCM) in the case of the quartic solution for the problem with Dirichlet boundary conditions at both extremities and interface locations $a=1$ and $b=2$. Left column: $\delta$-convergence with $m=2$. Right column: $m$-convergence with $\delta=1/8$.}
    \label{fig:SDBC-quartic-error}
\end{figure}

We first consider a symmetric problem with a symmetric quartic solution with respect to $x=3/2$. The computational domain is kept as $\Omega=(0,3)$ with interfaces located at $a=1$ and $b=2$. The manufactured displacement field is chosen as
\begin{equation}
\label{eq:DBC-manufacturedsol}
\uelast(x) 
= \frac{16}{81}\ x^2(3-x)^2 
= \frac{16}{9}x^2 - \frac{32}{27} x^3 + \frac{16}{81} x^4,
\end{equation}
such that the corresponding load is given by
\begin{equation}
f_b(x) 
= - \uelast''(x)
= - \frac{32}{9} + \frac{64}{9}x - \frac{64}{27} x^2.
\end{equation}
Moreover, we solved the problem with homogeneous Dirichlet boundary conditions at both extremities $x=0$ and $x=\ell=3$. For MDCM and MSCM, the maximal value of $v_D(x)$ is in this case achieved at $x=3/2$ due to the symmetry of the problem and is given by~\eqref{eq:vmaxDirichlet}. For VHCM, this value is modified as follows: \begin{equation}
    v_{D,\max} 
    = \frac{\lambda\delta^2}{48} (b^2 - a^2) \frac{4\ell^2-4b\ell+ b^2-a^2}{4\ell^2} 
    - \frac{\lambda\delta^3}{48} \frac{8a+3\delta}{6}
    = \frac{10}{81} \delta^2 - \frac{4}{243} \delta^3 (8+3\delta),
\end{equation}
where $\lambda = \uelast''''=128/27$. The quantity $v_{D,\max}$, as well as the numerical values of $\Delta_\text{max}$ and $\mathcal E_r$, are provided in Table~\ref{tab:SDBC-quartic-error} for several values of $\delta$ and $m$. Moreover, we show in Figure~\ref{fig:SDBC-quartic-error} some plots of $\Delta(x)$ to illustrate the $\delta$-convergence and $m$-convergence of the coupling approaches. We first observe that all functions $\Delta(x)$ are symmetric with respect to $x=3/2$, as expected. Moreover, the conclusions from the previous example all pertain in this case. 

\begin{figure}[tb!]
    \centering
    \includegraphics[width=0.485\textwidth]{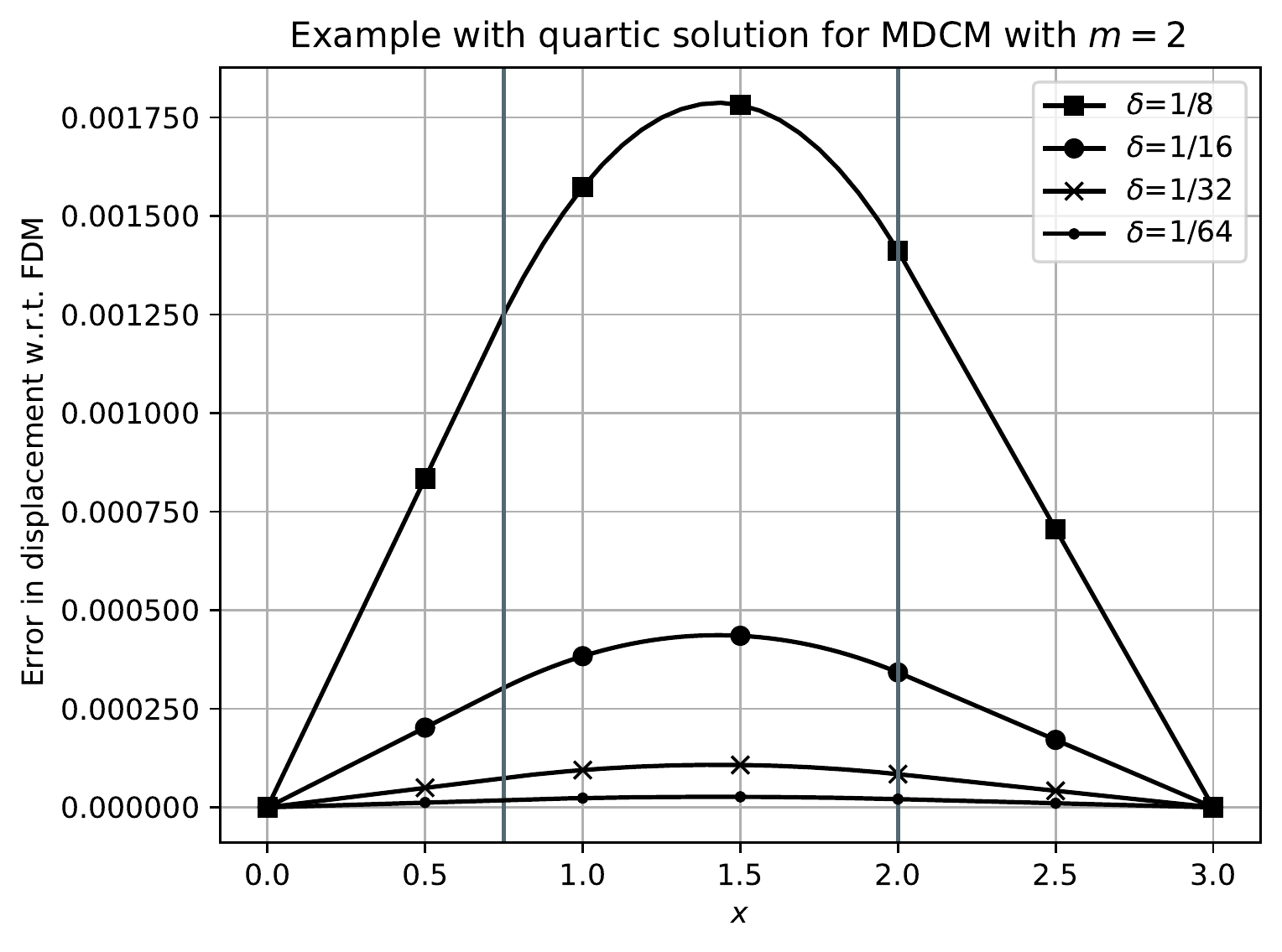}
    \hfill
    \includegraphics[width=0.485\textwidth]{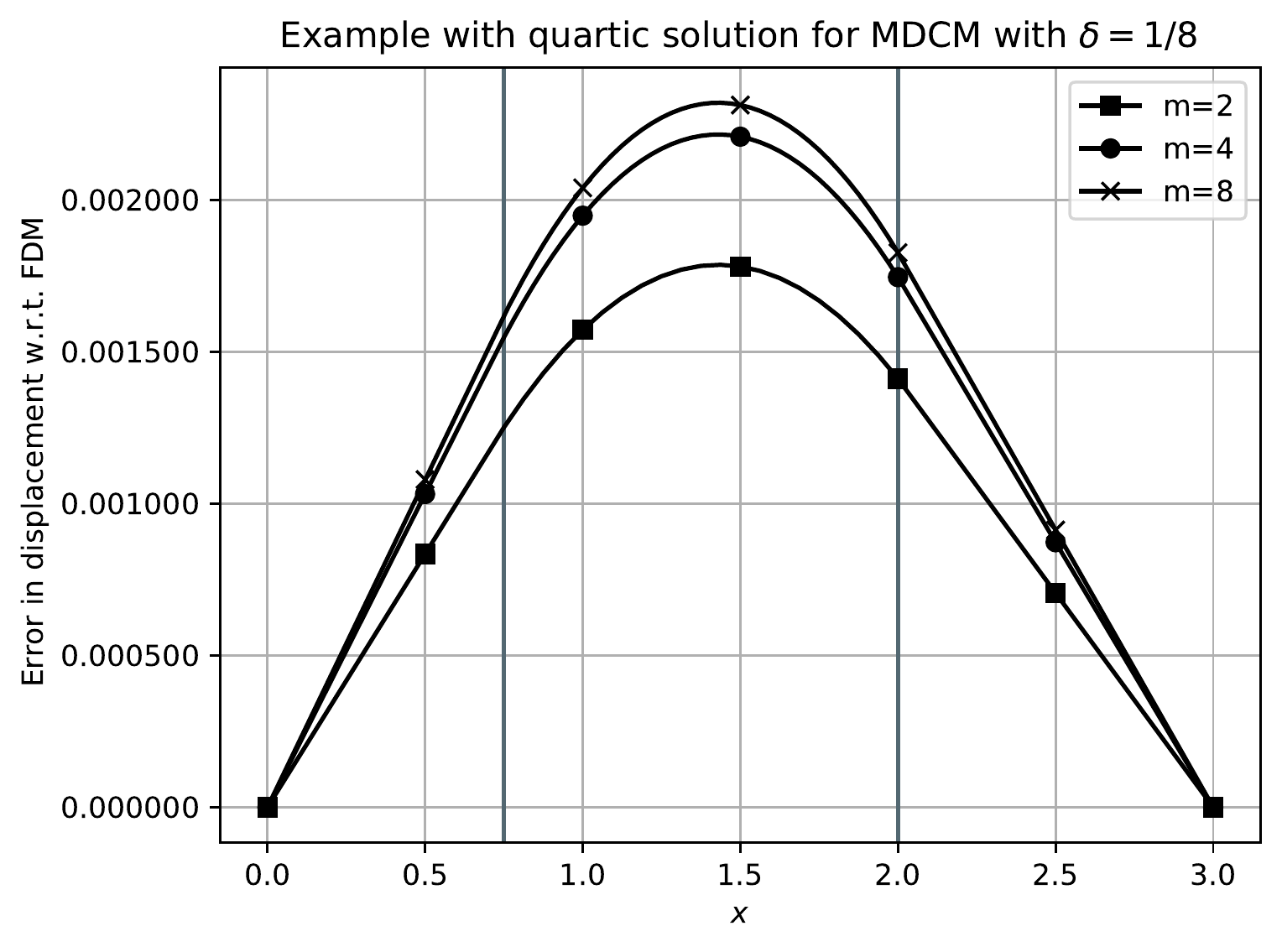}
    \\
    \includegraphics[width=0.485\textwidth]{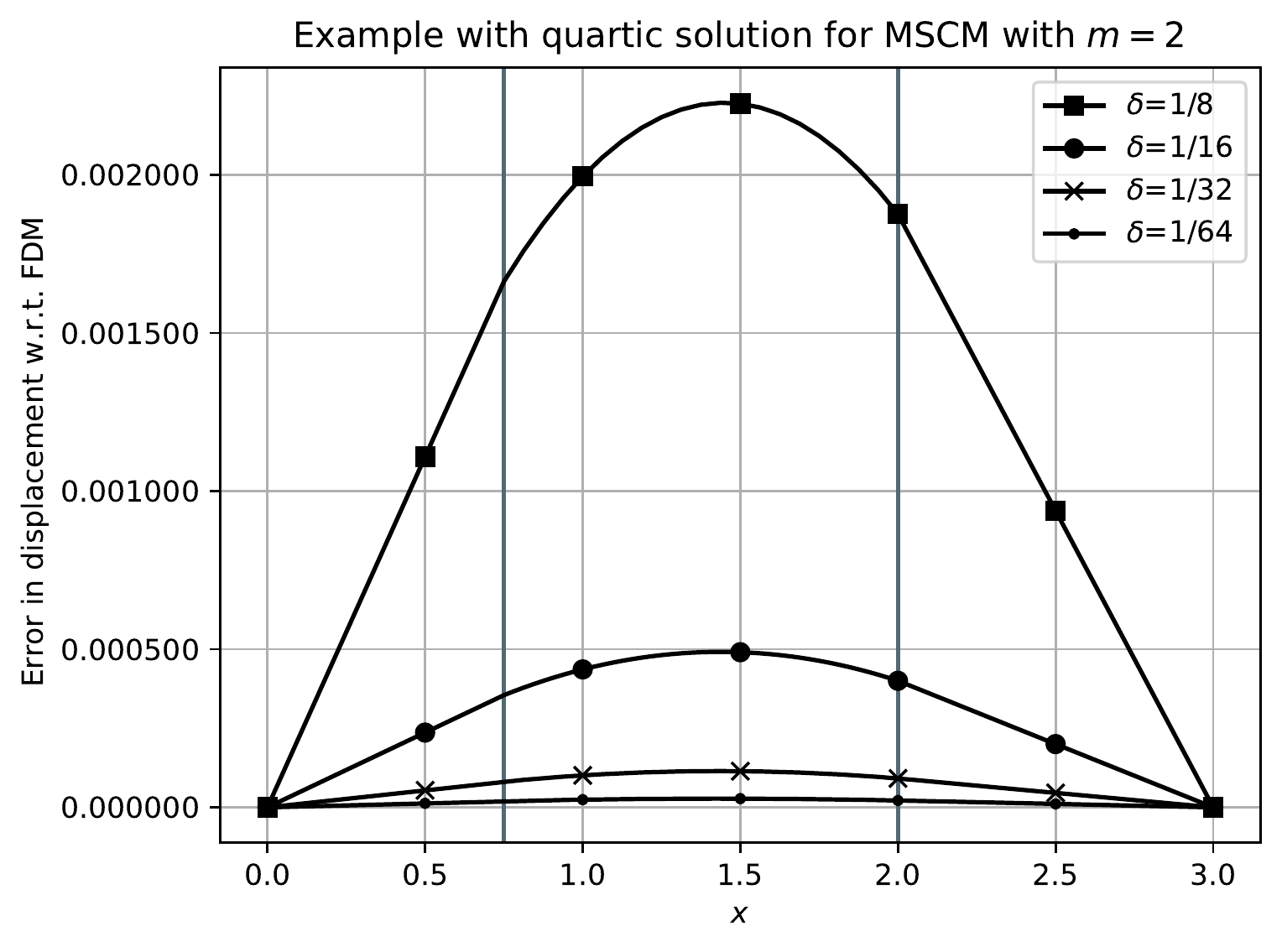}
    \hfill
    \includegraphics[width=0.485\textwidth]{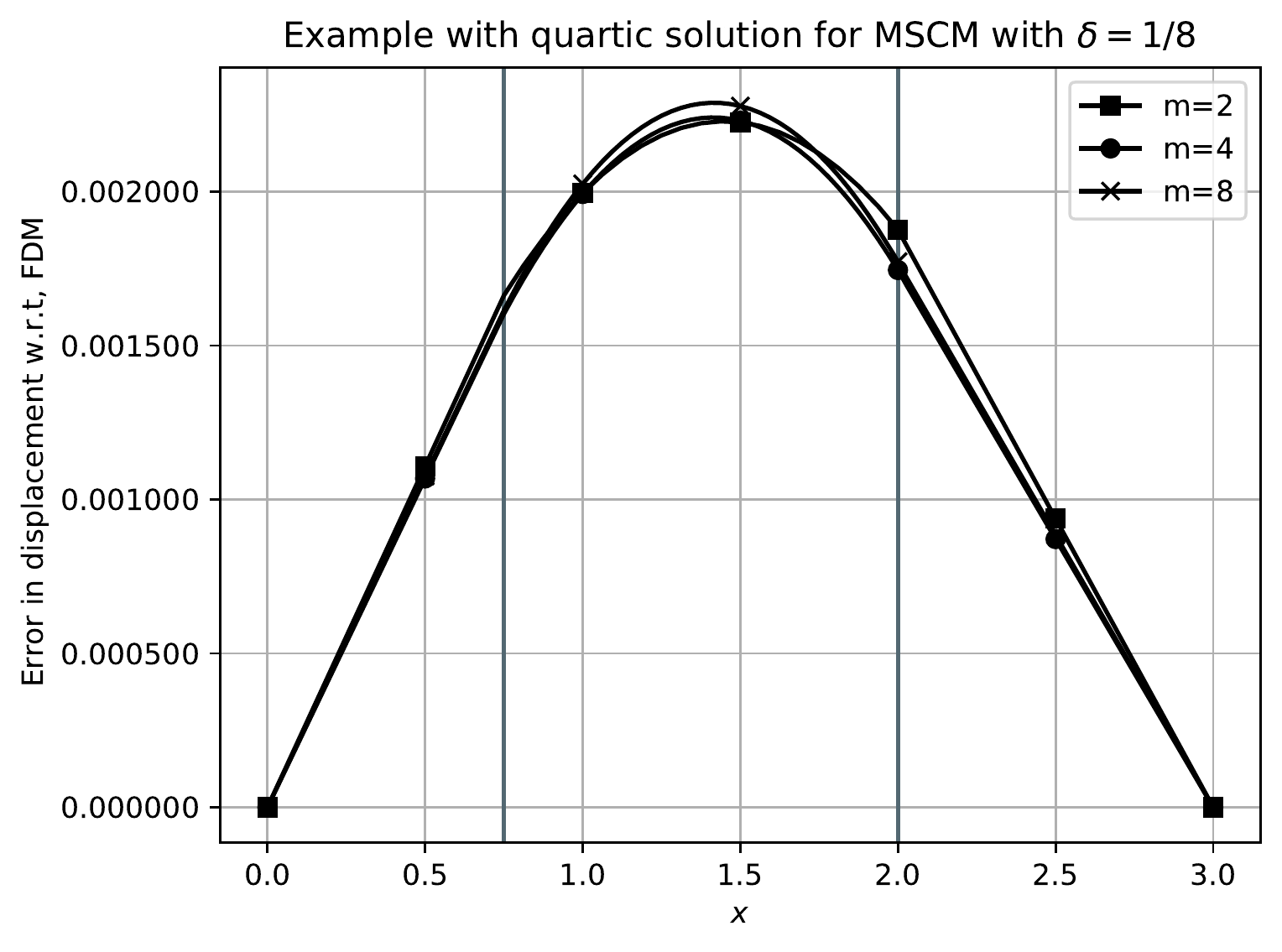}    
    \\
    \includegraphics[width=0.485\textwidth]{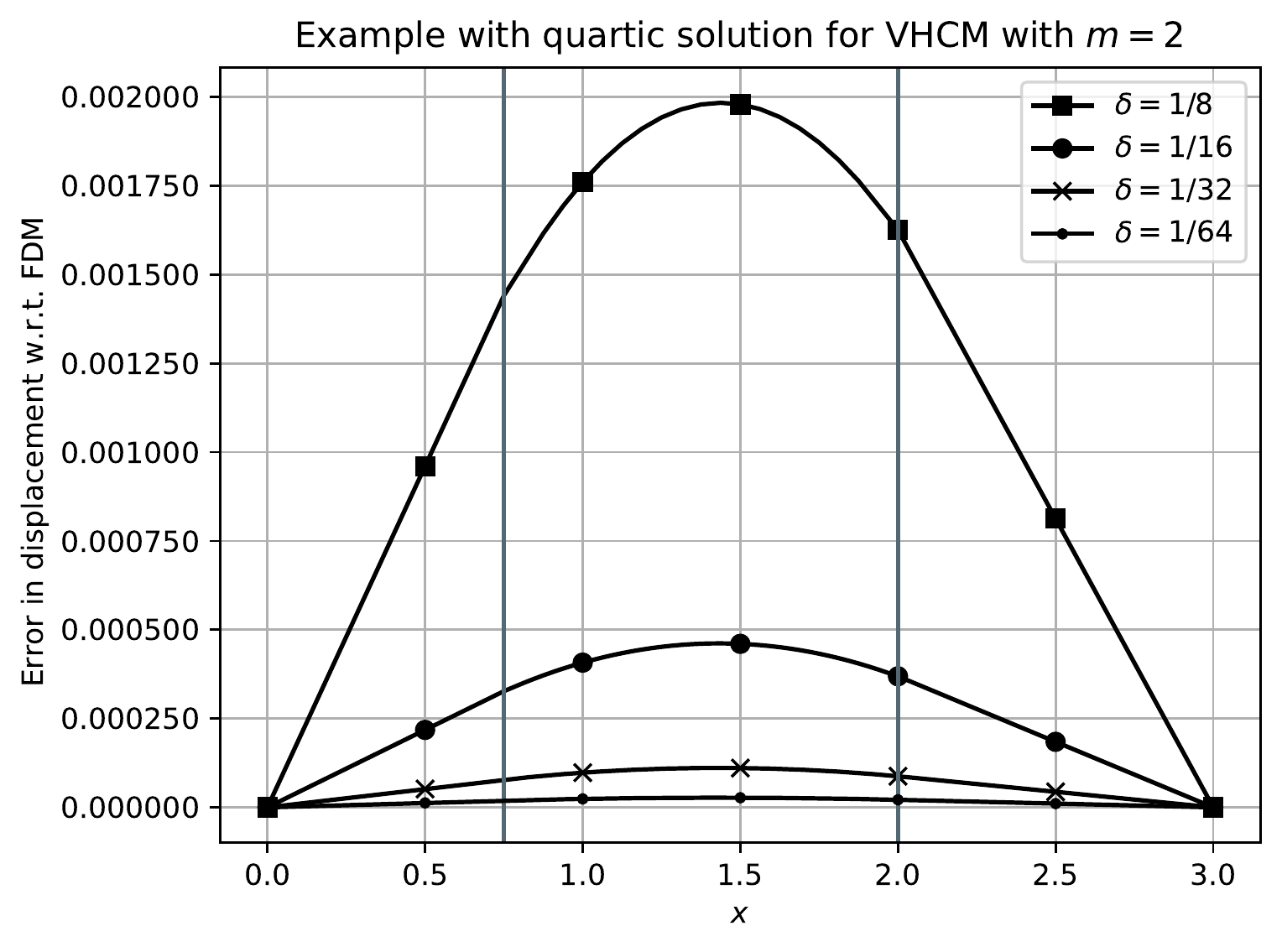}
    \hfill
    \includegraphics[width=0.485\textwidth]{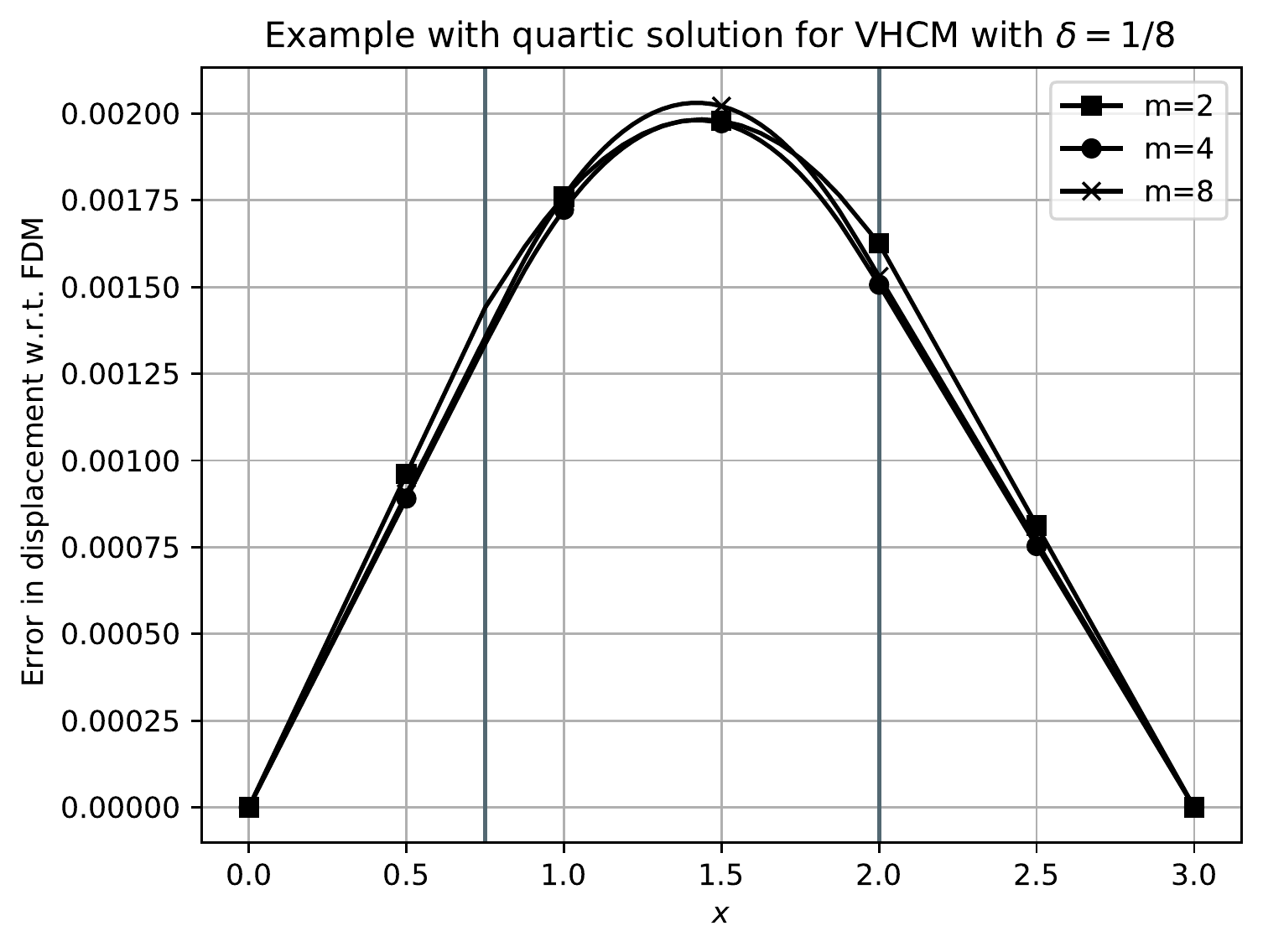}
    \caption{Error $\Delta(x)$ obtained by the three coupling methods (top row: MDCM, middle row: MSCM, and bottom row: VHCM) in the case of the quartic solution for the problem with Dirichlet boundary conditions at both extremities and interface locations $a=3/4$ and $b=2$. Left column: $\delta$-convergence with $m=2$. Right column: $m$-convergence with $\delta=1/8$.}
    \label{fig:ADBC-quartic-error}
\end{figure}

We now repeat the same exercise in the case of an asymmetric problem by moving the first interface to $a=3/4$ while keeping the second interface located at $b=2$. We again solve the problem with homogeneous Dirichlet boundary conditions at both ends defined such that it is satisfied by the manufactured solution~\eqref{eq:DBC-manufacturedsol}. The values of $v_{D,\max}$ for this new configuration are computed from their analytic expressions and shown in Table~\ref{tab:ADBC-quartic-error}. The expressions for $v_{D,\max}$ are omitted here due to their complexity, but the maximal value is attained for MDCM and MSCM at:
\begin{equation}
    x_{D,\max} 
    = \frac{2b\ell+a^2-b^2}{2\ell} \frac{137}{96} 
    \approx 1.427,
\end{equation}
and for VHCM at:
\begin{equation}
    x_{D,\max} 
    = \frac{2b\ell+a^2-b^2}{2\ell} + 2\delta \frac{a+b-\ell}{3\ell} 
    = \frac{137}{96} - \frac{\delta}{18}
    \approx 1.427 - 0.056 \times \delta.
\end{equation}
The numerical results for $\Delta_\text{max}$ and $\mathcal E_r$ are reported in Table~\ref{tab:ADBC-quartic-error} and plots of the function $\Delta(x)$ are shown in Figure~\ref{fig:ADBC-quartic-error}. We observe that the three coupling methods still exhibit a quadratic convergence with respect to the parameter $\delta$ and that the relative errors are similar to the previous case. However, we note that in the case $\delta=1/8$, see right column of Figure~\ref{tab:ADBC-quartic-error}, the function $\Delta(x)$ for MSCM and VHCM does not change much and its maximal value remains close to $v_{D,\max}$ as $m$ is increased, unlike for MDCM. In other words, the two coupling methods are less sensitive to the mesh size $h$ for a given value of $\delta$.



\begin{table}[tb!]
    \centering
    \sisetup{round-mode=places,round-precision=7,group-separator={}}
    \begin{tabular}{cc|ccc|ccc}
       & &  \multicolumn{3}{c|}{Error $\Delta_\text{max}$} & \multicolumn{3}{c}{Relative error $\mathcal E_r$} \\
       \midrule
       $\delta$ & $m$ & MDCM & MSCM & VHCM & MDCM & MSCM & VHCM  \\ 
       \midrule
       $\sfrac{1}{8}$ & 2  & \num{0.001540090595820276} & \num{0.0020472} & \num{0.0017665491607632422} &  \nums{0.2016170351267689} & \nums{.06127025368833336 } & \nums{0.06434373068048209 } \\
       & 4  & \num{0.0019048613622223431} & \num{0.0020226} & \num{0.0016349767907741786} &  \nums{0.012493519958} & \nums{0.04850043665385476} & \nums{0.014928462949345231} \\
       & 8  & \num{0.0019929165766250367} & \num{0.0020258} & \num{0.0016605351421095094} &  \nums{0.0331279533224191} & \nums{0.05016550895438604} & \nums{0.0004704127878776099 } \\
       \cmidrule{2-5}
       \multicolumn{2}{r|}{$v_{D,\max}$} & \num{0.001929012346} & \num{0.001929012346} & \num{0.001659754372} \\
       \midrule
       $\sfrac{1}{16}$ & 2  & \num{0.00037336835222134823} & \num{0.0004367} & \num{0.00040207528619728983} &  \nums{0.22578338483381227} & \nums{0.0944365624484305} & \nums{0.10520561104488288} \\
       & 4  & \num{0.000464187726735954} & \num{0.0004789} & \num{0.00043096906241935073} &  \nums{0.03746032984032576} & \nums{0.006970876479976996} & \nums{0.040904248273387206} \\
       & 8  & \num{0.0004864986200558974} & \num{0.0004906} & \num{0.0004455023757410359} &  \nums{0.008803538547908885} & \nums{0.017310979031270848} & \nums{0.00856123277456941} \\
       \cmidrule{2-5}
       \multicolumn{2}{r|}{$v_{D,\max}$} & \num{0.000482253086} & \num{0.000482253086} & \num{0.000449349360} \\
       \midrule
       $\sfrac{1}{32}$ & 2  & \num{0.0000919} & \num{0.0000998} & \num{0.0000955} &  \nums{0.23788540718167037} & \nums{0.17223620186125568} & \nums{0.1802703700196417} \\
       & 4  & \num{0.00011453899486557617} & \num{0.0001164} & \num{0.0001104} &  \nums{0.049967760986964954} & \nums{0.034728212000686} & \nums{0.05217655282362489 } \\
       & 8  & \num{0.00012015364556527963} & \num{0.0001207} & \num{0.0001151} &  \nums{0.0033976022233445646} & \nums{0.0008531730541790794} & \nums{0.012307868862550942} \\
       \cmidrule{2-5}
       \multicolumn{2}{r|}{$v_{D,\max}$} & \num{0.000120563272} & \num{0.000120563272} & \num{0.000116497401} \\
       \midrule
       $\sfrac{1}{64}$ & 2  & \num{0.0000228} & \num{0.0000238} & \num{0.0000232} &  \nums{0.24394105977135036} & \nums{0.21112255718462616} & \nums{0.21575674921153315} \\
       & 4  & \num{0.0000284} & \num{0.0000287} & \num{0.0000279} &  \nums{0.05623237116469676} & \nums{0.048612301320281354} & \nums{0.057449163616189465} \\
       & 8  & \num{0.0000299} & \num{0.0000299} & \num{0.0000292} &  \nums{0.009505537756285721} & \nums{0.007383440771263794} & \nums{0.014016858778185955} \\
       \cmidrule{2-5}
       \multicolumn{2}{r|}{$v_{D,\max}$} & \num{0.000030140818} & \num{0.000030140818} & \num{0.000029635527} \\
       \midrule
    \end{tabular}
    \caption{Error $\Delta_\text{max}$ and relative error $\mathcal E_r$ with respect to parameters $\delta$ and $m$ obtained by the three coupling methods in the case of the quartic solution for the problem with Dirichlet boundary conditions at both extremities and interface locations $a=1$ and $b=2$.}
    \label{tab:SDBC-quartic-error}
\end{table}

\begin{table}[tb!]
    \sisetup{round-mode=places,round-precision=7,group-separator={}}
    \centering
    \begin{tabular}{cc|ccc|ccc}
       & &  \multicolumn{3}{c|}{Error $\Delta_\text{max}$} & \multicolumn{3}{c}{Relative error $\mathcal E_r$} \\
       \midrule
       $\delta$ & $m$ & MDCM & MSCM & VHCM & MDCM & MSCM & VHCM  \\ 
       \midrule
       $\sfrac{1}{8}$ & 2  & \num{0.0017871} & \num{0.0022283} & \num{0.0019833} &  \nums{0.21438198477928305} & \nums{0.02043975351236474} & \nums{0.027478500421319264} \\
       & 4  & \num{0.0022163} & \num{0.0022409} & \num{0.0019805} &  \nums{0.025729017096367362} & \nums{0.014897390530104738} & \nums{0.02885725875859895} \\
       & 8  & \num{0.0023208} & \num{0.0022886} & \num{0.0020310} &  \nums{0.02023648049134907} & \nums{0.006087077080257216} & \nums{0.004068147175928762} \\
       \cmidrule{2-5}
       \multicolumn{2}{r|}{$v_{D,\max}$} & \num{0.002274794507} & \num{0.002274794507} & \num{0.002039324018} & & \\
       \midrule
       $\sfrac{1}{16}$ & 2  & \num{0.0004366} & \num{0.0004917} & \num{0.0004616} &  \nums{0.2321981894459193} & \nums{0.1353091784039366} & \nums{0.1452483176298157} \\
       & 4  & \num{0.0005436} & \num{0.0005466} & \num{0.0005147} &  \nums{0.044084026844382736} & \nums{0.03878637756285524} & \nums{0.046830792425876885} \\
       & 8  & \num{0.0005700} & \num{0.0005659} & \num{0.0005344} &  \nums{0.0023667456336188904} & \nums{0.004869687299632944} & \nums{0.010456991628599597} \\
       \cmidrule{2-5}
       \multicolumn{2}{r|}{$v_{D,\max}$} & \num{0.000568698627} & \num{0.000568698627} & \num{0.000540013685}\\
       \midrule
       $\sfrac{1}{32}$ & 2  & \num{0.0001079} & \num{0.0001148} & \num{0.0001110} &  \nums{0.24122767771286374} & \nums{0.19284022571017426} & \nums{0.1988957873557046} \\
       & 4  & \num{0.0001346} & \num{0.0001350} & \num{0.0001310} &  \nums{0.05343779881961292} & \nums{0.05084016380347633} & \nums{0.0547183990478962} \\
       & 8  & \num{0.0001412} & \num{0.0001407} & \num{0.0001368} &  \nums{0.006796241671615282} & \nums{0.010451276475890435} & \nums{0.012938404917037115} \\
       \cmidrule{2-5}
       \multicolumn{2}{r|}{$v_{D,\max}$} & \num{0.000142174657} & \num{0.000142174657} & \num{0.000138635843}\\
       \midrule
       $\sfrac{1}{64}$ & 2  & \num{0.0000268} & \num{0.0000277} & \num{0.0000272} &  \nums{0.24461012951104147} & \nums{0.2203915453556938} & \nums{0.22478469604961734} \\
       & 4  & \num{0.0000335} & \num{0.0000335} & \num{0.0000330} &  \nums{0.05672545870994404} & \nums{0.05543066589550721} & \nums{0.05867943995351454} \\
       & 8  & \num{0.0000351} & \num{0.0000351} & \num{0.0000346} &  \nums{0.009900291732917996} & \nums{0.011742162479606442} & \nums{0.014324200994094661} \\
       \cmidrule{2-5}
       \multicolumn{2}{r|}{$v_{D,\max}$} & \num{0.000035543664} & \num{0.000035543664} & \num{0.000035104238}\\
       \midrule
    \end{tabular}
    \caption{Error $\Delta_\text{max}$ and relative error $\mathcal E_r$ with respect to parameters $\delta$ and $m$ obtained by the three coupling methods in the case of the quartic solution for the problem with Dirichlet boundary conditions at both extremities and interface locations $a=3/4$ and $b=2$.}
    \label{tab:ADBC-quartic-error}
\end{table}

\subsection{Influence of the peridynamic material parameter}

We have seen from the last examples that the coupling methods cannot reproduce quartic solutions (or solutions that involve polynomials of degree at least four) to the classical linear elasticity model in the case of finite values of the horizon $\delta$, even if the mesh size $h$ goes to zero, due to the fact that the choice of the material parameter, i.e.\ $\kappa=2E/\delta^2$, provides a perfectly compatible peridynamic model only in the limit when $\delta$ vanishes. \edit{This theoretical result is certainly well known within the peridynamics community. However, we emphasize this point again as it does influence the construction and performance of the coupling approaches.} 

\begin{figure}[tb!]
    \centering
    \includegraphics[width=0.485\textwidth]{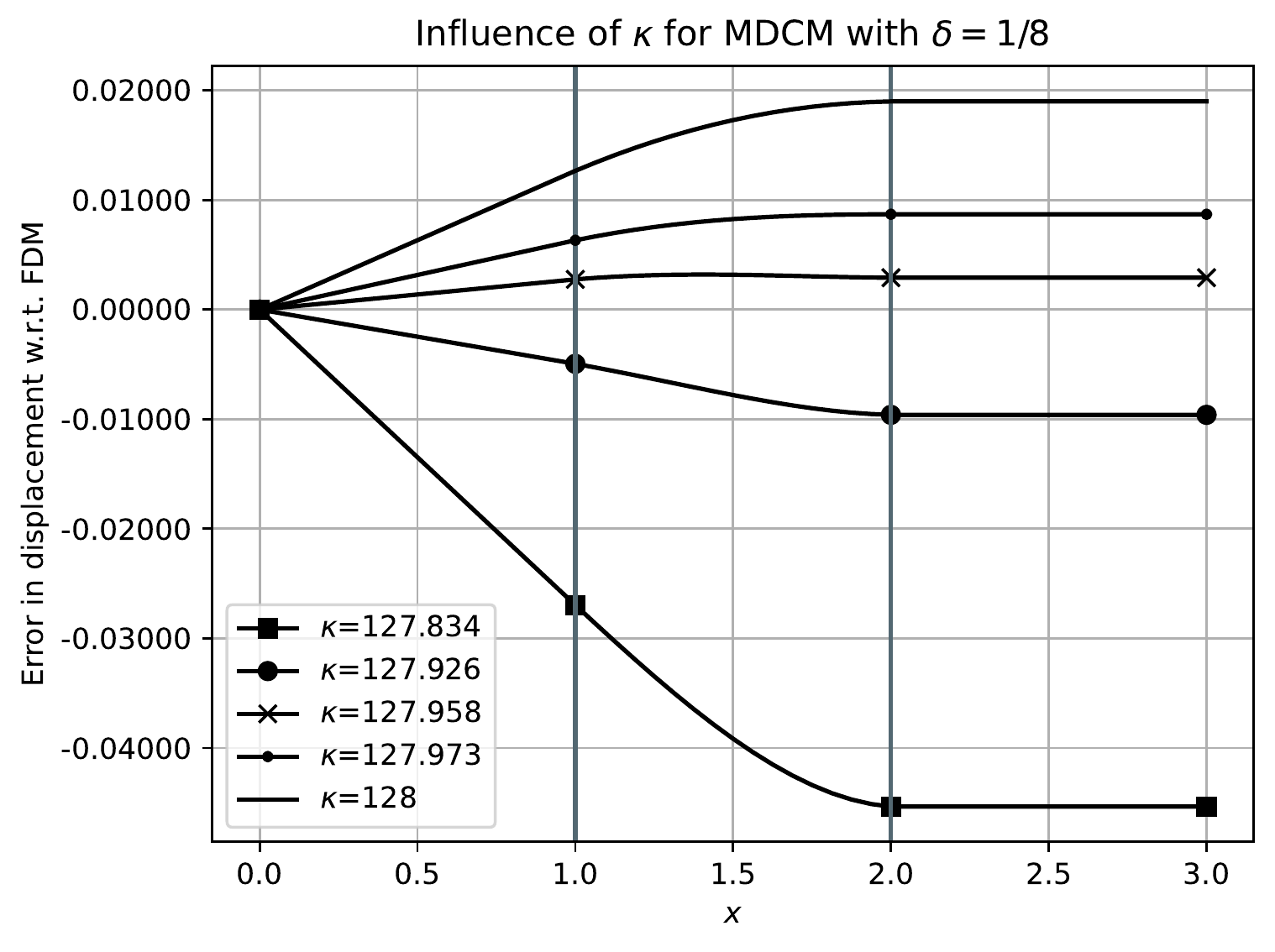}
    \hfill
    \includegraphics[width=0.485\textwidth]{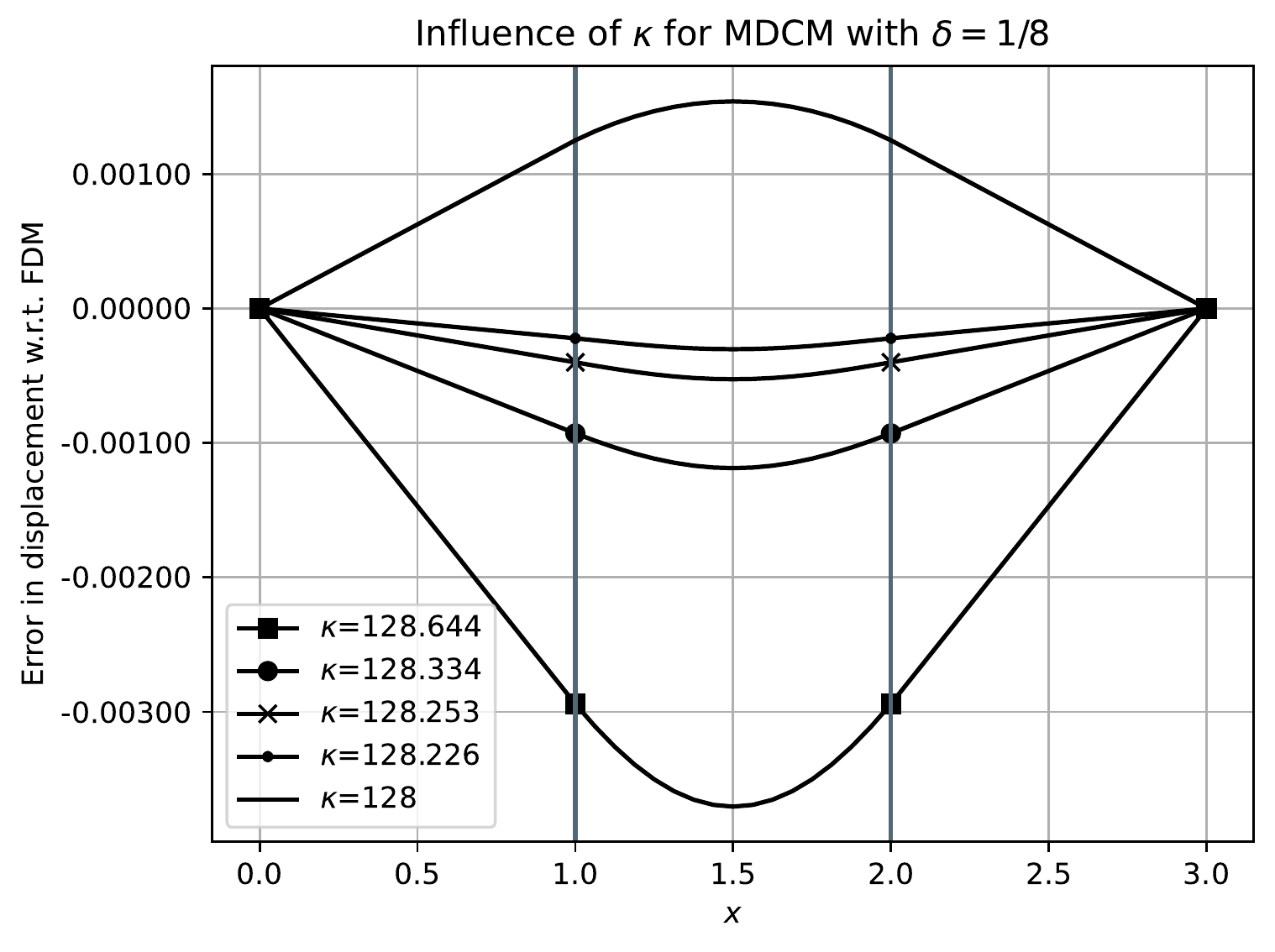}
    \\
    \includegraphics[width=0.485\textwidth]{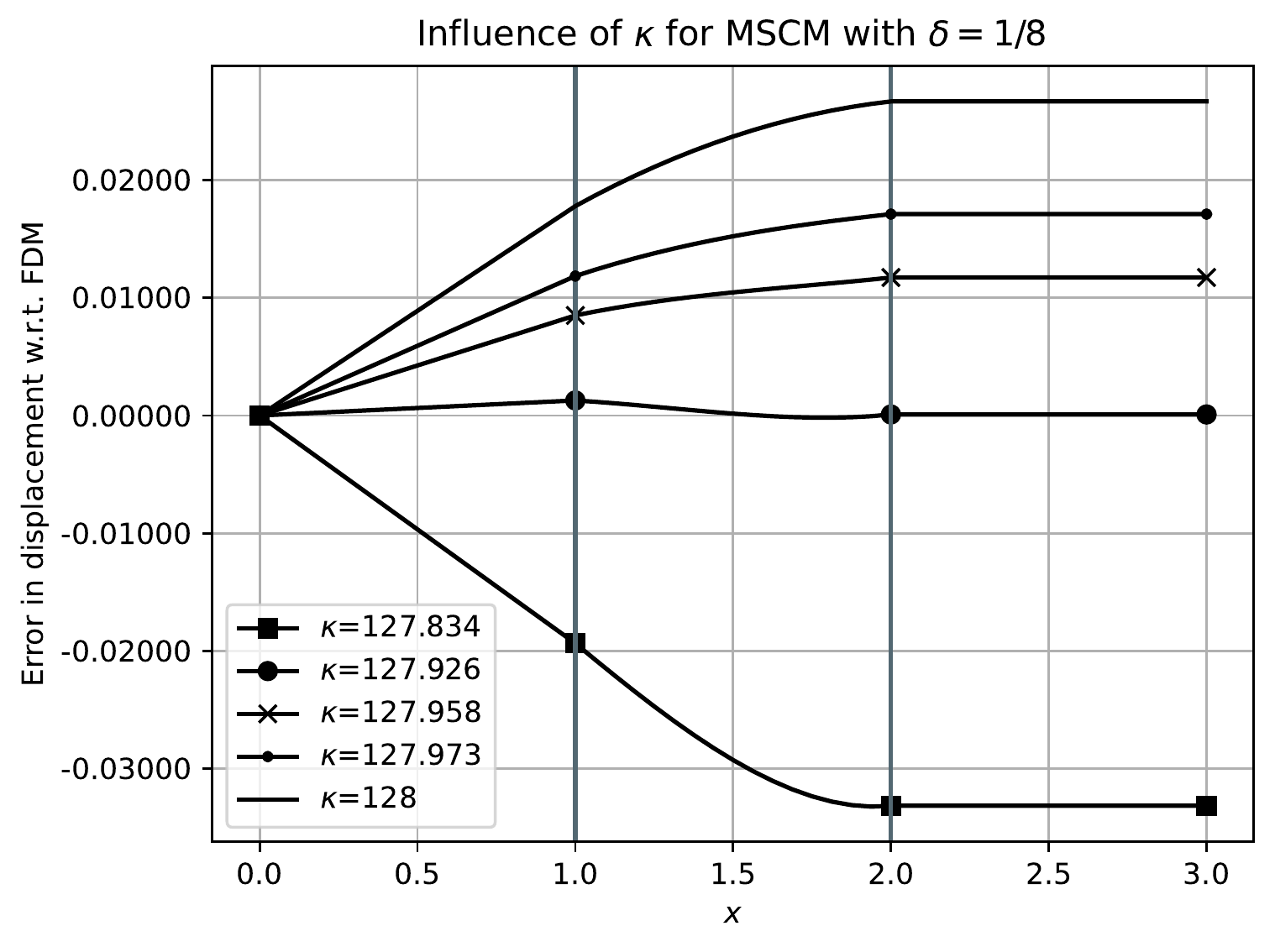}
    \hfill
    \includegraphics[width=0.485\textwidth]{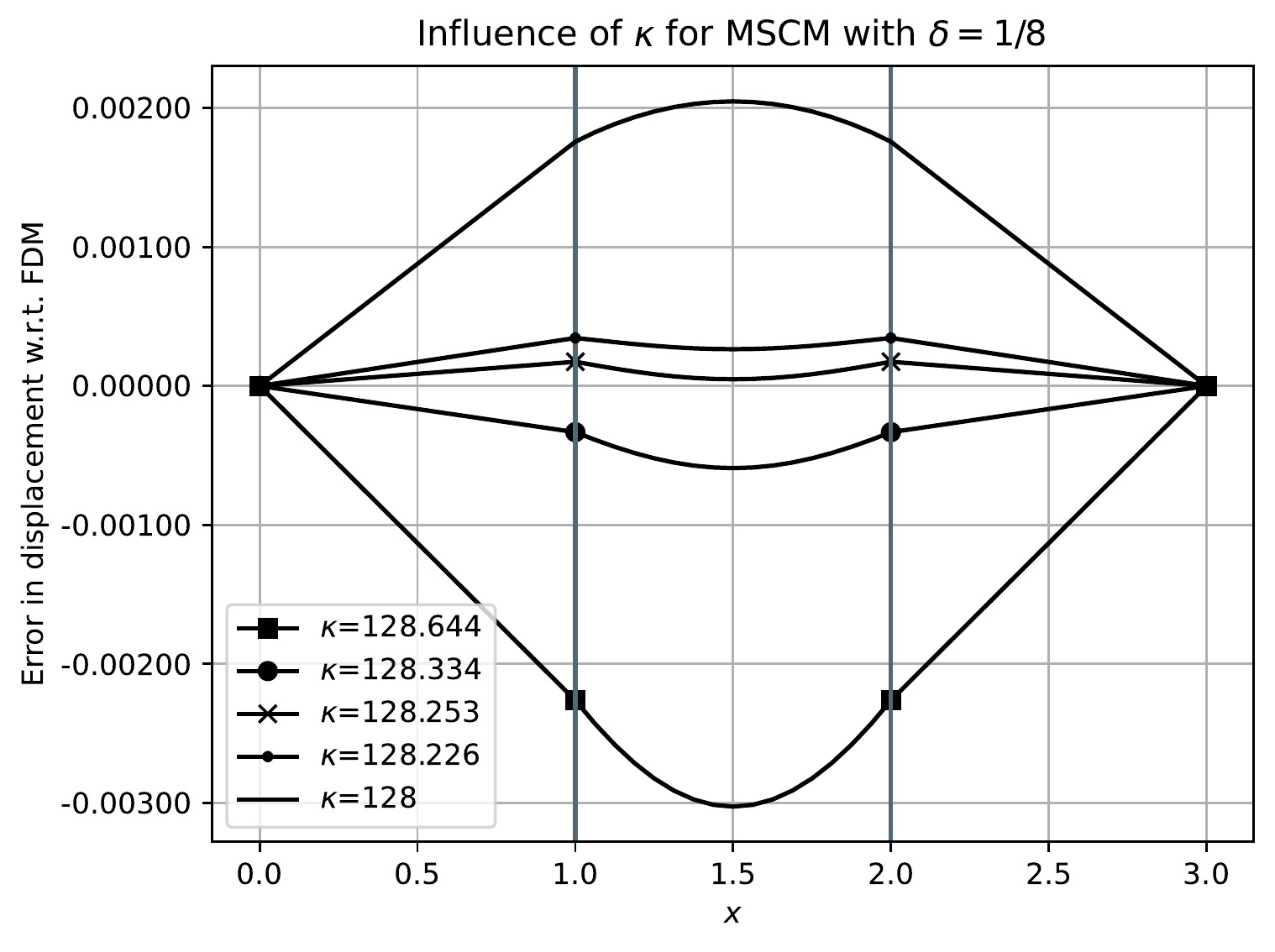}
    \caption{Influence of material parameter $\kappa$ on the error function $\Delta(x)$ for the coupling methods MDCM (top row) and MSCM (bottom row) in the case of the quartic solutions for the problem with mixed boundary conditions (left) and for the problem with Dirichlet boundary conditions at both extremities (right). The parameters of the simulations are $m=2$, $\delta=1/8$, $a=1$, $b=2$, and $\ell=3$.}
    \label{fig:kappa}
\end{figure}

We observe in Figures~\ref{fig:SNBC-quartic-error}, \ref{fig:SDBC-quartic-error}, and~\ref{fig:ADBC-quartic-error} that the error $\Delta(x)$ remains small in all cases when compared to the solution $\uelast(x)$, in the sense that it approximates the modeling error $v(x)$, which can be shown to be of the order $E\uelast''''\delta^2/24$, see Eq.~\eqref{eq:function-v}.
The challenge here is whether one could identify a value of $\kappa$ that decreases the modeling error. In fact, \edit{it is well known} that it is not possible to find a value that would allow one to exactly match the solution to the classical linear elasticity model as the material parameter would need in that case to depend on~$x$. We nevertheless test several values of the parameter, chosen as corrections of order $\delta^2$ of the nominal value $\kappa=2E/\delta^2$. We take here $m=2$ and $\delta=1/8$, so that the nominal value is given by $\kappa=128$. We consider the usual configuration with $\ell=3$ and interface locations $a=1$ and $b=2$. We show in Figure~\ref{fig:kappa} the error $\Delta(x)$ associated with the quartic solutions to the problems with mixed boundary conditions and with Dirichlet boundary conditions using MDCM and MSCM for several values of $\kappa$, including the nominal value. The main conclusion is that it is indeed possible to find a value of the parameter for which the error is overall much smaller than that with the nominal value. However, we also observe that the best value of $\kappa$ is very much dependent on the solution itself and the coupling method used. Moreover, the error is quite sensitive to the value of $\kappa$.

\subsection{Condition number of the coupled systems}

\begin{figure}
    \centering
    \includegraphics[width=0.485\textwidth]{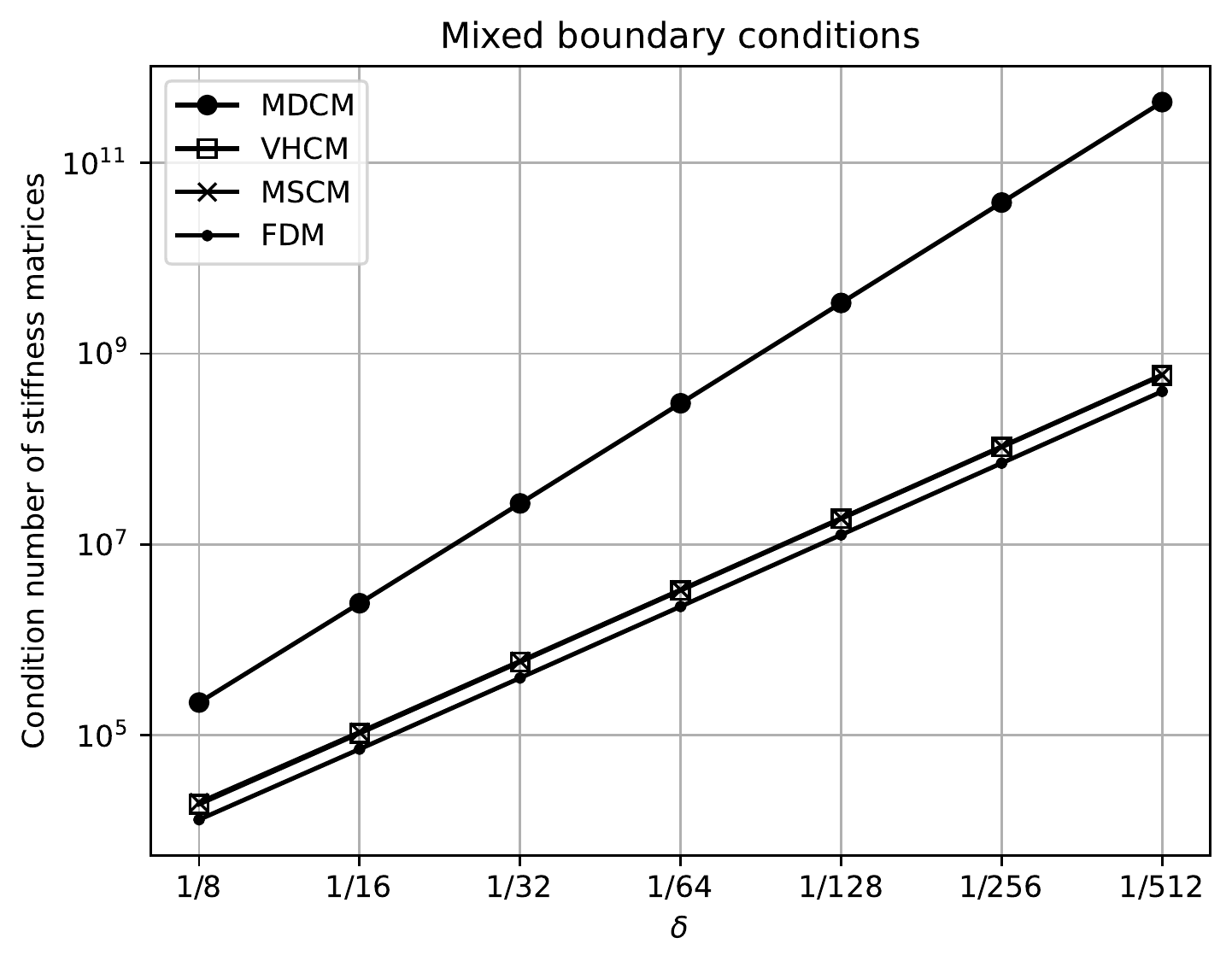}
    \hfill
    \includegraphics[width=0.485\textwidth]{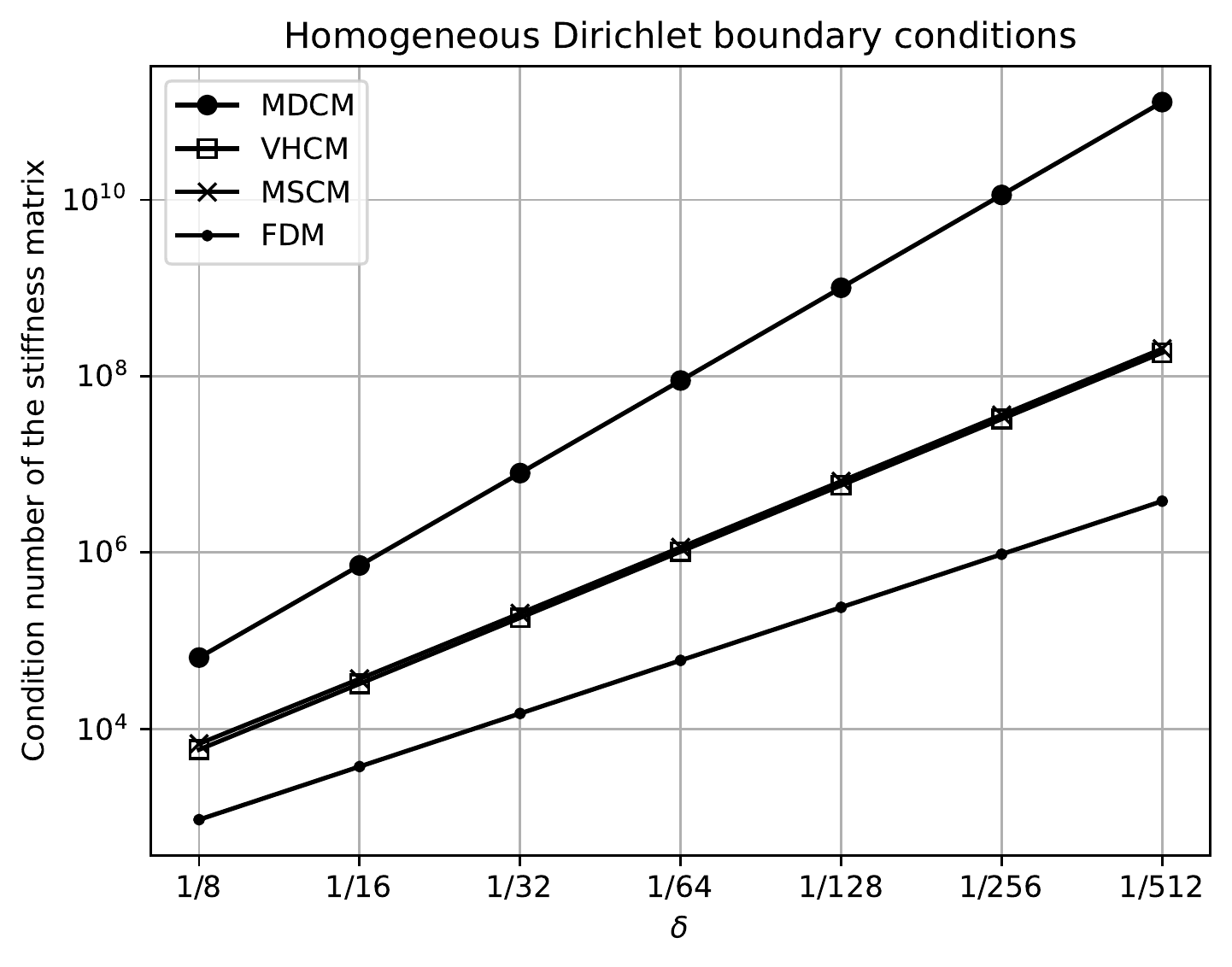}
    \caption{\edit{Condition number of the stiffness matrices with respect to the horizon $\delta$ for the problem with mixed boundary conditions (left) and the problem with homogeneous Dirichlet boundary conditions (right).}}
    \label{fig:con}
\end{figure}

\edit{We analyze in this section the condition number of the stiffness matrices $M$ resulting from the coupling methods. We used the Numpy function~\texttt{numpy.linalg.con}\footnote{\url{https://numpy.org/doc/stable/reference/generated/numpy.linalg.cond.html}} to compute the condition number $\text{Cond}(M)$ for each coupling approach. The condition number is computed here with respect to the $\ell^2$-norm as $\text{Cond}(M)={\| M \|_{\ell_2}}{\| M^{-1}\|_{\ell_2}}$~\cite{lay2016linear}. Figure~\ref{fig:con} shows the results for the two problems considered earlier, namely the problem with mixed boundary conditions and that with homogeneous Dirichlet boundary conditions. The only difference between the two problems lies in the last row of the matrix, which implements either the Neumann or Dirichlet boundary condition. We observe that, for both problems, the condition number associated with MDCM is larger than the ones obtained for MSCM, VHCM, or the Finite Difference Method (FDM) applied to~\eqref{eq:1dlinearelasticity}. For small values of $\delta$, the condition number can be two orders of magnitude or more. From a numerical point of view, systems of equations with lower condition numbers should be favored.}

\section{Conclusions}
\label{Sect:conclusions}

We have presented and compared in this paper three methods, namely the coupling method with matching displacements (MDCM), the coupling method with matching stresses (MSCM), and the variable horizon coupling method (VHCM), for coupling classical linear elasticity and peridynamic models. The methods were developed based on the general coupling formulation of classical linear elasticity models. We have provided the continuous formulations of the three corresponding problems and their corresponding discrete formulations using for instance the classical finite difference method. One challenge in comparing the three methods was to determine an adequate measure to assess the performance of each method. We have proposed to evaluate the maximum value of the difference between the solution of each coupling method and the solution to the classical linear elasticity problem. This difference actually corresponds to the modeling error between the two models. 

We have shown on one-dimensional examples with cubic solutions that the three methods were able to recover the exact solution of the classical linear elasticity problem as predicted by the theory. It follows that the interesting numerical test cases for the purpose of comparison are those that involve polynomial functions of degree at least four. We have therefore considered several one-dimensional examples whose manufactured solutions were defined in terms of quartic polynomial functions. The $\delta$- and $m$-convergence results have shown that the three coupling methods provide comparable errors when $\delta$ tends to zero, i.e.\ when the modeling error between the classical linear elasticity and peridynamic models become very small. 
\edit{However, the results show that the matching stress approach (MSCM) is less sensitive to the mesh size than the matching displacement approach (MDCM) whenever the value of the horizon parameter is not so small.}
This is an important result as most coupling approaches from the literature constrain the displacement fields to match in the coupling region. We have also observed that VHCM usually exhibits a similar behavior as MSCM. The original feature of VHCM is that, unlike the other two methods, it avoids introducing an overlap region between the two models. Moreover, we have indicated for finite values of the horizon $\delta$ that it was possible to reduce the modeling error by considering corrections in parameter $\delta$ of the order of~$\delta^2$.

\edit{The interfaces between the classical linear elasticity and peridynamic models in a coupling strategy should be preferably located in the regions where the solution to the problem has a local behavior. These regions can often be identified a priori in the case of steady-state problems but this could become trickier when considering dynamical systems.} In this case, one should choose a coupling approach that generates the smallest errors in order to avoid wave reflection phenomena at the interfaces. One could also appeal to error estimation and adaptivity to automatically determine the regions where the two models should be coupled, see e.g.~\cite{Prudhomme-Chamoin-2009,Bauman-Oden-Prudhomme-2009,BenDhia-Chamoin-2011}. This is left for future work. One also notes that MSCM could be used as an alternative candidate for the application of boundary conditions in peridynamic problems to those presented in~\cite{Prudhomme-Diehl-2020}. Nevertheless, this is a preliminary study of coupling methods and additional work needs to be carried out to confirm the performance of the three methods in more general situations. The performance of the coupling approaches should be assessed, for instance, when using other discretization methods, such as the finite element method, for their implementation. The methods should also be extended to two- and three-dimensional problems and to the case of state-based peridynamic modeling.

\section*{Supplementary materials}
\noindent
The Python code (using numpy~\cite{oliphant2006guide,5725236} and matplotlib~\cite{Hunter:2007}) used to generate the numerical results is available on GitHub\textsuperscript{\textregistered}\footnote{\url{https://github.com/diehlpk/paperCouplingAnalysis}} and on Zenodo\textsuperscript{\textregistered}~\cite{patrick_diehl_2021_4779820}.

\section*{Acknowledgements}
\noindent
Patrick Diehl thanks the LSU Center of Computation \& Technology at Louisiana State University for supporting this work.
Serge Prudhomme is grateful for the support by a Discovery Grant from the Natural Sciences and Engineering Research Council of Canada [grant number RGPIN-2019-7154]. 

\section*{Conflict of interest}
The authors declare that they have no conflict of interest.


\bibliographystyle{plain}
\bibliography{bibfile.bib}


\end{document}